%% file: closest_anisotropic.tex
\documentclass [12pt]{article}%

\usepackage{color}
\usepackage{hyperref}
\usepackage{graphicx}

\usepackage{amssymb,amsmath}
\usepackage{gastex}

\usepackage[latin1]{inputenc} 
\usepackage[hang,small,bf]{caption}
\usepackage[normalem]{ulem} 

\allowdisplaybreaks[1]
\newcommand{\beq}[1]{  \\ {\tiny ({#1})}   \begin{equation} \label{#1} }
\renewcommand{\beq}[1]{  \begin{equation} \label{#1} }    	
\newcommand{\eeq}{\end{equation}}		 

\renewcommand{\appendix}{\setcounter{section}{0}\renewcommand{\thesection}{\Alph{section}}  			\section*{Appendix} 
}

\def\Appendix#1{
 			\setcounter{equation}{0}
 			\renewcommand{\theequation}{\thesection.\arabic{equation}}
 			\section{#1}
 			}
 			
\newcommand{\rf}[1]{(\ref{#1})}

\def\bd#1{\mbox{\boldmath$\displaystyle\mathbf{#1}$} }
\def\sb#1{\text{\sffamily\bfseries #1}}

\def\cal{\mathcal}
\def\hbf#1{\hat{\bf #1}}                        
\def\whbf#1{\widehat{\bf #1}}										
\def\mat#1{{\bf #1}}														
\def\tens#1{\mathbb{\,#1}}											
\def\ds#1{\displaystyle{#1}}
\def\tr{\operatorname{tr}} 
\def\sinhc{\operatorname{sinhc}}

\def\Log{\operatorname{Log}}
\def\dis{\operatorname{d}}
\def\diag{\operatorname{diag}}
\def\det{\operatorname{det}}			

\def\singlespacing{\baselineskip=13pt}

\begin{document} 

\pagestyle{myheadings}\markright{{\sc Moakher and Norris }  ~~~~~~\today}
\singlespacing

\title{The closest elastic tensor of arbitrary symmetry \\to an elasticity tensor of  lower symmetry}  
\author{Maher Moakher\footnote{Laboratory for Mathematical and Numerical Modeling in Engineering Science, National Engineering School at Tunis, ENIT-LAMSIN, B.P. 37, 1002 Tunis Belv\'{e}d\`{e}re, Tunisia, maher.moakher@enit.rnu.tn}
\and
Andrew N. Norris\footnote{Rutgers University, Department of Mechanical and Aerospace Engineering, 98 Brett Road, Piscataway, NJ  08854-8058, norris@rutgers.edu}}
\maketitle

\begin{abstract}

 The closest tensors of  higher symmetry classes are derived in explicit form for a given  elasticity tensor of arbitrary symmetry.  The mathematical  problem is to minimize the elastic length or distance between the given tensor and the closest elasticity tensor of the specified symmetry.  
 Solutions are presented for  three  distance functions, with particular attention to the Riemannian and log-Euclidean distances.  These yield solutions  that are invariant under inversion, i.e., the same whether elastic stiffness or compliance are considered.   The Frobenius distance function, which corresponds to common notions of Euclidean length, is not invariant although it is simple to apply using projection operators.   A complete description of the Euclidean projection method is  presented.  
The  three metrics are considered at a level of detail far greater than heretofore, as we develop the general framework  to best fit a given set of moduli onto higher elastic symmetries.   The procedures for finding the closest elasticity tensor are illustrated by application to a set of 21 moduli with no underlying symmetry. 

\end{abstract}
%

\section{Introduction}\label{intro}

We address the question of finding the  elastic  moduli with a given material symmetry closest to an arbitrary set of elastic constants.  
There are several  reasons for reducing a set of  elastic constants in this way.  One might desire to fit a data set to  the {\it a priori} known symmetry of the material.  Alternatively, a seismic simulation might be best understood in terms of a model of the earth as a layered transversely isotropic medium, even though core samples indicate local anisotropy of a lower symmetry.  More commonly,  one might simply want to reduce the model complexity by decreasing the number of elastic parameters.  In each case a distance function  measuring  the difference between sets of elastic moduli is necessary to define an appropriate  closest set.  The most natural metric is  the  Euclidean norm in which the ``length" $\|{\tens C}\|$ of a set of moduli\footnote{Lower case Latin suffices take on the values 1, 2, and 3, and
the summation convention on repeated indices is assumed.} $C_{ijkl}$ is 
$\|{\tens C}\| = (C_{ijkl}C_{ijkl})^{1/2}$.   The Euclidean distance function is, however, not invariant under inversion of the elasticity tensor - one obtains a different result using compliance as compared with stiffness.   Alternative distance functions have been proposed which do not have this failing.  In particular, the Riemannian distance function of Moakher \cite{Moakher06} and the log-Euclidean metric of Arsigny {\it et al.\/}  \cite{Arsigny:MICCAI:05} are invariant under inversion, and as such are preferred general elastic norms.   


The problem of simplifying elastic moduli by increasing the elastic symmetry was apparently first 
considered by Gazis {\it et al.\/} \cite{Gazis63}.  They provided a general framework for defining and deriving the Euclidean projection, i.e., the closest elasticity using the Euclidean norm.  Several examples were given, including the closest isotropic material, which agrees with the isotropic approximant derived by Fedorov \cite{fed}.  Fedorov obtained   isotropic moduli using a different criterion: the isotropic material that best approximates the elastic wave velocities of the given moduli by minimizing the difference in the orientation averaged acoustical tensors.    It may be shown \cite{Norris05g} that the generalization of Fedorov's criterion to other symmetries is satisfied by the Euclidean projection of the stiffness tensor onto the elastic symmetry considered. 
The Euclidean projection is also equivalent to operating on the given elasticity tensor with the elements of the transformation group of the symmetry in question \cite{FV96}.  This approach has been used by Fran\c{c}ois {\it et al.\/} \cite{fgm} to find the closest moduli of trigonal and other symmetries for a set of ultrasonically measured stiffnesses.  The equivalence of the projection and group transformation averaging will be discussed in Section \ref{Euclideanprojection}. 
The Euclidean projection approach has received attention in the geophysical community from modelers interested in fitting rock data to particular elastic symmetries  \cite{Arts91,Helbig95,Cavallini99,Gangi00}. The work of Helbig \cite{Helbig95} is  particularly comprehensive.  He considers the problem both in terms of the 6$\times$6 matrix notation \cite{ry,cowin92} and in terms of 21-dimensional vectors representing the elastic moduli. The latter approach has been developed further by   Browaeys and Chevrot \cite{Browaeys04} who describe the projection operators for different elastic symmetries in the 21-dimensional viewpoint.  Dellinger \cite{Dellinger05} presents an algorithm for finding  the closest transversely isotropic medium by searching over all orientations of the symmetry axis.  Dellinger {\it et al.\/} \cite{Dellinger98}  proposed a distinct approach to  the problem of finding closest elasticity, based on the idea of generalized rotation of the 21-dimensional elasticity vector.

Although  the Euclidean projection  is commonly used in applications, it suffers from the fundamental drawback alluded to earlier, i.e., it  is not invariant under inversion of the elasticity tensor.  The projection found using the elastic stiffness is different from that obtained using the compliance tensor. This fundamental inconsistency arises from the dual physical properties of the elasticity tensor/matrix and its inverse.  The projection of one is clearly not the same as the inverse of the projection of the other, although both projections by definition possess the same elastic symmetry.  While there are circumstances in which the Euclidean projection is preferred, e.g., for the generalized Fedorov problem of finding the best acoustical approximant of a given symmetry \cite{Norris05g}, there is a clear need for a consistent technique to define a ``closest" elastic material of a given symmetry.   
The  solution to this quandary is to use an elastic distance function that is invariant under inversion.  Several have been proposed  of which we focus on two, the Riemannian distance function due to Moakher \cite{Moakher06} and the log-Euclidean length of Arsigny {\it et al.\/} 
\cite{Arsigny:MICCAI:05}, described  in Section \ref{distancefunctions}.  Moakher \cite{Moakher06} describes how the Riemannian distance function gives a consistent method for averaging elasticity tensors.  The only other application so far of these invariant distance functions to elasticity   is by Norris \cite{Norris05f} who discusses the closest isotropic elasticity.

\input{pic}   


The outline of the paper is as follows.  The three distance functions  are  introduced in  Section \ref{distancefunctions} after a brief review of notation in Section \ref{notation}.  
Section \ref{diffsyms} defines the different types of elastic symmetry using the algebraic tensor  decomposition of Walpole \cite{walpole84}. 
Euclidean projection is presented in Section \ref{Euclideanprojection}, with results for particular elastic symmetries summarized in Appendix \ref{closesteuclidean}.  
Before discussing the closest tensors using the logarithmic norms, methods to   evaluate the exponential and other functions of elasticity tensors are first described in Section \ref{expoftens}.
Section \ref{closestlog} discusses the closest tensors using the logarithmic norms, with particular attention given to isotropy and cubic symmetry as the target symmetries.   Numerical examples are given in Section \ref{numerical}, and final conclusions in  Section \ref{Conclusions}.

\section{Notation}\label{notation}

Elasticity tensors  relate stress ${\bf T}$ and infinitesimal strain $\bf E$ linearly 
according to 
\beq{00}
{\bf T} = \tens{C}{\bf E}, 
\qquad
{\bf E} = \tens{S}{\bf T} , 
\eeq
where $\tens{C}$ and  $\tens{S}$  denote the fourth-order stiffness and compliance tensors, respectively. They satisfy $\tens{C}\tens{S} =\tens{S}\tens{C} = \tens{I}$,  the identity.   
Although we are concerned primarily with fourth-order tensors in 3-dimensional space, calculation and presentation are sometimes better performed using  second-order tensors in 6-dimensions.  Accordingly  we   define ${\cal S}(d,r)$ as the space of symmetric tensors of order $r$ in $d-$dimensions.  Elasticity tensors, denoted by $\tens{E}{\rm la} \subset {\cal S}(3,4)$, are   positive definite, i.e., 
$\tens{A} \in \tens{E}{\rm la}$ if $\langle {\bf B}, \, \tens{A}{\bf B}\rangle > 0$ for all nonzero ${\bf B} \in {\cal S}(3,2)$.  
   Components are defined relative to the  basis  triad $\{ {\bf e}_1, {\bf e}_2,{\bf e}_3\}$;  thus, ${\bf a} = a_j {\bf e}_j$, ${\bf A} = A_{ij} \, {\bf e}_i
 \otimes {\bf e}_j$, and $\tens{A} = A_{ijkl} \, {\bf e}_i \otimes{\bf e}_j \otimes{\bf e}_k \otimes{\bf e}_l$, where the summation convention is assumed over $1,2,3$ for lower case subscripts.  Symmetry of second-order $(r=2)$ tensors implies   $A_{ij} = A_{ji}$, while for $\tens{A} \in {\cal S}(3,4)$  the elements satisfy 
\begin{align}\label{001}
  A_{ijkl}=  A_{jikl}=  A_{ijlk}, \qquad A_{ijkl}=   A_{klij}.
\end{align}
 The first pair of identities reflects the symmetry of the stress and the strain, while the last one is a consequence of the assumed existence of  a strain energy function, and consequently elasticity tensors have at most 21 independent components.  
 
 Throughout the paper lower case Latin, upper case Latin and ``ghostscript" indicate respectively $3-$dimensional vectors and tensors of order  2 and 4;  e.g., vector ${\bf b}$,  $\mat{A} \in {\cal S}(3,2)$,  $\tens{A} \in {\cal S}(3,4)$.
 The basis vectors are assumed   orthonormal, ${\bf e}_i \cdot {\bf e}_j =\delta_{ij}$, 
 so that products of tensors  are defined by summation over pairs of indices: 
$({\bf A} {\bf B})_{ij} = A_{ik}B_{kj}$ and $(\tens{A}\tens{B})_{ijkl} = A_{ijpq}B_{pqkl}$.  The inner product for tensors   is  defined as 
\beq{002}
\langle u,\, v \rangle = \tr (uv),  
\eeq
where $\tr {\bf A} = A_{ii}$, and for elasticity tensors, $\tr \tens{A} = A_{ijij}$.
The norm of a tensor is  
\beq{0021}
\left\| u\right\| \equiv 
\langle u,\, u \rangle^{1/2} \, . 
\eeq

We take advantage of the well known isomorphisms  between ${\cal S}(3,2)$ and ${\cal S}(6,1)$, and between ${\cal S}(3,4)$ and ${\cal S}(6,2)$.  
Thus, fourth-order elasticity tensors in 3 dimensions are 
equivalent to second-order symmetric tensor of 6 dimensions \cite{c3}, with properties
 $\tens{B} \mat{A} \leftrightarrow  \whbf{B} \hbf{a}$,  $\tens{A} \tens{B} \leftrightarrow  \whbf{A} \whbf{B}$, 
 and  $\langle \tens{A} , \tens{B} \rangle = \langle \whbf{A} , \whbf{B}\rangle $. 
  Vectors and second-order tensors in  six dimension  are distinguished by a hat, e.g., vector $\hbf{a}$, $\whbf{A} \in {\cal S}(6,2)$.  Components are defined relative to the orthonormal sextet $\{ \hat{{\bf e}}_I,\, I=1,2,\ldots, 6\}$,  $\hat{{\bf e}}_I \cdot \hat{{\bf e}}_J =\delta_{IJ}$, by  $\hbf{a} = a_I \hat{{\bf e}}_I$,   
     $\whbf{A} = \widehat{A}_{IJ} \hat{{\bf e}}_I \otimes \hat{{\bf e}}_J$, with the summation convention 
     over  $1,2,\ldots, 6$ for capital subscripts. Also, 
  $\tr \whbf{A} = \widehat{A}_{II}$.      
The connection between ${\cal S}(6,1)$ and ${\cal S}(3,2)$ is  made concrete by relating the basis vectors:
\beq{003}
\begin{matrix}
{\rm 6-vector}\\ \\ 
\hbf{e}_1 \quad \quad\hbf{e}_4 \\  \hbf{e}_2 \quad\quad \hbf{e}_5 \\ \hbf{e}_3 \quad\quad \hbf{e}_6  
\end{matrix}
\qquad \qquad
 \leftrightarrow  \qquad\qquad 
\begin{matrix}  
{\rm dyadic}\\ \\
{\bf e}_1\otimes{\bf e}_1 \quad\quad \frac1{\sqrt{2}}( {\bf e}_2\otimes{\bf e}_3 +{\bf e}_3\otimes{\bf e}_2)\\
  {\bf e}_2\otimes{\bf e}_2 \quad\quad \frac1{\sqrt{2}}( {\bf e}_3\otimes{\bf e}_1 +{\bf e}_1\otimes{\bf e}_3)\\
  {\bf e}_3\otimes{\bf e}_3 \quad\quad \frac1{\sqrt{2}}( {\bf e}_1\otimes{\bf e}_2 +{\bf e}_2\otimes{\bf e}_1) 
\end{matrix}
\eeq
This implies  a unique $\whbf{A} \in {\cal S}(6,2)$ for each $\tens{A} \in {\cal S}(3,4)$,  and vice versa.  Let $\tens{C}\in \tens{E}{\rm la}$  be the  tensor of  elastic stiffness, 
usually defined by 
 the Voigt notation:  $C_{ijkl} \equiv c_{IJ}$, where $I$ or $J=1,2,3,4,5,6$ correspond to $ij$ or $kl=11,22,33,23,13,12$, respectively, 
and $c_{IJ}$ are the elastic moduli in the Voigt notation \cite{Musgrave}, i.e., 
\beq{a1a}
\begin{pmatrix}
c_{11} & c_{12} & c_{13} & 
 c_{14} &  c_{15} & c_{16} 
\\ & & & & & \\
 & c_{22} & c_{23} & 
 c_{24} &  c_{25} &  c_{26} 
\\ & & & & & \\
 & & c_{33} & 
  c_{34} &  c_{35} &  c_{36} 
\\ & & & & & \\
 &   &   
 & c_{44} & c_{45} & c_{46}
\\ & & & & & \\
   S &Y  &M & & c_{55} & c_{56}
\\ & & & & & \\
&&&& & c_{66}
\end{pmatrix} . 
\eeq
The isomorphism  implies that the associated  matrix $\whbf{C}\in {\cal S}(6,2)$ is positive definite and has elements
\beq{a1}
\whbf{C}  \equiv \begin{pmatrix}
c_{11} & c_{12} & c_{13} & 
 2^{\frac12} c_{14} & 2^{\frac12} c_{15} & 2^{\frac12} c_{16} 
\\ & & & & & \\
c_{12} & c_{22} & c_{23} & 
 2^{\frac12} c_{24} & 2^{\frac12} c_{25} & 2^{\frac12} c_{26} 
\\ & & & & & \\
c_{13} & c_{23} & c_{33} & 
 2^{\frac12} c_{34} & 2^{\frac12} c_{35} & 2^{\frac12} c_{36} 
\\ & & & & & \\
 2^{\frac12} c_{14} & 2^{\frac12} c_{24} & 2^{\frac12} c_{34} 
 & 2c_{44} & 2c_{45} & 2c_{46}
\\ & & & & & \\
 2^{\frac12} c_{15} & 2^{\frac12} c_{25} & 2^{\frac12} c_{35} 
 & 2c_{45} & 2c_{55} & 2c_{56}
\\ & & & & & \\
 2^{\frac12} c_{16} & 2^{\frac12} c_{26} & 2^{\frac12} c_{36} 
 & 2c_{46} & 2c_{56} & 2c_{66}
\end{pmatrix} \, .  
\eeq

The spectral decomposition of elasticity tensors  
 can be expressed in both the fourth-order and second-order notations \cite{ry,c3},  
\beq{sp}
\tens{C} = \sum\limits_{I=1}^6
\Lambda_I \, \mat{N}_I\otimes\mat{N}_I\, \quad \leftrightarrow \quad
\whbf{C} = \sum\limits_{I=1}^6
\Lambda_I \, \hbf{n}_I\otimes\hbf{n}_I\, , 
\eeq
where  $\Lambda_I$ and $\hbf{n}_I$, $I=1,2, \ldots ,6$ are the  eigenvalues and eigenvectors of the  matrix $\whbf{C}$, $ \mat{N}_I$ are the associated dyadics, with   $\langle \hbf{n}_I , \hbf{n}_J \rangle=\langle \mat{N}_I, \mat{N}_J \rangle = \delta_{IJ}$.  Also,  $\Lambda_I > 0$ by virtue of the positive definite nature of the strain energy. 

The elastic moduli  $\tens C$ can also be expressed as a  vector with 21 elements \cite{Browaeys04}, 
{\small 
\beq{21v}
\begin{split}
X &= \left( x_1,x_2,x_3,x_4,x_5,x_6,x_7,x_8,x_9,x_{10},x_{11},x_{12},
 x_{13},x_{14},x_{15},x_{16}
							,x_{17},x_{18},x_{19},x_{20},x_{21}\right)^t
 \\ 
  &=\left(c_{11},c_{22},c_{33},
 \sqrt{2}c_{23},\sqrt{2}c_{13},\sqrt{2}c_{12},
 2c_{44},2c_{55},2c_{66},
 2c_{14},2c_{25},2c_{36}, \right.
 \\ 
 &  \qquad \qquad \qquad \qquad \left.
  2c_{34},2c_{15},2c_{26},
  2c_{24},2c_{35},2c_{16},
  2\sqrt{2}c_{56},2\sqrt{2}c_{46},2\sqrt{2}c_{45}\right)^t
 \\
 &=\left( 
\hat{c}_{11},\hat{c}_{22},\hat{c}_{33},
\sqrt{2}\hat{c}_{23},\sqrt{2}\hat{c}_{13},\sqrt{2}\hat{c}_{12},
\hat{c}_{44},\hat{c}_{55},\hat{c}_{66},
\sqrt{2}\hat{c}_{14},\sqrt{2}\hat{c}_{25},\sqrt{2}\hat{c}_{36}, \right.
 \\ 
 &  \qquad \qquad \qquad \qquad \left.
\sqrt{2}\hat{c}_{34},\sqrt{2}\hat{c}_{15},\sqrt{2}\hat{c}_{26},
\sqrt{2}\hat{c}_{24},\sqrt{2}\hat{c}_{35},\sqrt{2}\hat{c}_{16},
\sqrt{2}\hat{c}_{56},\sqrt{2}\hat{c}_{46},\sqrt{2}\hat{c}_{45}
\right)^t\, . 
\end{split}
\eeq
}
The $\sqrt{2}$'s  ensure that the inner product preserves the norm of the elastic moduli, whether it is a tensor, matrix or vector. Thus, the norm of an elasticity tensor can be expressed in a various ways depending on how  $\tens{C}$ is represented, 
\beq{l1}
\left\|
\tens{C} \right\|^2 \equiv \langle \tens{C},\, \tens{C} \rangle 
  =C_{ijkl}C_{ijkl} 
 =  \widehat{c}_{IJ}\widehat{c}_{IJ}  = \Lambda_I\Lambda_I = \|X\|^2\, . 
\eeq

\section{Elastic distance functions} \label{distancefunctions}

We consider three  metrics for $\tens{E}{\rm la}$:  the  Euclidean or Frobenius metric $\dis_F$, the log-Euclidean norm $\dis_L$ \cite{Arsigny:MICCAI:05} and the Riemannian metric $\dis_R$  \cite{Moakher06}.  They are defined for any pair of elasticity tensors as
\begin{subequations}\label{3}
\begin{align}\label{3a}
{\rm d}_F (\tens{C}_1, \tens{C}_2) &= \left\|\tens{C}_1 -\tens{C}_2 \right\|,
\\
{\rm d}_L (\tens{C}_1, \tens{C}_2) &= \left\| \Log(\tens{C}_1) - \Log(\tens{C}_2) \right\|,  
\label{3L}
\\
{\rm d}_R (\tens{C}_1, \tens{C}_2) &= \left\| \Log(\tens{C}_1^{-1/2} \tens{C}_2 \tens{C}_1^{-1/2}) \right\|\, .  
\label{3b}
\end{align}
\end{subequations}
 The Riemannian distance  ${\rm d}_R$ is  related to the exponential map   induced by the scalar product on the tangent space to $\tens{E}{\rm la}$ at $\tens C$ \cite{Moakher06,lang98}.
 The log-Euclidean metric  ${\rm d}_L$  can be motivated by the following definition of tensor  multiplication  \cite{Arsigny:MICCAI:05} :
\beq{5081}
{\tens C}_1 \odot {\tens C}_2 \equiv \exp \left( \Log({\tens C}_1) + \Log({\tens C}_2)\right)\, . 
 \eeq    
The $\odot$  product preserves symmetry and positive definiteness, unlike  normal multiplication. 
The three metrics   have the usual attributes of a distance function ${\rm d}$: 
\begin{enumerate}
\item  non-negative, ${\rm d} (\tens{C}_1,\tens{C}_2) \ge 0$ with equality iff $\tens{C}_1=\tens{C}_2$,
\item  symmetric in the arguments, ${\rm d} (\tens{C}_1,\tens{C}_2)= {\rm d} (\tens{C}_2,\tens{C}_1)$,
\item invariant under a change of basis, 
${\rm d} (\tens{C}_1',\tens{C}_2')  = {\rm d} (\tens{C}_1, \tens{C}_2)$ for all proper orthogonal  coordinate transformations $\{ {\bf e}_1, {\bf e}_2,{\bf e}_3\} \rightarrow \{ {\bf e}_1', {\bf e}_2',{\bf e}_3'\}$, and 
\item  it satisfies the triangle inequality,  ${\rm d} (\tens{C}_1,\tens{C}_3) \le {\rm d} (\tens{C}_1,\tens{C}_2) + {\rm d} (\tens{C}_2,\tens{C}_3)$.
\end{enumerate}  

The Riemannian and log-Euclidean distances possess the additional properties 
that they are invariant under inversion and (positive) scalar multiplication
\beq{51}
{\rm d}_{L,R} (\tens{C}_1^{-1}, \tens{C}_2^{-1}) = 
{\rm d}_{L,R} (\tens{C}_1, \tens{C}_2), 
\quad \text{ and } \quad
{\rm d}_{L,R} (a \tens{C}, a \tens{C}) = 
{\rm d}_{L,R} (\tens{C}_1, \tens{C}_2), \quad a>0.
\eeq 
The following inequality between the logarithmic distance functions is a consequence of the metric increasing property of the exponential \cite{bhatia}
\beq{50}
{\rm d}_{L} (\tens{C}_1, \tens{C}_2) 
\le 
{\rm d}_{R} (\tens{C}_1, \tens{C}_2) . 
 \eeq 
We also have (see \cite{bhatia} for the inequality)
\begin{subequations}
\begin{align}
{\rm d}_{L} (\tens{C}_1^b,\tens{C}_2^b) & =  |b|\, {\rm d}_{L} 
(\tens{C}_1,\tens{C}_2) \,,\quad b\in \mathbb{R} \\
{\rm d}_{R} (\tens{C}_1^b,\tens{C}_2^b) & \le  b\, {\rm d}_{R} 
(\tens{C}_1,\tens{C}_2) \,,\quad b\in [0,1], 
\end{align}
\end{subequations}
and the Riemannian distance is  invariant 
under congruent transformations, i.e.,
\beq{52}
{\rm d}_{R} (\tens{T} \tens{C}_1 \tens{T}^t, \tens{T} \tens{C}_2 \tens{T}^t) = 
{\rm d}_{R} (\tens{C}_1, \tens{C}_2), \quad \forall \tens{T} 
\text{ invertible}.
\eeq

 Distance functions satisfying \rf{51} are called bi-invariant, a property 
that makes the Riemannian and log-Euclidean distances consistent metrics for elasticity tensors.
   The Frobenius norm, not being invariant under inversion, gives different results depending on whether the stiffness or compliance is considered.  Other 
 metric functions may be considered.  For instance,  the 
 Kullback-Leibler metric \cite{Moakher06} is invariant under inversion and congruent transformation but does not satisfy the triangle inequality, which we take here as prerequisite for consideration as a distance function.   

Each distance function has a geometrical interpretation.  For instance,  the midpoint between  $\tens{C}_1$ and $\tens{C}_2$ is defined as the unique $\tens{C}_3$ such that 
 ${\rm d} (\tens{C}_1,\tens{C}_3)= {\rm d} (\tens{C}_3,\tens{C}_2)= \tfrac12 {\rm d} (\tens{C}_1,\tens{C}_2)$.  Using ${\rm d}_F$, the midpoint is the vector  halfway between the $21-$vectors $X_1$ and $X_2$  as defined by \rf{21v}, i.e., $\tens{C}_3 = \tfrac12 (\tens{C}_1+\tens{C}_2)$.  The midpoint using  ${\rm d}_L$ is
  $\tens{C}_3 = ({\tens C}_1 \odot {\tens C}_2)^{1/2}$.   
  The midpoint with ${\rm d}_R$ is \cite{Moakher05b} 
  $\tens{C}_3 =\tens{C}_1 (\tens{C}_1^{-1}\tens{C}_2)^{1/2}$ which can be expressed in several other ways, see \cite{Moakher06}.   More generally, the midpoint is  the value of    the geodesic $\tens{C}(t)$   at $t = \tfrac12$,  where 
 \begin{align}\label{mp}
 \tens{C}(t) = 
 \begin{cases} (1-t)\tens{C}_1+t \tens{C}_2 ,\qquad & {\rm Frobenius}, \\
  \exp \left( (1-t)\Log({\tens C}_1) + t\Log({\tens C}_2)\right), & {\rm log-Euclidean}, \\
  {\tens C}_1\, \exp\, \left(t \Log({\tens C}_1^{-1}\tens{C}_2 ) \right), & {\rm Riemannian},
 \end{cases} \qquad 0\le t\le 1. 
 \end{align}
 
\section{Elasticity tensors of the different symmetry classes}\label{diffsyms}

Elasticity tensors for the different symmetry classes are described in this section.  The choice of representation of $\tens{E}{\rm la}$  is important for  discussing the closest approximants, particularly for the bi-invariant metrics.  It should preferably be  independent of the coordinate system, although at the same time, practical application is normally in terms of the Voigt notation, so the format should not deviate too far from this.   We begin with a review of alternative representations for elastic moduli. 

\subsection{Representation of elasticity tensors}

There are many  representations of $\tens{E}{\rm la}$  in addition to the standard Voigt notation  of eq. \rf{a1a}. 
The $21-$dimensional vector format  
\cite{Browaeys04} is useful for some applications, including the Euclidean projection, as we show later.  However, these two representations depend upon the coordinate system.  Among the coordinate-free  forms of $\tens{E}{\rm la}$
that can be identified in the literature we distinguish (i) spectral decomposition,  (ii) algebraic decomposition,    (iii)   groups and reflection  symmetries, (iv) harmonic decomposition, (v) integrity bases.  

The first, (i),  spectral decomposition dates back to Kelvin \cite{kelvin}.
The idea is simple: the elasticity tensor operates on the 
six-dimensional space of symmetric second-order tensors, and therefore has a six-dimensional
spectral form. The associated eigenvalues are called
the Kelvin moduli \cite{ry,cowin92}.  Recent related developments 
are due to Rychlewski \cite{ry} and 
Mehrabadi and Cowin \cite{c3,cowin92}, who independently 
rediscovered and extended Kelvin's approach, see also \cite{sut,TS}. 
The associated six-dimensional tensor representation will be used here for practical implementation. 
(ii) Tensor functions operating on $\tens{E}{\rm la}$ 
can be greatly simplified using the irreducible tensor 
algebra proposed by Walpole  \cite{walpole84}, and independently by Kunin \cite{Kunin}.  This procedure  provides the most efficient means of representing elastic tensors of a given 
symmetry class, especially   those of high symmetry.   We adopt this method to develop most of the results here. 
(iii) The group of rotations associated with elastic symmetry provides an 
irreducible representation \cite{sur}.  There are various related ways of considering elasticity tensors in terms of rotational group properties of tensors, e.g.,  based on Cartan decomposition \cite{FV},  complex vectors and tensors \cite{Xiao95,Xiao98}, and subgroups of $O(3)$ \cite{Bona04,Bona04b}.  These ideas are closely related to definitions of 
 elastic symmetry in terms of a single symmetry element:  reflection about a plane.  The necessary algebra and the relationship to the more conventional crystallographic symmetry elements are described in detail by Cowin and Mehrabadi \cite{cowin95}.
Comparison between the rotation-based approach and that using symmetry planes is provided by Chadwick {\it et al.\/} \cite{CVC}.
(iv) Backus \cite{backus} proposed a representation of $\tens{E}{\rm la}$ in terms of harmonic  tensors.  These are based on an isomorphism between
the space of homogeneous harmonic polynomials of degree $q$ and 
the space of totally symmetric tensors of order $q$. 
There has been considerable interest in Backus's representation \cite{cowin89,baerheim,FV96,zz}.  For instance, Baerheim and  Helbig \cite{bh} provide an orthonormal 
decomposition of $\tens{E}{\rm la}$ in terms of harmonic tensors. 
(v) Elasticity decomposition via integrity bases  has been studied considerably \cite{Smith57,Tu68,Spencer71,Betten81,Xiao98b}.   An integrity basis is a set
of polynomials, each invariant under the group of symmetry transformations, such that any
polynomial function invariant under the group is expressible as a polynomial in
elements of the integrity basis.  For instance,  Tu \cite{Tu68}   used an integrity basis to construct five hierarchies of orthonormal
tensor bases which span the space of  elastic constants of all crystal systems. Any
elastic tensor of order four possessing certain  symmetry may be decomposed into a
sum of tensors of increasing symmetry.  The resulting decomposition has considerable similarity to the decomposition generated in Section \ref{Euclideanprojection} using Walpole's irreducible elements with Euclidean projection. 
 Among other methods for decomposing elasticity tensors, we mention the scheme of   Elata and Rubin \cite{er} who use a set of six vectors
related to a regular icosahedron to form a basis for $\tens{E}{\rm la}$.  This basis 
naturally splits a tensor into deviatoric and non-deviatoric parts.   

We use Walpole's \cite{walpole84} algebraic representation as it provides a  consistent and straightforward means to define projections of fourth-order tensors onto the given elastic symmetry.  This representation is better for our purposes than the spectral decomposition, since it is independent of the elastic moduli (i.e., no distributors \cite{cowin92}) and depends only on the crystallographic orientation.   It uses the notion of basis tensors, similar to but not the same as for a vector space, which makes it very suitable for Euclidean projection.  We work mainly with fourth-order tensors directly, although the associated six-dimensional matrix  notation is also provided.   The latter is simpler for purposes of computation, e.g., many of the matrix operations are easily implemented using MATLAB. 
 We  begin with the highest elastic symmetry.

\subsection{Isotropic system}

A general isotropic fourth-order tensor is given by
\beq{defa}
\tens{A} = a \tens{J} + b \tens{K},
\end{equation}
where $\tens{J}$ and $\tens{K}$ are the two linearly independent symmetric
tensors defined by 
\beq{def-JK}
\tens{J} = \frac13 \mat{I} \otimes \mat{I} \, ,   \qquad \tens{K}=\tens{I} - \tens{J}\, . 
\end{equation}
The component forms follow from   
$I_{ijkl} = \frac12(\delta_{ik} \delta_{jl} + \delta_{il}\delta_{jk})$ and 
$J_{ijkl} = \frac13 \delta_{ij} \delta_{kl}$.  
Note that the tensors $\tens{J}$ and $\tens{K}$  sum to the identity, 
\beq{J+K}
\tens{I} = \tens{J} + \tens{K}, 
\end{equation}
and they  satisfy the multiplication table
\beq{mul-JK}
\tens{J}^2 = \tens{J}, \quad \tens{K}^2 = \tens{K}, \quad 
\tens{JK} = \tens{KJ} = \tens{O}.
\end{equation}  
The  Euclidean lengths are  
 $\left\|\tens{J}\right\| = 1$, $\left\|\tens{K}\right\| = \sqrt{5}$. 
Tensors $\tens{J}$ and $\tens{K}$ form, respectively,  one and five-dimensional Kelvin subspaces \cite{cowin92}. 

An  important   result (see \cite{Gazis63,fed}) is that  the closest isotropic elasticity tensor in the Euclidean sense is 
\beq{gendec3}
\tens{C}_{\rm iso} =  3\kappa \tens{J} + 2\mu \tens{K}\, ,  
\eeq
where  
\beq{km}
\begin{split}
& 9 \kappa =  3 \langle \tens{C},  \tens{J} \rangle= 
\hat{c}_{11}+\hat{c}_{22}+\hat{c}_{33}+2\hat{c}_{12}+2\hat{c}_{13}+2\hat{c}_{23}, 
\\
& 30 \mu =  3\langle \tens{C},  \tens{K}\rangle = 2(\hat{c}_{11}+\hat{c}_{22}+\hat{c}_{33} - \hat{c}_{12}-\hat{c}_{23}-\hat{c}_{31}) +3(\hat{c}_{44}+\hat{c}_{55}+\hat{c}_{66}) \, . 
\end{split}
\eeq
Gazis {\it et al.\/} \cite{Gazis63} obtained \rf{gendec3} using methods similar to those we will generalize in Section \ref{Euclideanprojection}.  Fedorov actually found  $\kappa$ and $\mu$ using an apparently different approach - the minimization of the mean square difference in the acoustical or Christoffel matrices for the original and isotropic systems.  However, it can be shown \cite{Norris05g} that the generalization of Fedorov's approach to other symmetries is identical to the Euclidean minimization. 
The moduli \rf{km} will be derived later in the context of the general theory for Euclidean projection. 

\subsection{Cubic system}

Let $\bd{a}$, $\bd{b}$ and $\bd{c}$ be three mutually orthogonal unit vectors
that describe the three crystallographic directions of a cubic medium.
We introduce the second-order tensors
\beq{5345}
\begin{split} 
& \bd{U} = \tfrac1{\sqrt{2}}(\bd{a} \otimes \bd{c} + \bd{c} \otimes \bd{a}), \quad  
\bd{V} = \tfrac1{\sqrt{2}}(\bd{b} \otimes \bd{c} + \bd{c} \otimes \bd{b}),   \quad 
  \bd{W} = \tfrac1{\sqrt{2}}(\bd{a} \otimes \bd{b} + \bd{b} \otimes \bd{a}), 
  \\
& \bd{X} = \tfrac1{\sqrt{2}}(\bd{c} \otimes \bd{c} - \bd{a} \otimes \bd{a}),  \quad   
  \bd{Y} = \tfrac1{\sqrt{2}}(\bd{b} \otimes \bd{b} - \bd{c} \otimes \bd{c}),  \quad   
  \bd{Z} = \tfrac1{\sqrt{2}}(\bd{a} \otimes \bd{a} - \bd{b} \otimes \bd{b}), 
\end{split}
\eeq
and fourth-order tensors $\tens{L}$ and $\tens{M}$  defined by them,
\beq{def-LM}
\tens{L}  = \bd{U} \otimes \bd{U} +  \bd{V} \otimes \bd{V}
+  \bd{W} \otimes \bd{W}, 
\qquad 
\tens{M}  = \frac23 (\bd{X} \otimes \bd{X} +  \bd{Y} \otimes \bd{Y}
+ \bd{Z} \otimes \bd{Z}).
\eeq
The tensors $\tens{J}$, $\tens{L}$ and $\tens{M}$ sum up to the identity 
tensor, and  $\tens{L}$ and $\tens{M}$ partition $\tens{K}$, i.e., 
\beq{J+L+M}
\tens{I} = \tens{J} + \tens{L} + \tens{M},
\qquad
\tens{K} =  \tens{L} + \tens{M}\, . 
\eeq
The multiplication table for cubic tensors is
\beq{mul-JLM}
\tens{J}^2 = \tens{J}, \quad \tens{L}^2 = \tens{L}, \quad 
\tens{M}^2 = \tens{M}, \quad 
\tens{JL} = \tens{LJ} = \tens{JM} = \tens{MJ} = \tens{LM} =
\tens{ML} = \tens{O}, 
\end{equation}  
and the Euclidean lengths are  
 $\left\|\tens{L}\right\| = \sqrt{3}$, $\left\|\tens{M}\right\| =\sqrt{2}$.

A general symmetric fourth-order tensor for a cubic media is given by a 
linear combination of the three linearly independent tensors $\tens{J}$, 
$\tens{L}$ and $\tens{M}$
\beq{gencubic}
\tens{A} = a \tens{J} + b \tens{L} + c \tens{M}.
\end{equation}
The tensors $\tens{J}$, $\tens{L}$ and $\tens{M}$ form, respectively,  one, three  and two-dimensional Kelvin subspaces \cite{cowin92}. 

\subsection{Transversely isotropic system}

Assume that the unit vector $\bd{c}$ characterizes the preferred direction
of a transversely isotropic medium. Let $\bd{P}$ and $\bd{Q}$ 
be the second-order tensors
\beq{091}
\bd{P} = \bd{c} \otimes \bd{c}, \quad
\bd{Q} = \frac{1}{\sqrt{2}} (\mat{I} - \bd{c} \otimes \bd{c} ),  
\eeq
and  define the six linearly independent fourth-order elementary tensors
\beq{defE1E2E3E4}
\begin{split}
& \tens{E}_1 = \bd{P} \otimes \bd{P}, \quad
\tens{E}_2 = \bd{Q} \otimes \bd{Q},  \quad  
\tens{E}_3 = \bd{P} \otimes \bd{Q}, \quad
\tens{E}_4 = \bd{Q} \otimes \bd{P}, \\
& \tens{F} = \bd{W} \otimes \bd{W} + \bd{Z} \otimes \bd{Z} ,  \quad \quad\quad\,\
 \tens{G} = \bd{U} \otimes \bd{U} + \bd{V} \otimes \bd{V}.
\end{split}
\eeq
Both $\tens{E}_1$ and $\tens{E}_2$ are symmetric but the tensors   $\tens{E}_3$ 
and $\tens{E}_4$ are not, although they are the transpose of one another and so  their sum is
a symmetric tensor.  These asymmetric tensors are introduced because the set $\{\tens{E}_1, \ \tens{E}_2, \
\tens{E}_3 + \tens{E}_4 \}$ is not closed under tensor multiplication.
The multiplication table for the  six elementary tensors is 
\beq{mul-TI}
\begin{array}{c|cccccc}
& \tens{E}_1 & \tens{E}_2 & \tens{E}_3 & \tens{E}_4 & \tens{F} & \tens{G} \\
\noalign
{\hrule width 5.75cm}
\tens{E}_1 & \tens{E}_1 & \tens{O} & \tens{E}_3 & \tens{O} &\tens{O} &\tens{O}\\
\tens{E}_2 & \tens{O} & \tens{E}_2 & \tens{O} & \tens{E}_4 &\tens{O} &\tens{O}\\
\tens{E}_3 & \tens{O} & \tens{E}_3 & \tens{O} & \tens{E}_1 &\tens{O} &\tens{O}\\
\tens{E}_4 & \tens{E}_4 & \tens{O} & \tens{E}_2 & \tens{O} &\tens{O} &\tens{O}\\
\tens{F} & \tens{O} &\tens{O} &\tens{O} &\tens{O} &\tens{F} &\tens{O} \\
\tens{G} & \tens{O} &\tens{O} &\tens{O} &\tens{O} &\tens{O} &\tens{G}
\end{array} 
\end{equation}
The tensors $\tens{E}_1$, $\tens{E}_2$, $\tens{F}$ and $\tens{G}$ sum up to 
the identity, 
\beq{sum-TI}
\tens{I} = \tens{E}_1 + \tens{E}_2 + \tens{F} + \tens{G}.
\end{equation} 
The Euclidean lengths are 
 $\|\tens{E}_1\| = \|\tens{E}_2\| =1$, $\|\tens{E}_3+\tens{E}_4\| = \|\tens{F}\| =\|\tens{G}\| =\sqrt{2}$.

A general symmetric fourth-order tensor of transversely isotropic  symmetry is given by the 
5-parameter linear combination
\beq{defti}
\tens{A} = a \tens{E}_1 + b \tens{E}_2 + c (\tens{E}_3 + \tens{E}_4)
+ f \tens{F} + g \tens{G}.
\end{equation}
Note that the base tensors do not correspond to the Kelvin modes, although some can be identified as such, e.g., $\tens{G}$ is a two-dimensional Kelvin subspace \cite{cowin92}.   The three bases tensors  $\{ \tens{E}_{1}, \, \tens{E}_{2}, \, \tens{E}_{3}+\tens{E}_{4} \} $ together define a two-dimensional subspace.

\subsection{Tetragonal system} \label{tetragonal}

Nine base tensors are required: $\{ \tens{E}_1 , \, \tens{E}_2 , \, 
\tens{E}_3 , \, \tens{E}_4 , \, \tens{F}_1 , \, \tens{F}_2 , \, \tens{F}_3 , \, \tens{F}_4 , \, \tens{G} \}$, 
where the new tensors are a symmetric pair $ \tens{F}_{1}, \tens{F}_{2}$, and a pair that are 
mutual transposes, $ \tens{F}_{3}, \tens{F}_{4}$,
\beq{t6}
 \tens{F}_{1} =  \bd{W} \otimes\bd{W} , \qquad
\tens{F}_{2} =   \bd{Z} \otimes\bd{Z}, \qquad
 \tens{F}_{3} =  \bd{W} \otimes\bd{Z}, \qquad
\tens{F}_{4} =  \bd{Z} \otimes\bd{W}   \, . 
\eeq
In comparison with the transversely isotropic  system, $\tens{F}$ is replaced by the symmetric pair 
of  tensors, i.e., $ \tens{F} = \tens{F}_{1} + \tens{F}_{2} $, 
and  the decomposition of  the identity  is
\beq{sum-Tet}
\tens{I} = \tens{E}_1 + \tens{E}_2 + \tens{F}_1 + \tens{F}_2 + \tens{G}.
\end{equation} 
The multiplication table of the nine elementary tensors is 
\beq{mul-Tet}
\begin{array}{c|ccccccccc}
& \tens{E}_1 & \tens{E}_2 & \tens{E}_3 & \tens{E}_4 
& \tens{F}_1 & \tens{F}_2 & \tens{F}_3 & \tens{F}_4 & \tens{G} \\
\noalign
{\hrule width 8.5cm}
\tens{E}_1 & \tens{E}_1 & \tens{O} & \tens{E}_3 & \tens{O} &\tens{O} &\tens{O}
& \tens{O} &\tens{O} &\tens{O} \\
\tens{E}_2 & \tens{O} & \tens{E}_2 & \tens{O} & \tens{E}_4 &\tens{O} &\tens{O}
& \tens{O} &\tens{O} &\tens{O} \\
\tens{E}_3 & \tens{O} & \tens{E}_3 & \tens{O} & \tens{E}_1 &\tens{O} &\tens{O}
& \tens{O} &\tens{O} &\tens{O} \\
\tens{E}_4 & \tens{E}_4 & \tens{O} & \tens{E}_2 & \tens{O} &\tens{O} &\tens{O}
& \tens{O} &\tens{O} &\tens{O} \\
\tens{F}_1 & \tens{O} &\tens{O} &\tens{O} &\tens{O} &\tens{F}_1 &\tens{O}
&\tens{F}_3 &\tens{O} &\tens{O} \\
\tens{F}_2 & \tens{O} &\tens{O} &\tens{O} &\tens{O} &\tens{O} &\tens{F}_2 
&\tens{O} &\tens{F}_4 &\tens{O} \\
\tens{F}_3 & \tens{O} &\tens{O} &\tens{O} &\tens{O} &\tens{O} &\tens{F}_3 
&\tens{O} &\tens{F}_1 &\tens{O} \\
\tens{F}_4 & \tens{O} &\tens{O} &\tens{O} &\tens{O} &\tens{F}_4 &\tens{O}
&\tens{F}_2 &\tens{O} &\tens{O} \\
\tens{G} & \tens{O} &\tens{O} &\tens{O} & \tens{O} &\tens{O} &\tens{O} 
&\tens{O} &\tens{O} &\tens{G}
\end{array} 
\end{equation}
Under multiplication the $\tens{E}_i$'s 
and $\tens{F}_i$'s  decouple from one another and from $\tens{G}$. 
Furthermore, the algebra of the $\tens{F}_i$'s is similar to that of
the $\tens{E}_i$'s. The Euclidean lengths of the new  tensors are 
$\|\tens{F}_1\| = \|\tens{F}_2\| =1$, 
$\|\tens{F}_3+\tens{F}_4\|  =\sqrt{2}$.

A general symmetric fourth-order tensor of tetragonal symmetry is given by the 
$6-$parameter  combination
\beq{deftet}
\tens{A} = a \tens{E}_1 + b \tens{E}_2 + c (\tens{E}_3 + \tens{E}_4)
+ p \tens{F}_1 + q \tens{F}_2 + r (\tens{F}_3 + \tens{F}_4) 
+ g \tens{G}. 
\end{equation}

Tetragonal symmetry is often represented by 6 rather than 7 independent parameters.  Fedorov \cite{fed} pointed out that transformation by rotation about the ${\bd c}$ axis by angle $\theta$ yields zero for the transformed coefficient $r$ if (see eq. (9.7) in \cite{fed})
\beq{ifa}
\tan 4 \theta = r/(q-p)\, . 
\eeq
The transformation \rf{ifa} depends on  knowledge of the coefficients.  However, it is assumed here that  the only properties of the tetragonal symmetry available {\em a priori} are the orthogonal  crystal axes ${\bd a}$,  ${\bd b}$ and ${\bd c}$.  The reduction from 7 to 6 parameters can be achieved after the effective  tetragonal material is found, but the axes of the final 6-parameter material depend upon the initial moduli.  
Hence,  it is important to retain the seven-parameter representation \rf{deftet}.

\subsection{Trigonal system}

Trigonal symmetry is characterized by three planes of reflection symmetry with normals coplanar and at 120$^\circ$ to one another. This symmetry class is also known as hexagonal \cite{walpole84} but we refer to it as trigonal since hexagonal symmetry is often used synonymously with transverse isotropy.  It is advantageous to choose  non-orthogonal basis vectors $\bd{a}',\bd{b}',\bd{c}$ with $\bd{a}'$ and $\bd{b}'$  normals to two of the planes and  $\bd{c}$  
perpendicular to them (the pair $\bd{a}'$ and $\bd{b}'$ may also be represented in terms of the orthonormal vectors via, for instance, $\bd{a}'=\bd{a}$, $\bd{b}' = -\tfrac12 \bd{a} + {\tfrac{\sqrt{3}}{2}}\bd{b}$).   In order to describe symmetric fourth-order tensors of trigonal symmetry, following \cite{walpole84} we  introduce the second-order tensors 
\begin{subequations} \label{trig1}
\begin{align}
&\bd{S} = \sqrt{\tfrac23}(\bd{a}' \otimes \bd{b}' + \bd{b}' \otimes \bd{a}' +\bd{a}' \otimes \bd{a}'), & &
\bd{T} = \sqrt{\tfrac23}(\bd{a}' \otimes \bd{b}' + \bd{b}' \otimes \bd{a}' + \bd{b}' \otimes \bd{b}'),
\\
&\bd{U}' = \tfrac1{\sqrt{2}}(\bd{a}' \otimes \bd{c} + \bd{c} \otimes \bd{a}'),  
&  &
 \bd{V}' = -\tfrac1{\sqrt{2}}(\bd{b}' \otimes \bd{c} + \bd{c} \otimes \bd{b}') ,  
\end{align}
\end{subequations}
 the two symmetric fourth-order tensors
\begin{subequations} \label{hex1}
\begin{align}
\tens{R}_{1} &=  \tfrac43 (\bd{S} \otimes \bd{S} + \bd{T} \otimes \bd{T} 
- \tfrac12 \bd{S} \otimes \bd{T} - \tfrac12  \bd{T} \otimes \bd{S})\, , \\
\tens{R}_{2} &=  \tfrac43(\bd{U}' \otimes \bd{U}' + \bd{V}' \otimes \bd{V}' 
- \tfrac12 \bd{U}' \otimes \bd{V}' - \tfrac12 \bd{V}' \otimes \bd{U}')\, ,
\end{align}
\end{subequations}
and  two pairs of  mutually transpose tensors:
\begin{subequations}\label{hex2}
\begin{align}
\tens{R}_{3} & = \tfrac4{3} (\bd{S} \otimes \bd{U}' + 
\bd{T} \otimes \bd{V}' - \tfrac12 \bd{T} \otimes \bd{U}' - \tfrac12
\bd{S} \otimes \bd{V}' )\, , 
\\
\tens{R}_{4} & = \tfrac4{3} (\bd{U}' \otimes \bd{S} + 
\bd{V}' \otimes \bd{T} - \tfrac12 \bd{U}' \otimes \bd{T} - \tfrac12 \bd{V}' \otimes \bd{S})\, , 
\intertext{and}  
\tens{R}_{5} & = \tfrac2{\sqrt{3}} (\bd{S} \otimes \bd{V}' - 
\bd{T} \otimes \bd{U}') \, , \quad
\tens{R}_{6} = \tfrac2{\sqrt{3}} (\bd{V}' \otimes \bd{S} -
\bd{U}' \otimes \bd{T}) \,.
\end{align}
\end{subequations}
To make the $\tens{R}_{i}$'s closed under multiplication, we need to
further introduce  two skew-symmetric tensors 
\beq{hex3}
\tens{R}_{7} = \tfrac2{\sqrt{3}} (\bd{T} \otimes \bd{S} - 
\bd{S} \otimes \bd{T}) \, , \quad 
\tens{R}_{8} = \tfrac2{\sqrt{3}}(\bd{U}' \otimes \bd{V}' - 
\bd{V}' \otimes \bd{U}') \, .
\eeq
Thus,  $\tens{E}_i \tens{R}_j = \tens{R}_j \tens{E}_i = \tens{O}$ 
for $i=1,2,\ldots, 4$ and $j=1,2, \ldots,8$,
and the multiplication table 
is\footnote{The multiplication table in \cite{walpole84} contains some typographical errors; specifically the  elements (3,8), (4,5), (5,8), (6,3), (8,4) and (8,6) in the original need to be multiplied by $-1$.} 
\beq{mul-Hex}
\begin{array}{c|cccccccc}
& \tens{R}_1 & \tens{R}_2 & \tens{R}_3 & \tens{R}_4 & 
\tens{R}_5 & \tens{R}_6 & \tens{R}_7 & \tens{R}_8 \\
\noalign
{\hrule width 10cm}
\tens{R}_1 &\tens{R}_1 &\tens{O} &\tens{R}_3 &\tens{O} & \tens{R}_5 & 
\tens{O} &\tens{R}_7 &\tens{O} \\
\tens{R}_2 & \tens{O} &\tens{R}_2 &\tens{O} &\tens{R}_4 &\tens{O} & 
\tens{R}_6 & \tens{O} &\tens{R}_8 \\
\tens{R}_3 &\tens{O} &\tens{R}_3 &\tens{O} &\tens{R}_1 &\tens{O} & 
\tens{R}_7 & \tens{O} &  \tens{R}_5 \\
\tens{R}_4 & \tens{R}_4 &\tens{O} &\tens{R}_2 & \tens{O} & \tens{R}_8 & 
\tens{O} & \tens{R}_6 &\tens{O} \\
\tens{R}_5 & \tens{O} &\tens{R}_5 &\tens{O} & -\tens{R}_7 & \tens{O} & 
\tens{R}_1 &\tens{O} & -\tens{R}_3 \\
\tens{R}_6 & \tens{R}_6 &\tens{O} &-\tens{R}_8 &\tens{O} &\tens{R}_2 & 
\tens{O} & -\tens{R}_4 &\tens{O} \\
\tens{R}_7 &\tens{R}_7 &\tens{O} &-\tens{R}_5 &\tens{O} &\tens{R}_3 & 
\tens{O} & -\tens{R}_1 &\tens{O} \\
\tens{R}_8 & \tens{O} &\tens{R}_8 &\tens{O} & -\tens{R}_6 & \tens{O} & 
\tens{R}_4 &\tens{O} & -\tens{R}_2 
\end{array} 
\end{equation}
The algebra of the $\tens{R}_i$'s is equivalent to that of $2\times 2$
complex matrices \cite{walpole84}.

We note that $\tens{R}_{1}= \tens{F}$ and $\tens{R}_{2}= \tens{G}$, and hence the decomposition for the identity tensor is 
\beq{sum-hex}
\tens{I} = \tens{E}_1 + \tens{E}_2 + \tens{R}_1 + \tens{R}_2.
\end{equation} 
The Euclidean lengths of the new tensors are 
$\|\tens{R}_1\| = \|\tens{R}_2\| =\sqrt{2}$, 
$\|\tens{R}_3+\tens{R}_4\| = \|\tens{R}_5+\tens{R}_6\| =2$. 

A general symmetric fourth-order tensor of trigonal symmetry is given by the 
linear combination
\beq{defhex}
\tens{A} = a \tens{E}_1 + b \tens{E}_2 + c (\tens{E}_3 + \tens{E}_4)
+ p \tens{R}_1 + q \tens{R}_2 + r (\tens{R}_3 + \tens{R}_4) 
+ s (\tens{R}_5 + \tens{R}_6). 
\end{equation}

\subsection{Rhombic,  Monoclinic and Triclinic systems}

The algebra for orthorhombic (equivalently orthotropic) symmetry is the same as that of $3\times 3$ real matrices plus three real numbers. For monoclinic symmetry it is the same as the algebra of pairs of $4\times 4$ and $2 \times 2$ real matrices.   The lowest symmetry, triclinic or no symmetry, has algebra the same as that of $6\times 6$ real matrices.   These low symmetries are probably better defined in terms of their group properties, which also serves as efficient means for Euclidean projection, as described in Section \ref{Euclideanprojection}.

\subsection{Six-dimensional representation}

An equivalent 6-dimensional matrix format is described for the various elementary tensors.  For each symmetry class  the 6-dimensional tensors possess the same algebraic properties as the fourth-order tensors, with the same multiplication tables and Euclidean lengths. 

The 6-dimensional vectors  $\{\hbf{a},\hbf{b},\hbf{c} \}$ represent the second-order tensors
$\{\bd{a} \otimes \bd{a}$, $\bd{b} \otimes \bd{b}$, $\bd{c} \otimes \bd{c} \}$ 
that correspond to the axes  $\{\bd{a}, \bd{b}, \bd{c}\}$, where according to eq. \rf{003}, 
\beq{0031}
\bd{a} = a_1\bd{e}_1+a_2\bd{e}_2+a_3\bd{e}_3
\ \Rightarrow \
\hbf{a} = \left( a_1^2,\, a_2^2,\, a_3^2,\, \sqrt{2}a_2a_3,\,\sqrt{2}a_3a_1,\,\sqrt{2}a_1a_2\right)^t, \quad \text{etc}.
\eeq
 The associated $6\times6$ base matrices are given next. 

\subsubsection{Isotropic system}
The six-dimensional analogs of $\tens{I}$, $\tens{J}$ and $\tens{K}$ are 
$\whbf{I}$, the $6\times6$ identity matrix,   and $ \whbf{J}$, $ \whbf{K}$, where 
\beq{wh2}
\whbf{J}= \frac13 (\hbf{e}_1+\hbf{e}_2+\hbf{e}_3)(\hbf{e}_1+\hbf{e}_2+\hbf{e}_3)^t ,\qquad \whbf{K} = \whbf{I} - \whbf{J} \, .  
\eeq

\subsubsection{Cubic system}

The analogs of $\tens{L}$ and  $\tens{M}$ are
\beq{911}
\whbf{L} =  \whbf{I}
-\hbf{a}\hbf{a}^t - \hbf{b}\hbf{b}^t-\hbf{c}\hbf{c}^t,
\qquad  \whbf{M} = \whbf{K}-\whbf{L}.
\eeq

\subsubsection{Transversely isotropic system}

Define the $6-$vectors 
\beq{t43}
\begin{split}
& \hbf{p}= \hbf{c},\quad
\hbf{q}= \tfrac1{\sqrt{2}} (\hbf{a}  + \hbf{b}) ,\quad
\hbf{z}= \tfrac1{\sqrt{2}} (\hbf{a} - \hbf{b}) , 
\\
& \hbf{w}=\left( \sqrt{2}a_1b_1,\, \sqrt{2}a_2b_2,\, a\sqrt{2}_3b_3,\,  (a_2b_3+a_3b_2),\,(a_3b_1+a_1b_3),\, (a_1b_2+a_2b_1) \right)^t, 
\end{split}
\eeq
then the 6-dimensional matrices for the tensors  $\tens{E}_1$, $\tens{E}_2$, $\tens{E}_3$, $\tens{E}_4$,
$\tens{F}$  and $\tens{G}$ are  
\beq{t42}
\begin{split}
 &\whbf{E}_{1}  = \hbf{p}  \hbf{p}^t  , \quad 
\whbf{E}_{2}  =\hbf{q} \hbf{q}^t ,  \quad
\whbf{E}_{3}=    \hbf{p}   \hbf{q}^t , \quad
\whbf{E}_{4}=   \hbf{q}   \hbf{p}^t , 
\\ & 
\whbf{F}  =  \hbf{w}   \hbf{w}^t +  \hbf{z}   \hbf{z}^t , \quad
\whbf{G} =  \whbf{L} - \hbf{w}   \hbf{w}^t   .  
\end{split}
\eeq

\subsubsection{Tetragonal system}

The six-dimensional representation is as for TI, with the addition
\beq{t901}
\whbf{F}_{1} = \hbf{w}   \hbf{w}^t\, , \qquad
\whbf{F}_{2} =  \hbf{z}   \hbf{z}^t  \, \qquad 
\whbf{F}_{3} =  \hbf{w}  \hbf{z}^t\, , \qquad
\whbf{F}_{4} =  \hbf{z}  \hbf{w}^t  \,. 
\eeq

The trigonal  system can be treated in the same manner using equations similar to \rf{0031} for $\hbf{a}$ and \rf{t43} for $\hbf{w}$, but the details are omitted for brevity.

\section{Euclidean projection}\label{Euclideanprojection}
Euclidean projection is an essential ingredient for determining the closest tensors in both the Euclidean and the log-Euclidean metrics.  It is  defined  in abstract terms in the next subsection, with the remainder of the section  devoted to explicit procedures for  the projection. 

\subsection{Definition of the projection} 

We wish  to find  the   tensor $\tens{C}_{\rm sym}$ of a specific symmetry class
which minimizes the Euclidean  distance $\|\tens{C} - \tens{C}_{\rm sym}\|$
of  an elasticity tensor of arbitrary symmetry, $\tens{C}$, 
from the particular symmetry.  The solution is a Euclidean  decomposition
\beq{dec1}
\tens{C} =  \tens{C}_{\rm sym} + \tens{C}_{\perp \rm sym}\, ,
\eeq
where $\tens{C}_{\rm sym}$ possesses the symmetries appropriate to the symmetry class considered.  The complement,  or residue \cite{Gazis63},  is orthogonal to $\tens{C}_{\rm sym}$, 
\beq{dec1a}
\langle\tens{C}_{\rm sym} ,  \tens{C}_{\perp \rm sym}\rangle =0\, ,  
\eeq
and hence 
\begin{subequations}\label{dec1b}
\begin{align}
\|\tens{C} \|^2 &= \|\tens{C}_{\rm sym}\|^2 + \|\tens{C}_{\perp \rm sym}\|^2\, , \\
\|\tens{C} -\tens{C}_{\rm sym}\|^2 & = \|\tens{C}_{\perp \rm sym}\|^2\, . 
\end{align}
\end{subequations}
We illustrate  the recursive nature of the Euclidean projection  for the special case of isotropy. 
Any elasticity may be partitioned into orthogonal components as 
 \begin{align}\label{iso02}
 \tens{C} &= \tens{C}_{\rm iso} + \tens{C}_{\perp \rm iso} 
 \nonumber \\ 
 &=  \tens{C}_{\rm iso} + \tens{C}_{\rm cub/iso} + \tens{C}_{\rm tet/cub} + \tens{C}_{\rm ort/tet} +\tens{C}_{\rm mon/ort} +\tens{C}_{\perp \rm mon},
 \end{align} 
where 
\beq{132}
 \tens{C}_{\rm sym\, B/A} \equiv  \tens{C}_{\rm sym\, B} - \tens{C}_{\rm sym\, A}, 
 \eeq
with  sym A   of higher symmetry than sym B. 
Equation 
\rf{iso02} partitions $\tens{C}$  using the path from no symmetry to isotropy via cubic symmetry, see Figure \ref{fig1}.     An alternative decomposition via hexagonal, gives
\beq{iso03}
 \tens{C} = \tens{C}_{\rm iso} + \tens{C}_{\rm hex/iso} + \tens{C}_{\rm tet/hex} + \tens{C}_{\rm ort/tet} +\tens{C}_{\rm mon/ort} +\tens{C}_{\perp \rm mon}.
 \eeq
 The  different paths in Figure  \ref{fig1} were identified by Gazis {\it et al.\/} \cite{Gazis63} and discussed more recently in greater detail by Chadwick {\it et al.\/} \cite{CVC}.       The idea of sequential decomposition in the Euclidean norm is not new, e.g., \cite{Tu68,Browaeys04}, but the explicit projection and complement have not been represented previously.   
 
Details of the  recursive Euclidean partition of $\tens{C}$  are given in the Apendix.  For instance,  eq. \rf{leng} represents $\|\tens{C} \|^2$ using three different routes from triclinic to isotropic.   In particular, the length of the isotropic projection is given by 
\beq{053}
\|\tens{C}_{\rm iso} \|^2 = 9\kappa^2 + 20 \mu^2, 
\eeq
where $\kappa$ and $\mu$ are defined in \rf{km}, 
and the distance of $\tens{C}$ from isotropy is given by
\beq{3.6}
\left\|\tens{C}_{\perp \rm iso} \right\|^2 
= \left\|\tens{C}\right\|^2 - 9\kappa^{2} - 20\mu^{2}\, . 
\eeq
It appears that Fedorov \cite{fed} was the first to show that the distance from isotropy is 
minimized  if the isotropic tensor  is as given in \rf{gendec3}.

The remainder of this Section details three alternative projection methods  using basis tensors, using symmetry planes or groups,  and using  21-dimensional vectors.  The projection operators are valid for all symmetries, but the method based on symmetry planes becomes more complicated for the higher symmetries.  Conversely, the projection operator using basis tensors becomes simpler at the higher symmetries.  The 21-dimensional vector procedure is straightforward but is coordinate dependent. 

\subsection{Projection using a tensor basis}

We assume that  the set of tensors $\{\tens{V}_i\in \tens{E}{\rm la} \}$  form a linearly independent basis for the symmetry sym in the sense that any elasticity tensor of that symmetry may be expressed uniquely in terms of $N$ linearly independent tensors ${\tens V}_1, {\tens V}_2, 
\ldots {\tens V}_N$, where $2\le N\le 13 $ is the dimension of the  space for the material symmetry.   We have seen the explicit form of several sets of basis tensors in Section 
\ref{diffsyms}, e.g., $N=2$ for isotropic elasticity.   $N=13$ corresponds to monoclinic, which is the lowest symmetry apart from triclinic   (technically $N=21$) which is no symmetry. 
The $N$ elements of the orthogonal basis  are assumed to be independent of the elasticity itself, i.e., they do  not require     ``elasticity distributors'', which are necessary for the spectral decomposition of  $\tens{E}{\rm la}$ \cite{cowin92}.  They depend only on the choice of the planes and/or axes which define the group $G$ of symmetry preserving transformations (see below). 
The precise form of the basis tensors is irrelevant, all that is required is that  they be linearly independent in the sense of elements of a vector space, and consequently any tensor with the desired symmetry can be expressed   
\beq{dec2}
 \tens{C}_{\rm sym}  = \sum\limits_{i=1}^N a_i \tens{V}_i\, .  
\eeq
Minimizing ${\rm d}_F^2 (\tens{C}, \tens{C}_{\rm sym})$ with respect to the coefficients $a_i$  implies
\beq{527}
\frac{\partial }{\partial a_i}\left\|\tens{C} - \tens{C}_{\rm sym} \right\|^2 
= 2 \langle \tens{C} , \tens{V}_i \rangle - 2 \langle \tens{C}_{\rm sym} , \tens{V}_i \rangle
=  0. 
\eeq
Hence, 
\beq{528}
\sum\limits_{j=1}^N a_j \langle \tens{V}_j , \tens{V}_i\rangle   = \langle \tens{C} , \tens{V}_i \rangle . 
\eeq
Let  $\bd{D}$ be the $N\times N$  symmetric matrix  with elements   
 \beq{fed7}
  D_{ij}   \equiv  \langle{\tens V}_i , \,  {\tens V}_j \rangle \, .   
 \eeq
This is  invertible by virtue of the linear independence of the basis tensors, and so
\beq{m1}
a_i = \sum\limits_{j=1}^N \, D_{ij}^{-1} \langle { {\tens V}_j ,\, \tens C} \rangle
    \, .  
\eeq
Noting that $ \tens{C}_{\perp \rm sym} = \tens{C} -  \tens{C}_{\rm sym}$, we have  
\begin{align}\label{nq1}
\langle \tens{C}_{\perp \rm sym} ,  \tens{C}_{\rm sym} \rangle &= \langle 
 \tens{C} -  \sum\limits_{i=1}^N a_i  \tens{V}_i    , 
 \sum\limits_{j=1}^N a_j \tens{V}_j \rangle
 \nonumber \\
&= \sum\limits_{j=1}^N  a_j \, \langle \tens{C} , \tens{V}_j\rangle
- \sum\limits_{i=1}^N \sum\limits_{j=1}^N  a_i  a_j \,  \langle\tens{V}_i , \tens{V}_j\rangle
= 0.
\end{align} 
 Hence,  the partition  \rf{dec2} satisfies the fundamental projection property \rf{dec1}. 
 It is a simple exercise to show that $\tens{C}_{\rm sym} $ of eq. \rf{dec2} minimizes the Euclidean distance $\| \tens{C}- \tens{C}_{\rm sym} \|$ iff the coefficients are given by eq. \rf{m1}. 

The Euclidean projection can also be expressed in terms of tensors of order eight, $P_{\rm sym}$, such that 
\beq{031}
\tens{C}_{\rm sym} = P_{\rm sym}\tens{C}, 
\qquad 
\Leftrightarrow 
\qquad 
\tens{C}^{\rm sym}_{ijkl} = P^{\rm sym}_{ijklmnrs}C_{mnrs}.  
\eeq
The projector can be expressed in terms of dyadics of  basis tensors, 
\beq{0312}
 P_{\rm sym} = 
  \sum\limits_{i, j=1}^N  \, D_{ij}^{-1} \, \tens{V}_i \otimes \tens{V}_j \, .  
\eeq
We now apply these ideas to particular symmetries, focusing only on the higher symmetries  since the lower ones (monoclinic, orthorhombic) are relatively trivial  and can be handled using the projector of eq. \rf{gp1}.

\subsection{Projection operators for particular symmetries}

The projector for the  symmetries described in Section \ref{diffsyms} follow  naturally from the vectorial formulation.  In particular, we note that the $\bd D$ is diagonal in each case, and 
\begin{subequations} \label{032}
\begin{align}
P_{\rm iso} &= \tens{J}\otimes \tens{J} + \tfrac15 \tens{K}\otimes \tens{K},
\label{032a}
\\
P_{\rm cub} &= \tens{J}\otimes \tens{J} + \tfrac13 \tens{L}\otimes \tens{L}
+ \tfrac12 \tens{M}\otimes \tens{M}, \label{032b}
\\
P_{\rm hex} &= \tens{E}_1\otimes \tens{E}_1 
+ \tens{E}_2\otimes \tens{E}_2 
+ \tfrac12 \left( \tens{E}_3+\tens{E}_4\right)\otimes \left( \tens{E}_3+\tens{E}_4\right)
\nonumber \\
& \quad 
+ \tfrac12 \tens{F}\otimes \tens{F}+ \tfrac12 \tens{G}\otimes \tens{G},
\label{032c}
\\  
P_{\rm tet} &= \tens{E}_1\otimes \tens{E}_1 + \tens{E}_2\otimes \tens{E}_2 
+ \tfrac12 \left( \tens{E}_3+\tens{E}_4\right)\otimes \left( \tens{E}_3+\tens{E}_4\right)
\nonumber \\  & \quad 
+\tens{F}_1\otimes \tens{F}_1 + \tens{F}_2\otimes \tens{F}_2 
+ \tfrac12 \left( \tens{F}_3+\tens{F}_4\right)\otimes \left( \tens{F}_3+\tens{F}_4\right)
+ \tfrac12 \tens{G}\otimes \tens{G},
\\  
P_{\rm trig} &= \tens{E}_1\otimes \tens{E}_1 + \tens{E}_2\otimes \tens{E}_2 
+ \tfrac12 \left( \tens{E}_3+\tens{E}_4\right)\otimes \left( \tens{E}_3+\tens{E}_4\right)
+\tfrac12 \tens{R}_1\otimes \tens{R}_1 
\nonumber \\  & \quad 
+ \tfrac12\tens{R}_2\otimes \tens{R}_2 
+ \tfrac14 \left( \tens{R}_3+\tens{R}_4\right)\otimes \left( \tens{R}_3+\tens{R}_4\right)
+ \tfrac14 \left( \tens{R}_5+\tens{R}_6\right)\otimes \left( \tens{R}_5+\tens{R}_6\right).
\end{align}
\end{subequations}
Expressions can be given for the lower symmetries, but as discussed, the number of elementary tensors involved is  much larger.   The form of the projectors in \rf{032} are independent of the crystal axes, and provide a coordinate-free scheme for Euclidean projection.  They can be easily programmed using the 6-dimensional matrix representation.

If sym A $>$ sym B, we may convert $P_{\rm sym B}$ into a form that yields $\tens{C}_{\rm sym\, B} $ and  the complement $\tens{C}_{\rm sym\, B/A} $ 
of \rf{132}.  A  general procedure for achieving this uses the fact that  basis tensors for a given symmetry are not unique, and accordingly, alternate forms of $P_{\rm sym}$ may be found.  The decomposition
\beq{1331}
P_{\rm sym \, B} = P_{\rm sym \, A} + P_{\rm sym \, B/A}, 
\eeq
can be found by a Gram-Schmid type of process.  As an example, let  sym A$=$ iso and sym B$=$ cub,  and consider the tensor basis for cubic symmetry 
$\{ {\tens V}_1,{\tens V}_2,{\tens V}_3\} =  \{ {\tens J}, {\tens K}, {\tens L}- a{\tens M}\}$, where $a$ is a free parameter.   Requiring that $\bd D$ is diagonal implies $a=3/2$, and consequently we obtain 
\beq{1335}
P_{\rm cub}  = \tens{J}\otimes \tens{J} + \tfrac15 \tens{K}\otimes \tens{K}
+ \tfrac1{30} (2\tens{L}-3\tens{M})\otimes (2\tens{L}-3\tens{M})
= P_{\rm iso}  + P_{\rm cub/iso}\, . 
\eeq
The analogous partition for isotropy and transverse isotropy is described in Appendix A, with 
$P_{\rm hex/iso}$ given in  eq. \rf{0613}.

\subsection{Projection using the transformation group}
  
Let $G$ be the group of transformations for the material symmetry.    $G\subset SO(3)$,  the space of 3-dimensional special (or proper) orthogonal transformations.  
Under the change of basis associated with $\mat{Q}\in SO(3)$, an elasticity tensor transforms  $\tens{C} \rightarrow \tens{C}' $ where $C_{ijkl}'= Q_{ip}Q_{jq}Q_{kr}Q_{ls}C_{pqrs}$.
The projection is  
\beq{gp1}
{C}^{\rm sym}_{ijkl} = \frac{1}{n} \sum\limits_{m=1}^n\, 
Q_{ip}^{(m)}Q_{jq}^{(m)}Q_{kr}^{(m)}Q_{ls}^{(m)}C_{pqrs},   
\eeq
where $n$ is the order of the group  $G$, and 
 $\mat{Q}^{(m)}$,  are  elements of the group.  
 The order of the groups are as follows
(see Table 6 of \cite{cowin95}): triclinic $1$, monoclinic $2$, trigonal $6$, orthorhombic $4$,  tetragonal $8$, transverse isotropy $\infty +2$, and isotropy $\infty^3$.
The projection defined by \rf{gp1} was discussed by, among others, Gazis {\it et al.\/} \cite{Gazis63}, and has been applied to ultrasonic data by Fran\c{c}ois  {\it et al.\/} \cite{fr}.

The six-dimensional version of the projection \rf{gp1} is
\beq{gp1p}
\whbf{C}_{\rm sym} = \frac{1}{n} \sum\limits_{m=1}^n\, 
\whbf{Q}^{(m)} \whbf{C} \whbf{Q}^{{(m)}^t}\, , 
\eeq
where $\whbf{Q}^{(m)}  \in SO(6)$ is the orthogonal second-order tensor  corresponding to $\mat{Q}^{(m)}$, satisfying $\whbf{Q}\whbf{Q}^t = \whbf{Q}^t\whbf{Q} = \whbf{I}$. 
 A necessary and sufficient condition that $\whbf{C}_{\rm sym}$ has the correct symmetry is that $\whbf{Q}^{(j)} \whbf{C}_{\rm sym}\whbf{Q}^{{(j)}^t}$, corresponding to the transformation  $\mat{Q}^{(j)}$, is also a member of the symmetry class. Thus, for $j=1,2,\ldots, n$, 
\beq{p1a}
\whbf{Q}^{(j)} \whbf{C}_{\rm sym} \whbf{Q}^{{(j)}^t} = \frac{1}{n} \sum\limits_{i=1}^n\, 
\whbf{Q}^{(j)}\whbf{Q}^{(i)} \whbf{C} \whbf{Q}^{{(i)}^t} \whbf{Q}^{{(j)}^t}
=  \frac{1}{n} \sum\limits_{k=1}^n\, \whbf{Q}^{(k)} \whbf{C} \whbf{Q}^{{(k)}^t} \, . 
\eeq
The latter identity follows from  $\whbf{Q}^{(j)}\whbf{Q}^{(i)} = \whbf{Q}^{(k)}$ and the fact that the $\whbf{Q}^{(i)}$ form a group of order $n$.  This proves that $\tens{C}_{\rm sym}$  defined in eq. \rf{gp1} is indeed an element of the symmetry class. 

The projection formula \rf{gp1} or \rf{gp1p} may be  effected in several ways, e.g., by referring to the extensive lists of the $6\times 6$ matrices $\whbf{Q}^{(i)}$ provided by Cowin and Mehrabadi \cite{cowin95}.   We note that  each $\mat{Q}^{(i)}$ can be decomposed into a product of reflection operators \cite{cowin95}.  The fundamental  operator is $\mat{R} ({\bf n}) \equiv \mat{I} - 2 {\bf n} \otimes{\bf n} $, where  the unit vector ${\bf n}$ is normal to the plane.    Chadwick {\it et al.\/} \cite{CVC} provide  a complete description of these sets and groups.  It is also possible to represent the group of transformations in terms of elements each corresponding to a single rotation \cite{cowin95,CVC}. The equivalence  between  reflection operator and rotations follows from the identity $\mat{Q}( {\bf e} , \pi) = 
- \mat{R} ({\bf e} ) $,  
where $\mat{Q} ({\bf n} , \theta )\in SO(3)$ defines rotation about ${\bf n}$ by angle $\theta$.

We illustrate this general procedure with an example. 

\subsubsection{Example: Projection onto monoclinic  symmetry}
There is only one plane of symmetry \cite{CVC}, with normal $\bd c$, say, and  
 $n=2$ group elements $G = \{ \mat{I}, -\mat{R} ({\bd c}) \}$.      Using \rf{gp1} we have 
 \beq{9375}
{C}^{\rm mon}_{ijkl} = \tfrac{1}{2} C_{ijkl} + \tfrac{1}{2}
R_{ip}({\bd c}) R_{jq}({\bd c}) R_{kr}({\bd c}) R_{ls}({\bd c}) C_{pqrs} \, . 
\eeq
For instance, let  ${\bd c}= {\bd e}_3$, and using the explicit form of $\whbf{R}({\bd e}_3)$ from eq. (54) of \cite{cowin95} (where it is called $\whbf{R}^{(3)}$), 
\beq{9376}
\whbf{R}({\bd e}_3) = \diag (1,\,1,\,1,\,-1,\,-1,\,1 ) \, , 
\eeq
then  \rf{gp1p} gives
\beq{9377}
\whbf{C}_{\rm mon} = \frac12 \whbf{C}  + \frac12  \whbf{R}({\bd e}_3) \whbf{C}\whbf{R}({\bd e}_3)
= \begin{pmatrix}
 \hat{c}_{11} &  \hat{c}_{12} &  \hat{c}_{13} & 0 & 0 &  \hat{c}_{16} \\
 &  &  &  &  &   \\ 
     &  \hat{c}_{22} &  \hat{c}_{23} & 0 & 0 &  \hat{c}_{26} \\ 
 & ~ & ~ & ~ & ~ & ~  \\ 
   ~  &   ~     &  \hat{c}_{33} & 0 & 0 &  \hat{c}_{36} \\ 
~ & ~ & ~ & ~ & ~ & ~  \\ 
  ~  &  ~ & ~ &  \hat{c}_{44} &  \hat{c}_{45}  &0  \\
~ & ~ & ~ & ~ & ~ & ~  \\ 
 {\rm S}  &  {\rm Y} & {\rm M} & ~&  \hat{c}_{55} &0 \\
~ & ~ & ~ & ~ & ~ & ~  \\ 
~ & ~ & ~ & ~ & ~ &  \hat{c}_{66}
\end{pmatrix}\, . 
\eeq

The example illustrates that   the group projection method   is practical if the order $n$ of the group $G$ is not large.  This is the case  for monoclinic, orthorhombic and perhaps trigonal systems, but even for these symmetries it is simpler to implement the algorithm by machine.  It is not practical for isotropy and transverse isotropy, each with an infinity of symmetry planes.  Application of \rf{gp1}  for cubic isotropy is unwieldy although has been performed \cite{fr}.   The explicit orthogonal basis method of \rf{dec2} and \rf{m1}  provides a simpler procedure for the higher symmetries.

\subsection{Projections using 21-dimensional vectors}

The projection is simplest for this representation of $\tens{E}{\rm la}$ 
 because the elastic moduli are explicitly represented as a  vector.  The operator is a $21\times 21$ matrix $P_{\rm sym}$, and the closest elasticity is 
\beq{270}
X_{\rm sym} = P_{\rm sym} X,
\eeq
where $X$ is defined in \rf{21v}.  The various projection matrices can be read off from the formulas of Appendix A.   Thus,  the projector for cubic symmetry is 
\beq{1x}
P_{\rm cub} = \begin{pmatrix} p_{\rm cub}  & 0_{9\times 12} 
\\ & \\
0_{12\times 9} & 0_{12\times 12} 
\end{pmatrix}, \qquad
p_{\rm cub} = 
\begin{pmatrix}
 \frac13 & \frac13 & \frac13 & 0& 0&0&0&0& 0\\
 \frac13 & \frac13 & \frac13 & 0& 0&0&0&0& 0\\
 \frac13 & \frac13 & \frac13 & 0& 0&0&0&0& 0\\
 0& 0&0&     \frac13 & \frac13 & \frac13 & 0& 0&0\\ 
 0& 0&0&     \frac13 & \frac13 & \frac13 & 0& 0&0\\ 
 0& 0&0&     \frac13 & \frac13 & \frac13 & 0& 0&0\\ 
 0& 0&0 &0& 0&0&     \frac13 & \frac13 & \frac13  \\ 
 0& 0&0 &0& 0&0&     \frac13 & \frac13 & \frac13  \\ 
 0& 0&0 &0& 0&0&     \frac13 & \frac13 & \frac13  
 \end{pmatrix}. 
\eeq
Browaeys and Chevrot \cite{Browaeys04} derived  $P_{\rm sym}$ for isotropy, transverse isotropy (hexagonal symmetry), tetragonal, orthorhombic and monoclinic symmetry.   They found that in all cases except  monoclinic that  the non-zero part of $P_{\rm sym}$ reduces to a $9\times 9$ matrix 
$p_{\rm sym}$, as in \rf{1x}.  However, it should be noted that this is not true for the tetragonal projector derived here, see eq. \rf{tetra}, for reasons discussed in  Section \ref{tetragonal}.  Nor it is true for the trigonal projector, see eq. \rf{Trig}.

\section{Exponential, logarithm and square root of symmetric fourth-order tensors}\label{expoftens}

Unlike the Euclidean projection for the Frobenius norm,  finding minimizers for the  logarithmic and Riemannian distance functions involves  evaluation of analytic functions with symmetric fourth-order tensor arguments. 
In this section  we  derive the necessary  expressions  with emphasis on exponential,
logarithm and square root  functions that will be used to obtain the closest 
elastic tensors in section \ref{closestlog}.    
For this purpose, let $h$ denote the
analytic function given by its power series
\[
h(x) = \sum_{m=0}^\infty a_m x^m.
\]
The higher symmetries are simpler, and are considered first. 

\subsection{Isotropic system}

Using (\ref{J+K}) and (\ref{mul-JK}), for $\tens{A}$ defined in \rf{defa} 
we have
\begin{align}
\label{h-ISO}
h(\tens{A}) & = \sum_{m=0}^\infty a_m (a \tens{J} + b \tens{K})^m
= a_0 \tens{I} + \sum_{m=1}^\infty a_m (a^m \tens{J} + b^m \tens{K}) 
\nonumber \\  
& = a_0 \tens{I} + \sum_{m=0}^\infty a_m (a^m \tens{J} + b^m \tens{K}) - 
a_0 (\tens{J} + \tens{K}) = h(a) \tens{J} + h(b) \tens{K}.
\end{align}
It follows that 
\beq{exp-ISO}
\exp \tens{A} = e^a \tens{J} + e^b \tens{K}.
\eeq

When $a \ne 0$ and $b \ne 0$, the symmetric tensor $\tens{A}$ is invertible 
and its inverse is  
\beq{inv-ISO}
\tens{A}^{-1} = \frac1a \tens{J} + \frac1b \tens{K}.
\end{equation} 
The  tensor $\tens{A}$ is positive 
definite if $a>0$ and $b>0$, with  square root  given by
\beq{sqrt-ISO}
\tens{A}^{1/2} = \sqrt{a} \tens{J} + \sqrt{b} \tens{K},
\end{equation} 
and  logarithm   
\beq{log-ISO}
\log \tens{A} = \ln a \tens{J} + \ln b \tens{K}.
\end{equation}

\subsection{Cubic system}

Using $\rf{J+L+M}_1$ and (\ref{mul-JLM}), for $\tens{A}$ defined in 
\rf{gencubic} we have
\beq{h-CUB}
\begin{split}
h(\tens{A}) &= \sum_{m=0}^\infty a_0 (a \tens{J} + b \tens{L} + 
c \tens{M})^m = a_0 \tens{I} + \sum_{m=1}^\infty a_m (a^m \tens{J} + 
b^m \tens{L} + c^m \tens{M})  \\  
&= a_0 \tens{I} + \sum_{m=0}^\infty a_m (a^m \tens{J} + b^m \tens{L} +
c^m \tens{M}) - a_0(\tens{J} + \tens{L} + \tens{M}) = 
h(a) \tens{J} + h(b) \tens{L} + h(c) \tens{M}.
\end{split}
\eeq
In particular,
\beq{exp-CUB}
\exp(\tens{A}) = e^{a} \tens{J} + e^{b} \tens{L} + e^c \tens{M}.
\eeq

When $a \ne 0$, $b \ne 0$ and $c \ne 0$, the symmetric tensor $\tens{A}$ is 
invertible and its inverse is given by
\beq{inv-CUB}
\tens{A}^{-1} = \frac1a \tens{J} + \frac1b \tens{L} + \frac1c \tens{M}.
\end{equation} 
If $a>0$, $b>0$ and $c>0$ then $\tens{A}$ is positive 
definite. Its square root is 
\beq{sqrt-CUB}
\tens{A}^{1/2} = \sqrt{a} \tens{J} + \sqrt{b} \tens{L} + \sqrt{c} \tens{M},
\end{equation} 
and its logarithm is given by
\beq{log-CUB}
\Log \tens{A} = \ln a \tens{J} + \ln b \tens{L} + \ln c \tens{M}.
\end{equation}

\subsection{Transversely isotropic system}

Using (\ref{sum-TI}) and (\ref{mul-TI}), for $\tens{A}$ defined in \rf{defti} 
we have
\begin{align}\label{tih}
h(\tens{A}) &= \sum_{m=0}^\infty a_m \left[
a \tens{E}_1 + b \tens{E}_2 + c (\tens{E}_3 + \tens{E}_4) + f \tens{F} 
+ g \tens{G} \right]^m 
\nonumber \\
& = a_0 \tens{I} + \sum_{m=1}^\infty a_m \left[(a \tens{E}_1 + b \tens{E}_2 
+ c (\tens{E}_3 + \tens{E}_4))^m + f^m \tens{F} + g^m \tens{G} \right] 
 \nonumber  \\  
& = \sum_{m=0}^\infty a_m (a \tens{E}_1 + b \tens{E}_2 +
c (\tens{E}_3 + \tens{E}_4))^m + h(f) \tens{F} + h(g) \tens{G} - a_0 
(\tens{F} + \tens{G})    
\nonumber \\
& = h(0)(\tens{E}_1 + \tens{E}_2 -\tens{I}) + h(a \tens{E}_1 + b \tens{E}_2 + 
c (\tens{E}_3 + \tens{E}_4)) +  h(f) \tens{F} + h(g) \tens{G}. 
\quad
\end{align} 
Contrary to the two previous classes of material symmetries, the upper left
$4 \times 4$ multiplication table for the elementary tensors of a transversely 
isotropic medium is not diagonal. This fact makes the algebra of transversely 
isotropic tensors a bit more difficult. Fortunately, Walpole \cite{walpole84} 
showed that the algebra for $\tens{E}_1$, $\tens{E}_2$ $\tens{E}_3$ and 
$\tens{E}_4$ is equivalent to the algebra of $2 \times 2$ matrices.
The details are described in Appendix \ref{appexp}. 

For the sake of simplicity of notation, we use the symbol $\sb{a}$
to denote the quadruple $(a,b,c,d)$ and introduce the functions
\beq{defabcd}
\begin{split}
\alpha(\sb{a}) & = \frac12(a+b), \\
\beta(\sb{a}) & = \frac12(a-b), \\
\gamma(\sb{a}) & = \frac12\sqrt{(a-b)^2 + 4 (c^2 + d^2)}, \\
\delta(\sb{a}) & = \sqrt{ab - c^2-d^2} \quad \text{(if } ab - c^2 - d^2 \ge 0).
\end{split}
\eeq
The variable $d$ is not needed in this subsection, so we take $d=0$.
We used it to keep the number of auxiliary functions to a minimum.

The procedure is to diagonalize the  compound tensor argument in \rf{tih}, 
\beq{diag1}
\tens{B} \equiv  a \tens{E}_1+b\tens{E}_2 + c( \tens{E}_3+ \tens{E}_4) = 
(\alpha(\sb{a}) + \gamma(\sb{a}))  \tens{E}_+ + 
(\alpha(\sb{a}) - \gamma(\sb{a}))  \tens{E}_- ,
\eeq
where 
\beq{diag2} 
\tens{E}_\pm  = \frac12 (  \tens{E}_1+\tens{E}_2 )
\pm \frac1{2} [\cos \psi(\sb{a}) (  \tens{E}_1  -\tens{E}_2 ) + 
\sin\psi (\sb{a}) ( \tens{E}_3+ \tens{E}_4) ],
\eeq
with $\psi(\sb{a})$  defined by
\[
\begin{cases}
\cos \psi(\sb{a}) = \dfrac{\beta(\sb{a})}{\gamma(\sb{a})}, \quad \sin \psi(\sb{a}) = \dfrac{c}{\gamma(\sb{a})}
& \text{ if } \gamma(\sb{a}) > 0, \\
\psi(\sb{a}) = 0, & \text{ if } \gamma(\sb{a}) = 0.
\end{cases}
\]
The derivation is apparent from the results below, and explained in detail in Appendix \ref{appexp}.  The central feature of this decomposition is that $ \tens{E}_\pm $ satisfy $\tens{E}_\pm \tens{E}_\pm  = \tens{E}_\pm $, $\tens{E}_+\tens{E}_- = \tens{O}$, $\tens{E}_+ + \tens{E}_- = \tens{E}_1  + \tens{E}_2$, and consequently any function, $h$, of the tensor $\tens{B}$ may be expressed 
\beq{diag3} h( \tens{B})  = h( \alpha(\sb{a}) + \gamma(\sb{a}))  \tens{E}_+
+ h(\alpha(\sb{a}) - \gamma(\sb{a}))  \tens{E}_-  + h(0)(\tens{I} - \tens{E}_1 - \tens{E}_2) \, , 
\eeq
and hence
\beq{diag4} h( \tens{A})  = h( \alpha(\sb{a}) + \gamma(\sb{a}))  \tens{E}_+
+ h(\alpha(\sb{a}) - \gamma(\sb{a}))  \tens{E}_-  + h(f) \tens{F} + h(g) \tens{G} \, . 
\eeq
Converting back to the $\tens{E}_i$'s gives
\beq{diag5} 
h(\tens{A}) = h^+(\sb{a}) (\tens{E}_1 + \tens{E}_2) 
+ h^-(\sb{a}) \left[\beta(\sb{a}) (\tens{E}_1 - \tens{E}_2) 
+ c (\tens{E}_3 + \tens{E}_4) \right] + h(f) \tens{F} + h(g) \tens{G} \,,
\eeq
where 
\[
h^+(\sb{a}) = \dfrac{h(\alpha(\sb{a}) + \gamma(\sb{a})) + 
h(\alpha(\sb{a}) - \gamma(\sb{a}))}2, 
\]
and
\[
h^-(\sb{a}) = \begin{cases}
\dfrac{h(\alpha(\sb{a}) + \gamma(\sb{a})) - 
h(\alpha(\sb{a}) - \gamma(\sb{a}))}{2\gamma(\sb{a})}, & \text{ if }\gamma(\sb{a}) > 0, \\
h'(\alpha(\sb{a})), & \text{ if }\gamma(\sb{a}) = 0.
\end{cases}
\]

The results of Appendix \ref{appexp} and above imply
\beq{exp-TI}
\exp(\tens{A}) = \mathfrak{e}_1(\sb{a}) \tens{E}_1 + \mathfrak{e}_2(\sb{a}) 
\tens{E}_2 + \mathfrak{e}_3(\sb{a}) (\tens{E}_3 + \tens{E}_4) +  e^f \tens{F} 
+ e^g \tens{G}, 
\end{equation}
where
\beq{exp123}
\begin{split}
\mathfrak{e}_1(\sb{a}) & = \exp(\alpha(\sb{a})) [\cosh (\gamma(\sb{a}))
+ \beta(\sb{a}) \sinhc (\gamma(\sb{a}))], 
\\ 
\mathfrak{e}_2(\sb{a}) & = \exp (\alpha(\sb{a})) [\cosh (\gamma(\sb{a}))
- \beta(\sb{a}) \sinhc (\gamma(\sb{a}))], 
 \\ 
\mathfrak{e}_3(\sb{a}) & = c \exp (\alpha(\sb{a})) \sinhc (\gamma(\sb{a})),
\end{split}
\eeq
and $\sinhc(\cdot)$ is the hyperbolic sine cardinal function 
defined by
\beq{defsinhc}
\sinhc(x) = \begin{cases} 1 & \text{ if } x = 0,\\ \ds{\frac{\sinh (x)}x}
& \text{ otherwise}. \end{cases}
\eeq
Note that $\sinhc(x)$ is continuous at 0.

When $ab -c^2 \ne 0$, $f \ne 0$ and $g \ne 0$, the symmetric tensor 
$\tens{A}$ is invertible and 
\beq{inv-TI}
\tens{A}^{-1} = \frac{1}{ab-c^2}\left[b \tens{E}_1 + a \tens{E}_2 - 
c (\tens{E}_3 + \tens{E}_4) \right] + \frac1f \tens{F} + \frac1g \tens{G}.
\end{equation} 
$\tens{A}$ is positive definite if $a>0$, $b>0$, $ab-c^2>0$, $f>0$ and $g>0$, with square root  
\beq{sqrt-TI}
\tens{A}^{1/2} = \frac1{\sqrt{2(\alpha(\sb{a})+\delta(\sb{a}))}} \left[ (a 
+ \delta(\sb{a})) \tens{E}_1 + 
(b + \delta(\sb{a})) \tens{E}_2 + c (\tens{E}_3 + \tens{E}_4) \right] + 
\sqrt{f} \tens{F} + \sqrt{g} \tens{G},
\end{equation} 
and   logarithm  
\beq{log-TI}
\Log \tens{A} = \mathfrak{l}_1(\sb{a}) \tens{E}_1 +  \mathfrak{l}_2(\sb{a}) 
\tens{E}_2 + \mathfrak{l}_3(\sb{a}) (\tens{E}_3 + \tens{E}_4) + 
\ln f \tens{F} + \ln g \tens{G}, 
\end{equation} 
where
\beq{defls}
\begin{split}
 \mathfrak{l}_1(\sb{a}) & = \ln \delta(\sb{a}) + \beta(\sb{a}) \ell(\sb{a}),\\ 
 \mathfrak{l}_2(\sb{a}) & = \ln \delta(\sb{a}) - \beta(\sb{a}) \ell(\sb{a}),\\
 \mathfrak{l}_3(\sb{a}) & = c\; \ell(\sb{a}),
\end{split}
\eeq
with the function  $\ell(\cdot)$  defined by
\beq{deflf}
\ell(\sb{a}) = \begin{cases} \ds{\frac1{\alpha(\sb{a})}} & \text{ if } 
\gamma(\sb{a}) = 0,\\
\ds{\frac1{2 \gamma(\sb{a})}\ln \frac{\alpha(\sb{a}) + \gamma(\sb{a})}{\alpha(\sb{a}) - \gamma(\sb{a})}} & 
\text{ otherwise}. 
\end{cases}
\eeq
Note that for $\gamma(\sb{a})$ we have $\frac1{2 \gamma(\sb{a})}\ln \frac{\alpha(\sb{a}) + \gamma(\sb{a})}{\alpha(\sb{a}) - \gamma(\sb{a})} = \frac{1}{\gamma(\sb{a})} \tanh^{-1}\frac{\gamma(\sb{a})}{\alpha(\sb{a})}$ and that the limit of this
expression is $1/{\alpha(\sb{a})}$ as $\gamma(\sb{a})$ goes to zero.

\subsection{Tetragonal system}

Using  the decomposition (\ref{sum-Tet}) and the multiplication
table (\ref{mul-Tet}), an analysis similar to that used for the transversely isotropic 
system implies that for a tetragonal tensor $\tens{A}$ as defined in 
\rf{deftet},  
\begin{align} \label{tet1}
h(\tens{A}) & = h^+(\sb{a}) (\tens{E}_1 + \tens{E}_2) 
+ h^-(\sb{a}) \left[\beta(\sb{a}) (\tens{E}_1 - \tens{E}_2) 
+ c (\tens{E}_3 + \tens{E}_4) \right] \nonumber \\ & \quad +  
h^+(\sb{p}) (\tens{F}_1 + \tens{F}_2) 
+ h^-(\sb{p}) \left[\beta(\sb{p}) (\tens{F}_1 - \tens{F}_2) 
+ r (\tens{F}_3 + \tens{F}_4) \right] 
+ h(g) \tens{G} \, .
\end{align}
As before, $\sb{a} = (a,b,c,d)$ (with $d=0$) and $\sb{p} = (p,q,r,s)$
(with $s=0$).   

Thus, 
\begin{align} \label{Tet2}
\exp(\tens{A}) & = \mathfrak{e}_1(\sb{a}) \tens{E}_1 + \mathfrak{e}_2(\sb{a}) 
\tens{E}_2 + \mathfrak{e}_3(\sb{a}) (\tens{E}_3 + \tens{E}_4) \nonumber \\ 
& \quad + \mathfrak{e}_1(\sb{p}) \tens{F}_1 + \mathfrak{e}_2(\sb{p}) 
\tens{F}_2 + \mathfrak{e}_3(\sb{p}) (\tens{F}_3 + \tens{F}_4)  
+ e^g \tens{G}. 
\end{align}
The tensor 
$\tens{A}$ is invertible if $ab -c^2 \ne 0$, $pq -r^2 \ne 0$ and $g \ne 0$, and 
\beq{inv-Tet}
\tens{A}^{-1}  = \frac{1}{ab-c^2}\left[b \tens{E}_1 + a \tens{E}_2 - 
c (\tens{E}_3 + \tens{E}_4) \right] 
 + \frac{1}{pq-r^2}\left[q \tens{F}_1 + p \tens{F}_2 - 
r (\tens{F}_3 + \tens{F}_4) \right]
+ \frac1g \tens{G}.
\eeq
$\tens{A}$ is positive definite if $a>0$, $b>0$, $ab-c^2>0$, $p>0$, $q>0$, $pq -r^2 >0$ and $g>0$, 
with  square root  
\begin{align}
\label{sqrt-Tet}
\tens{A}^{1/2} &= \frac1{\sqrt{2(\alpha(\sb{a})+\delta(\sb{a}))}} 
\left[ (a + \delta(\sb{a})) \tens{E}_1 + (b + \delta(\sb{a})) \tens{E}_2 + 
c (\tens{E}_3 + \tens{E}_4) \right] \nonumber \\
& \quad + \frac1{\sqrt{2(\alpha(\sb{p})+\delta(\sb{p}))}} 
\left[ (p + \delta(\sb{p})) \tens{F}_1 + (q + \delta(\sb{p})) \tens{F}_2 + 
r (\tens{F}_3 + \tens{F}_4) \right] +
\sqrt{g} \tens{G},
\end{align} 
and  logarithm  
\begin{align}
\label{log-Tet}
\Log \tens{A} & = \mathfrak{l}_1(\sb{a}) \tens{E}_1 +  \mathfrak{l}_2(\sb{a}) 
\tens{E}_2 + \mathfrak{l}_3(\sb{a}) (\tens{E}_3 + \tens{E}_4) \nonumber \\
& \quad + \mathfrak{l}_1(\sb{p}) \tens{F}_1 +  \mathfrak{l}_2(\sb{p}) 
\tens{F}_2 + \mathfrak{l}_3(\sb{p}) (\tens{F}_3 + \tens{F}_4) +
\ln g \tens{G}.
\end{align}

\subsection{Trigonal system}

The decomposition (\ref{sum-hex}) and the multiplication table (\ref{mul-Hex}) yield  for $\tens{A}$ defined 
in \rf{defhex} 
\begin{align} \label{Hex1}
h(\tens{A}) & = h^+(\sb{a}) (\tens{E}_1 + \tens{E}_2) 
+ h^-(\sb{a}) \left[\beta(\sb{a}) (\tens{E}_1 - \tens{E}_2) 
+ c (\tens{E}_3 + \tens{E}_4) \right] \nonumber \\ & \quad +  
h^+(\sb{p}) (\tens{R}_1 + \tens{R}_2) 
+ h^-(\sb{p}) \left[\beta(\sb{p}) (\tens{R}_1 - \tens{R}_2) 
+ r (\tens{R}_3 + \tens{R}_4) + s (\tens{R}_5 + \tens{R}_6) \right] \,,
\end{align}
where, as before, $\sb{a} = (a,b,c,d)$ (with $d=0$) and $\sb{p} = (p,q,r,s)$
(here $s$ need not be zero).

Using these results  we have
\begin{align} \label{Hex2}
\exp(\tens{A}) & = \mathfrak{e}_1(\sb{a}) \tens{E}_1 + \mathfrak{e}_2(\sb{a}) 
\tens{E}_2 + \mathfrak{e}_3(\sb{a}) (\tens{E}_3 + \tens{E}_4) \nonumber \\ 
& \quad + \mathfrak{e}_1(\sb{p}) \tens{R}_1 + \mathfrak{e}_2(\sb{p}) 
\tens{R}_2 + \mathfrak{e}_3(\sb{p}) (\tens{R}_3 + \tens{R}_4) 
+ \mathfrak{e}_4(\sb{p}) (\tens{R}_5 + \tens{R}_6) , 
\end{align}
where $\mathfrak{e}_i(\cdot)$, $i=1,3$ are as defined in (\ref{exp123}) and 
$\mathfrak{e}_4(\sb{p}) = s\exp (\alpha(\sb{p})) \sinhc (\gamma(\sb{p}))$.
The tensor 
$\tens{A}$ is invertible if $ab -c^2 \ne 0$ and $pq -r^2-s^2 \ne 0$, with 
\begin{align} \label{inv-Hex}
\tens{A}^{-1} & = \frac{1}{ab-c^2}\left[b \tens{E}_1 + a \tens{E}_2 - 
c (\tens{E}_3 + \tens{E}_4) \right] \nonumber \\
& \quad + \frac{1}{pq-r^2-s^2}\left[q \tens{R}_1 + p \tens{R}_2 - 
r (\tens{R}_3 + \tens{R}_4) - s (\tens{R}_5 + \tens{R}_6)\right].
\end{align} 
$\tens{A}$ is positive definite if $a>0$, $b>0$, $ab-c^2>0$, $p>0$, $q>0$ and $pq -r^2-s^2 >0$, 
with  square root 
\beq{sqrt-Hex}
\begin{split}
\tens{A}^{1/2} =& \frac1{\sqrt{2(\alpha(\sb{a})+\delta(\sb{a}))}} 
\left[ (a + \delta(\sb{a})) \tens{E}_1 + (b + \delta(\sb{a})) \tens{E}_2 + 
c (\tens{E}_3 + \tens{E}_4) \right] 
 \\
& + \frac1{\sqrt{2(\alpha(\sb{p})+\delta(\sb{p}))}} 
\left[ (p + \delta(\sb{p})) \tens{R}_1 + (q + \delta(\sb{p})) \tens{R}_2 + 
r (\tens{R}_3 + \tens{R}_4) + s (\tens{R}_5 + \tens{R}_6) \right],
\end{split} 
\eeq
and   logarithm  
\begin{align}
\label{log-Hex}
\Log \tens{A} & = \mathfrak{l}_1(\sb{a}) \tens{E}_1 +  \mathfrak{l}_2(\sb{a}) 
\tens{E}_2 + \mathfrak{l}_3(\sb{a}) (\tens{E}_3 + \tens{E}_4) \nonumber \\
& \quad + \mathfrak{l}_1(\sb{p}) \tens{R}_1 +  \mathfrak{l}_2(\sb{p}) 
\tens{R}_2 + \mathfrak{l}_3(\sb{p}) (\tens{R}_3 + \tens{R}_4) +
\mathfrak{l}_4(\sb{p}) (\tens{R}_5 + \tens{R}_6),
\end{align} 
where $\mathfrak{l}_i(\cdot)$, $i=1,3$ are as defined in (\ref{defls}) and 
$\mathfrak{l}_4(\sb{p}) = s \ell(\sb{p})$.

\subsection{Rhombic, monoclinic and triclinic systems}

There are no practical analytical results for the exponential, logarithm and square root for these low symmetries.    The purely numerical route is recommended.

\section{The closest tensors using logarithmic norms}\label{closestlog}

We now consider the problem of finding the closest elasticity tensors using the Riemannian and the logarithmic distance functions.  The solutions use the machinery developed in the previous Section for evaluating functions of tensors.  We begin with the higher symmetries. 

\subsection{The closest isotropic tensor using the Riemannian norm}

We want to find the closest isotropic
tensor $\tens{C}^{\rm iso}_R = 3 \kappa_R \tens{J} + 2 \mu_R \tens{K}$
to a given elasticity tensor  $\tens{C}$, i.e., find $\kappa_R > 0$
and $\mu_R > 0$ such that the Riemannian distance
$\dis_R(\tens{C},\tens{C}^{\rm iso}_R)$ 
is minimized, where $\dis_R$ is defined in eq. \rf{3b}. 
To find the optimality conditions, we note that minimizing the Riemannian 
distance  is equivalent 
to minimizing the square of this distance, and  recall the following
result \cite[Prop.~2.1]{Moakher05b}
\[
\frac{d}{dt} \tr \left[ \Log^2 \bd{X}(t) \right] =
2\tr \left[ \Log \bd{X}(t) \bd{X}^{-1}(t) \frac{d}{dt} \bd{X}(t) \right]. 
\]
Here $\bd{X}(t)$ is a real matrix-valued function of the real variable $t$
such that, for all $t$ in its domain, $\bd{X}(t)$ is an invertible
matrix which does not have eigenvalues on the closed negative real line.
This result applies to fourth-order elasticity tensors via the
isomorphism (\ref{a1a}) and \rf{a1}.

Using this general result implies two conditions, 
\begin{subequations}
\label{opti-R}
\begin{align}
\dfrac{\partial \dis_R^2}{\partial \kappa_R} (\tens{C},\tens{C}^{\rm iso}_R) 
& = \dfrac{2}{\kappa_R} \tr \left[ \Log (\tens{C}^{-1} \tens{C}^{\rm iso}_R) 
\tens{J} \right], \\
\dfrac{\partial \dis_R^2}{\partial \mu_R} (\tens{C},\tens{C}^{\rm iso}_R)& =
\dfrac{2}{\mu_R} \tr \left[ \Log (\tens{C}^{-1} \tens{C}^{\rm iso}_R) 
\tens{K} \right].
\end{align}
\end{subequations}
Hence, the optimality conditions for $\tens{C}^{\rm iso}_R$ are
\beq{opti-R2}
\tr \left[ \Log (\tens{C}^{-1} \tens{C}^{\rm iso}_R) 
\tens{J} \right] = 0, \qquad
 \tr \left[ \Log (\tens{C}^{-1} \tens{C}^{\rm iso}_R) 
\tens{K} \right] = 0.
\eeq

In view of the partition (\ref{J+K}), addition of these two equations yields
\[
\tr \left[ \Log (\tens{C}^{-1} \tens{C}^{\rm iso}_R) 
\right] = 0, 
\]
which can be rewritten as\footnote{Here and throughout, the determinant of 
a fourth-order elasticity tensor $\tens{C}$ is $\det \tens{C} = \prod_{I=1}^6
\Lambda_I$, where the $\Lambda_I$'s are the eigenvalues of $\tens{C}$. From (\ref{sp})
we have $\det \tens{C} = \det \whbf{C}$, where $\whbf{C}$ is
the associated six-dimensional second-order tensor.}
(see Appendix \ref{appexp})
\[
\det (\tens{C}^{-1} \tens{C}^{\rm iso}_R) = 1, 
\]
or, equivalently,
\beq{det-iso}
\det \tens{C} = \det \tens{C}^{\rm iso}_R. 
\eeq
This is the counterpart of the fact that, for the Euclidean distance,
the closest isotropic tensor has the same trace as the given elasticity 
tensor. 

\subsubsection{The closest isotropic tensor to a given cubic tensor}

For $\tens{C} = a \tens{J} +  b \tens{L} +  c \tens{M}$,
the conditions (\ref{opti-R2}a) and (\ref{det-iso}), after some algebra, 
reduce to 
\beq{iso-cub}
\ln \dfrac{3\kappa_R}{a} = 0, \qquad 
a b^3 c^2 = 3\kappa_R (2\mu_R)^5, 
\eeq
which can readily be solved to yield
\[
3\kappa_R = a, \qquad 2\mu_R = (b^3 c^2)^{1/5}. 
\]

\subsubsection{The closest isotropic tensor to a given transversely isotropic tensor}

For $\tens{C} = a \tens{E}_1 +  b \tens{E}_2 +  c (\tens{E}_3 + \tens{E}_4)
+ f  \tens{F} +  g \tens{G}$, the conditions (\ref{opti-R2}a) and
(\ref{det-iso}) yield
\beq{iso-TI}
3 \ln \delta(\sb{x}) - \ell(\sb{x})(\beta(\sb{x}) - 2 \sqrt2 z) = 0,
\qquad (a b - c^2) f^2 g^2 = 3\kappa_R (2\mu_R)^5,
\eeq
where $\sb{x} = (x, y, z, 0)$ with
\beq{xyz}
\begin{pmatrix} x & z \\ z & y \end{pmatrix}
= \bd{\Sigma}^t \bd{\Lambda} \begin{pmatrix} A & C \\ C & B \end{pmatrix} 
\bd{\Lambda} \bd{\Sigma}, 
\qquad 
\begin{pmatrix} A & C \\ C & B \end{pmatrix} =
 \bd{\Sigma} \begin{pmatrix} a & c \\ c & b 
\end{pmatrix} \bd{\Sigma}^t , 
\eeq
where 
\beq{037b} 
\bd{\Sigma} = \tfrac1{\sqrt{3}}
\begin{pmatrix}
1 & \sqrt{2} \\ - \sqrt{2} & 1
\end{pmatrix}, \quad 
\bd{\Lambda}=\diag\left(\frac1{\sqrt{3\kappa_R}}, \frac1{\sqrt{2\mu_R}}\right).
\eeq
Since $\bd{\Sigma}$ is an orthogonal transformation, it follows that  
$AB-C^2 = ab-c^2$, $A+B= a+b$, $xy-z^2=(ab-c^2)/(6 \kappa_R \mu_R)$ and
$x+y = (a+b)/(6 \kappa_R \mu_R)$.

Equation (\ref{iso-TI})$_2$ can be used to eliminate $\kappa_R$ in equation
(\ref{iso-TI})$_1$ and hence we obtain a single nonlinear equation
for a single unknown. An appropriate choice for this unknown is
the positive variable $\xi$ defined by the ratio
\[
\xi^2 = \dfrac{3\kappa_R}{2\mu_R} = \dfrac{1+\nu_R}{1-2\nu_R},
\]
where $\nu_R$ is the Poisson's ratio of the closest isotropic tensor
with respect to the Riemannian distance.
Then (\ref{iso-TI}) yields the following equation for $\xi$
\beq{eqxi}
(\bar{A} - \bar{B} \xi^{2}) \ln \frac{\bar{A} + \bar{B} \xi^2 + R(\xi)}
{\bar{A} + \bar{B} \xi^2 - R(\xi)} - R(\xi) 
\ln \frac{\xi^{4/3}}{\bar{A} \bar{B} - \bar{C}^2} = 0,
\eeq
where
\[
R(\xi) = \sqrt{(\bar{A} - \bar{B}\xi^2)^2 + 4 \bar{C}^2 \xi^2},
\]
and $\bar{A}=A/(\det \tens{C})^{1/6}$, $\bar{B}=B/(\det \tens{C})^{1/6}$,
$\bar{C}=C/(\det \tens{C})^{1/6}$.
Note here that for this type of symmetry $\det \tens{C} = (ab-c^2) f^2 g^2$.

The isotropic moduli therefore depend upon  three unidimensional
combinations of the transversely isotropic moduli, $\bar{A}$, $\bar{B}$ and $\bar{C}$.
Once the solution $\xi$ of  eq. \rf{eqxi} is found, the isotropic
moduli are given by
\beq{kmTI}
3 \kappa_R = (\det \tens{C})^{1/6} \xi^{5/3}, \qquad
2 \mu_R = (\det \tens{C})^{1/6} \xi^{-1/3}.
\eeq

\subsubsection{The closest isotropic tensor to a given tetragonal tensor}

For $\tens{C} = a \tens{E}_1 +  b \tens{E}_2 +  c (\tens{E}_3 + \tens{E}_4)
+ p \tens{F}_1 +  q \tens{F}_2 +  r (\tens{F}_3 + \tens{F}_4)  
+  g \tens{G}$, the conditions (\ref{opti-R2}a) and 
(\ref{det-iso}) yield
\beq{iso-tet}
3 \ln \delta(\sb{x}) - \ell(\sb{x})(\beta(\sb{x}) - 2 \sqrt2 z) = 0, 
\qquad (a b - c^2) (pq - r^2) g^2 = 3\kappa_R (2\mu_R)^5, 
\eeq
where again $\sb{x} = (x, y, z, 0)$ is defined by (\ref{xyz}).
The solution is obtained in similar manner to   that for 
 transverse isotropy, thus, $\kappa_R$ and $\mu_R$ are given by
(\ref{kmTI}) but in this case $\det \tens{C} = (ab-c^2) (pq -r^2) g^2$.

\subsubsection{The closest isotropic tensor to a given trigonal tensor}

For $\tens{C} = a \tens{E}_1 +  b \tens{E}_2 +  c (\tens{E}_3 + \tens{E}_4)
+ p \tens{R}_1 +  q \tens{R}_2 +  r (\tens{R}_3 + \tens{R}_4)  
+  s (\tens{R}_5 + \tens{R}_6)$, the conditions (\ref{opti-R2}a) and 
(\ref{det-iso}) yield
\beq{iso-tri}
3 \ln \delta(\sb{x}) - \ell(\sb{x})(\beta(\sb{x}) - 2 \sqrt2 z) = 0, 
\qquad \tfrac12 (a b - c^2) (pq - r^2 - s^2)^2 = 3\kappa_R (2\mu_R)^5, 
\eeq
with $\sb{x} = (x, y, z, 0)$ of eq.  (\ref{xyz}).   
The isotropic elastic moduli are obtained in similar fashion as for 
of transverse isotropy, i.e., $\kappa_R$ and $\mu_R$ are given by
(\ref{kmTI}) but now
$\det \tens{C} = \tfrac12 (ab-c^2)(pq - r^2 - s^2)^2$.

\subsection{The closest cubic tensor using the Riemannian norm}

The optimality conditions that define the closest cubic tensor $\tens{C}^{\rm cub}_R = 3\kappa_R \tens{J} + 2\mu_R \tens{L} + 
2\eta_R \tens{M}$ to a given elasticity tensor $\tens{C}$ of lower symmetry, 
are obtained by minimizing the Riemannian 
distance $\dis_R(\tens{C},\tens{C}^{\rm cub}_R)$:
\begin{subequations}
\label{opti-cub}
\begin{align}
\dfrac{\partial \dis_R^2}{\partial \kappa_R} (\tens{C},\tens{C}^{\rm cub}_R)& =
\dfrac{2}{\kappa_R} \tr \left[ \Log (\tens{C}^{-1} \tens{C}^{\rm cub}_R) 
\tens{J} \right] = 0, \\
\dfrac{\partial \dis_R^2}{\partial \mu_R} (\tens{C},\tens{C}^{\rm cub}_R)& =
\dfrac{2}{\mu_R} \tr \left[ \Log (\tens{C}^{-1} \tens{C}^{\rm cub}_R) 
\tens{L} \right] = 0, \\
\dfrac{\partial \dis_R^2}{\partial \eta_R} (\tens{C},\tens{C}^{\rm cub}_R)& =
\dfrac{2}{\eta_R} \tr \left[ \Log (\tens{C}^{-1} \tens{C}^{\rm cub}_R) 
\tens{M} \right] = 0.
\end{align}
\end{subequations}
As $\kappa_R > 0$, $\mu_R>0$ and $\eta_R > 0$, these conditions are 
equivalent to
\beq{opti-cub2}
 \tr \left[ \Log (\tens{C}^{-1} \tens{C}^{\rm cub}_R) \tens{V} \right] = 0,\qquad
\tens{V}  = \tens{J},\, \tens{L},\, \tens{M}.
\eeq
In view of the partition $\rf{J+L+M}_1$, combining these three equations 
yields, in the same manner as for \rf{det-cub}, 
\beq{det-cub}
\det \tens{C} = \det \tens{C}^{\rm cub}_R = 3\kappa_R (2\mu_R)^3 (2\eta_R)^2. 
\eeq

If $\tens{C} = a \tens{E}_1 + b \tens{E}_2 + c (\tens{E}_3 + \tens{E}_4)
+ p \tens{F}_1 + q \tens{F}_2 + r (\tens{F}_3 + \tens{F}_4) + g \tens{G}$,
i.e., is of tetragonal symmetry, then equation (\ref{opti-cub2}a) reduces to 
\beq{cub-tet}
3 \ln \delta(\tilde{\sb{x}}) - \ell(\tilde{\sb{x}})(\beta(\tilde{\sb{x}}) - 
2 \sqrt2 \tilde{z}) = 0,
\eeq
where $\tilde{\sb{x}} = (\tilde{x}, \tilde{y}, \tilde{z}, 0)$ with
\beq{037c}
\begin{pmatrix} \tilde{x} & \tilde{z} \\ \tilde{z} & \tilde{y} \end{pmatrix}
= \bd{\Sigma}^t \begin{pmatrix} \tilde{A} & \tilde{C} \\ \tilde{C} & \tilde{B} 
\end{pmatrix} \bd{\Sigma} =
\bd{\Sigma}^t \tilde{\bd{\Lambda}} \bd{\Sigma} \begin{pmatrix} a & c \\ c & b 
\end{pmatrix} \bd{\Sigma}^t \tilde{\bd{\Lambda}} \bd{\Sigma}, 
\eeq
where $\bd{\Sigma}$ is as defined in (\ref{037b}) and
\[
\tilde{\bd{\Lambda}}=\diag\left(\frac1{\sqrt{3\kappa_R}},\frac1{\sqrt{2\eta_R}}\right).
\]
Again, since $\bd{\Sigma}$ is an orthogonal transformation, it follows that  
$\tilde{A}\tilde{B}-\tilde{{C}}^2 = ab-c^2$, $\tilde{A}+\tilde{B}= a+b$, 
$\tilde{x}\tilde{y}-{\tilde{z}}^2=(ab-c^2)/(6 \kappa_R \eta_R)$, and 
$\tilde{x}+\tilde{y}=(a+b)/(6 \kappa_R \eta_R)$.
Equation (\ref{det-cub}) becomes
\beq{444}
(ab-c^2)(pq-r^2)g^2=3\kappa_R (2\mu_R)^3 (2\eta)^2.
\eeq

Let $\zeta$ and $\sigma$ be the nondimensional positive variables defined by 
\[
\zeta^2 = \dfrac{3\kappa_R}{2\eta_R}, \qquad \sigma^2 = \dfrac{\mu_R}{\eta_R},
\]
then (\ref{cub-tet}) and (\ref{opti-cub2}b) yield the following coupled 
two equations for $\zeta$ and $\sigma$
\begin{subequations}
\label{eqzeta}
\begin{align}
(\tilde{A} - \tilde{B} \zeta^{2}) \ln \frac{\tilde{A} + \tilde{B} \zeta^2 + 
Q_1(\zeta)}{\tilde{A} + \tilde{B} \zeta^2 - Q_1(\zeta)} - Q_1(\zeta) 
\ln \frac{\zeta^{4/3} \sigma^{-2} (\det \tens{C})^{1/3}}{ab - c^2} = 0, \\
(p - q \sigma^2) \ln \frac{p + q \sigma^2 + Q_2(\sigma)}{p + q \sigma^2 - 
Q_2(\sigma)} - Q_2(\sigma) \ln \frac{\zeta^{-2} \sigma^{-1} (ab  - c^2)}{g^2} 
= 0.
\end{align}
\end{subequations}
where
\[
Q_1(\zeta) = \sqrt{(\tilde{A} - \tilde{B}\zeta^2)^2 + 4 {\tilde{C}}^2 \zeta^2}, 
\qquad Q_2(\sigma) = \sqrt{(p - q\sigma)^2 + 4 r^2 \sigma^2}.
\]
The three  moduli of the closest cubic elasticity tensor are then given by 
\[
3\kappa_R = \zeta^{5/3} \sigma^{-1} (\det \tens{C})^{1/6}, \quad
2\mu_R = \zeta^{-1/3} \sigma (\det \tens{C})^{1/6}, \quad
2\eta_R = \zeta^{-1/3} \sigma^{-1} (\det \tens{C})^{1/6}.
\]

\subsection{The closest transversely isotropic tensor using the Riemannian norm}

We want to find the closest, in the Riemannian metric, transversely isotropic
tensor $\tens{C}^{\rm hex}_R = a \tens{E}_1 + b \tens{E}_2 + c (\tens{E}_3 +
\tens{E}_4) + f \tens{F} + g \tens{G}$
to a given elasticity tensor $\tens{C}$ of lower symmetry, i.e., find 
$a > 0$, $b>0$ $f>0$, $g>0$ and $c$ with $ab - c^2>0$ such that the 
Riemannian distance $\dis_R(\tens{C},\tens{C}^{\rm hex}_R)$ is minimized. 

The optimality conditions are
\begin{subequations}
\label{opti-TI}
\begin{align}
\dfrac{\partial \dis_R^2}{\partial a} (\tens{C},\tens{C}^{\rm hex}_R)& =
\dfrac{2}{ab-c^2} \tr \left[ \Log (\tens{C}^{-1} \tens{C}^{\rm hex}_R) 
(b \tens{E}_1 - c \tens{E}_4) \right] = 0, \\
\dfrac{\partial \dis_R^2}{\partial b} (\tens{C},\tens{C}^{\rm hex}_R)& =
\dfrac{2}{ab-c^2} \tr \left[ \Log (\tens{C}^{-1} \tens{C}^{\rm hex}_R) 
(a \tens{E}_2 - c \tens{E}_3) \right] = 0, \\
\dfrac{\partial \dis_R^2}{\partial c} (\tens{C},\tens{C}^{\rm hex}_R)& =
\dfrac{2}{ab-c^2} \tr \left[ \Log (\tens{C}^{-1} \tens{C}^{\rm hex}_R) 
(-c \tens{E}_1 - c \tens{E}_2 + b \tens{E}_3 + a \tens{E}_4) \right] = 0, \\
\dfrac{\partial \dis_R^2}{\partial f} (\tens{C},\tens{C}^{\rm hex}_R)& =
\dfrac{2}{f} \tr \left[ \Log (\tens{C}^{-1} \tens{C}^{\rm hex}_R) 
\tens{F} \right] = 0, \\
\dfrac{\partial \dis_R^2}{\partial g} (\tens{C},\tens{C}^{\rm hex}_R)& =
\dfrac{2}{g} \tr \left[ \Log (\tens{C}^{-1} \tens{C}^{\rm hex}_R) 
\tens{G} \right] = 0.
\end{align}
\end{subequations}
As $a > 0$, $b>0$, $ab - c^2 > 0$, $f>0$ and $g>0$, these five conditions are 
equivalent to
\beq{opti-TI2}
\tr \left[ \Log (\tens{C}^{-1} \tens{C}^{\rm hex}_R) 
\tens{V} \right] = 0, 
\qquad
\tens{V}  = 
b\tens{E}_1-c\tens{E}_4 ,\, a\tens{E}_2-c\tens{E}_3 ,\,  
\tens{E}_1+\tens{E}_2 ,\,  \tens{F},\, \tens{G}. 
\eeq
Combining the  conditions for $\tens{E}_1+\tens{E}_2 $, $  \tens{F}$ and$ \tens{G}$, and 
  partition (\ref{sum-TI}) implies the constraint
\[
\det \tens{C} = \det \tens{C}^{\rm hex}_R. 
\]

\subsection{The closest tensors using the log-Euclidean norm}

The closest elasticity tensors according to the log-Euclidean metric follow by 
minimizing $\dis_L^2 (\tens{C}, \tens{C}_{\rm sym})$.  The stationarity condition implies that 
$(\Log \tens{C} - \Log \tens{C}_{\rm sym}) $ is orthogonal (in the Euclidean sense) to the symmetry class, and hence 
\beq{231}
\Log\tens{C}_{\rm sym} = P_{\rm sym} \Log\tens{C},  
\eeq
where the Euclidean projector is defined in \rf{0312}. 
We may therefore write 
\beq{232}
\tens{C}_{\rm sym} = \exp \left( P_{\rm sym} \Log\tens{C} \right) .   
\eeq
This may be evaluated for the particular symmetries using the explicit expressions for $P_{\rm sym}$ in \rf{032}, combined with the  formulas for the logarithm and exponential of elasticity tensors in 
Section \ref{expoftens}. 
We note in particular the Euclidean  property, 
\beq{739}
\dis_L^2 \left( \tens{C},\, \tens{C}_{\rm sym\, A}\right) 
= \dis_L^2 \left( \tens{C},\, \tens{C}_{\rm sym\, B}\right)  + 
\dis_L^2 \left( \tens{C}_{\rm sym\, B},\, \tens{C}_{\rm sym\, A}\right) ,
\qquad {\rm sym\, A \ge sym\, B}.
\eeq

We note the following identity, which is a consequence of the requirement that 
$(\Log \tens{C} - \Log \tens{C}^{\rm sym}) $ is orthogonal to each of the basis tensors,  
\beq{517}
\tr [ \tens{V}_i   \Log \tens{C}  ]
 = \tr [\tens{V}_i \Log \tens{C}_{\rm sym} ] . 
\eeq
In the remainder of this Section we apply this to the particular cases of isotropy and cubic isotropy, for which we derive explicit formulas for the closest moduli. 

\subsubsection{The closest isotropic tensor using the log-Euclidean norm}
There are two basis tensors and therefore the general  conditions \rf{517} become, using \rf{log-ISO} with $\tens{C}^{\rm iso}_L= 3\kappa_L\tens{J} + 2\mu_L \tens{K}$, 
\beq{518}
\ln 3\kappa_L = \tr [\tens{J} \Log \tens{C}] , \qquad \ln 2\mu_L  = \tfrac15\tr [\tens{K} \Log \tens{C} ].
\eeq
Hence, we obtain explicit formulas
\beq{519}
3\kappa_L = \exp \left( \tr [\tens{J} \Log \tens{C} ]\right) , \qquad 2\mu_L  = \exp \left(\tfrac15\tr [\tens{K} \Log \tens{C} ]\right) .
\eeq
Using $\tens{I}= \tens{J}+\tens{K}$ and eq. \rf{1a15}, it follows that 
\beq{5191}
3\kappa_L (2\mu_L)^5 = \det  \tens{C} .
\eeq

\subsubsection{The closest isotropic tensor to a given cubic tensor}

For $\tens{C} = a \tens{J} +  b \tens{L} +  c \tens{M}$, we use \rf{log-CUB} and $\tens{K}=\tens{L}+\tens{M}$ to simplify  
the conditions \rf{518}.  After some algebra we find \cite{Norris05f}
\[
3\kappa_L = a, \qquad 2\mu_L = (b^3 c^2)^{1/5}. 
\]
These moduli coincide with those obtained for the Riemannian norm. This is a
consequence of the fact that when two tensors $\tens{C}_1$ and  $\tens{C}_2$
commute under multiplication we have $\Log (\tens{C}_1 \tens{C}_2^{-1}) =
\Log \tens{C}_1 - \Log \tens{C}_2$.

\subsubsection{The closest isotropic tensor to a given transversely isotropic tensor}

For $\tens{C} = a \tens{E}_1 +  b \tens{E}_2 +  c (\tens{E}_3 + \tens{E}_4)
+ f  \tens{F} +  g \tens{G}$, we use eqs. \rf{log-TI}, $\rf{518}_1$ and $\rf{t91}_1$
to get
\beq{51123}
 \ln 3\kappa_L = \frac12 \ln (ab-c^2) + \frac{b-a+4\sqrt{2}c}{12 \gamma(\sb{a})}
 \ln \frac{a+b + 2\gamma(\sb{a})}{a+b - 2\gamma(\sb{a})}\, . 
\eeq
Thus, 
\beq{5112}
 3\kappa_L =\sqrt{ab-c^2} \, \left(\frac{a+b + \sqrt{(a-b)^2 +4c^2} }{a+b - \sqrt{(a-b)^2 +4c^2}}
 \right)^{ \frac{b-a+4\sqrt{2}c}{6\sqrt{(a-b)^2 +4c^2}} }, 
 \eeq
and the shear modulus then follows from \rf{5191}  as
\beq{51913}
 2\mu_L = \left[ \frac1{3\kappa_L} {(ab-c^2)f^2g^2}\right]^{1/5}.
\eeq
Note that the equation \rf{51123} for $\kappa_L$ can be cast in the form of eq. $\rf{kmTI}_1$, where 
$\xi$ satisfies  an equation similar to \rf{eqxi},
\beq{eqxiL}
(\bar{A} - \bar{B}) \ln \frac{\bar{A} + \bar{B}  + R(1)}
{\bar{A} + \bar{B}  - R(1)} - R(1) 
\ln \frac{\xi^{10/3}}{\bar{A} \bar{B} - \bar{C}^2} = 0.
\eeq

\subsubsection{The closest isotropic tensor to a given tetragonal tensor}

For $\tens{C} = a \tens{E}_1 +  b \tens{E}_2 +  c (\tens{E}_3 + \tens{E}_4)
+ p \tens{F}_1 +  q \tens{F}_2 +  r (\tens{F}_3 + \tens{F}_4)  
+  g \tens{G}$, we again use eqs. \rf{log-TI}, $\rf{518}_1$ and the identity
$\rf{t91}_1$ to derive \rf{5112} for $\kappa_L$.  The shear modulus  is
\beq{5194}
 2\mu_L = \left[ \frac1{3\kappa_L} {(ab-c^2)(pq-r^2)g^2}\right]^{1/5}.
\eeq

\subsubsection{The closest isotropic tensor to a given trigonal tensor}

For $\tens{C} = a \tens{E}_1 +  b \tens{E}_2 +  c (\tens{E}_3 + \tens{E}_4)
+ p \tens{R}_1 +  q \tens{R}_2 +  r (\tens{R}_3 + \tens{R}_4)  
+  s (\tens{R}_5 + \tens{R}_6)$, the condition \rf{5112} for $\kappa_L$ is again recovered, while $\mu_L$ follows from 
\beq{5195}
 2\mu_L = \left[ \frac1{3\kappa_L}{(ab-c^2)(pq-r^2-s^2)^2}\right]^{1/5}.
\eeq

\subsubsection{The closest cubic tensor using the log-Euclidean norm}
Proceeding in  the same way as for the isotropic case,  in this case with three basis tensors,   $\tens{C}^{\rm cub}_L= 3\kappa_L\tens{J} + 2\mu_L \tens{L}+ 2\eta_L \tens{M}$, we find that the closest cubic tensor has explicit solution 
\beq{849}
3\kappa_L = \exp \left( \tr [\tens{J} \Log \tens{C} ]\right) , 
\quad 2\mu_L  = \exp \left(\frac13\tr [\tens{L} \Log \tens{C} ]\right)  , 
\quad 2\eta_L  = \exp \left(\frac12\tr [\tens{M} \Log \tens{C} ]\right) . \quad
\eeq
The moduli satisfy the same determinant constraint as for the Riemannian norm, in this case
\beq{850}
3\kappa_L (2\mu_L)^3(2\eta_L)^2 = \det  \tens{C} .
\eeq

\subsubsection{The closest cubic tensor to a given tetragonal tensor}

For $\tens{C} = a \tens{E}_1 +  b \tens{E}_2 +  c (\tens{E}_3 + \tens{E}_4)
+ p \tens{F}_1 +  q \tens{F}_2 +  r (\tens{F}_3 + \tens{F}_4)  
+  g \tens{G}$, 
the bulk modulus   $\kappa_L$ is again given by  \rf{5112}.
We use 
\beq{792}
\tens{L} =  \tens{F}_1 +   \tens{G}, 
\eeq
to obtain 
\beq{583}
 2\mu_L = (pq-r^2)^{1/6} \, g^{2/3}\, 
 \left( \frac{ p+q+\sqrt{(p-q)^2+4r^2} }{ p+q-\sqrt{(p-q)^2+4r^2} }\right)^{
 \frac{ p-q }{ 2\sqrt{(p-q)^2+4r^2} } }\, . 
\eeq
   The shear modulus $\eta_L$  follows from 
\beq{51944}
 2\eta_L = \left[ \frac{(ab-c^2)(pq-r^2)g^2}{3\kappa_L ( 2\mu_L )^3} \right]^{1/2}.
\eeq

\subsubsection{The closest transversely isotropic tensor using the log-Euclidean norm}

In this case there are  five  basis tensors,   $\tens{C}^{\rm hex}_L= a \tens{E}_1 +  b \tens{E}_2 +  c (\tens{E}_3 + \tens{E}_4) + f  \tens{F} +  g \tens{G} $.  Using  eqs. \rf{log-TI} and \rf{517}, we have 
\beq{643}
f =   \exp \left(\frac12\tr [\tens{F} \Log \tens{C} ]\right)  ,
\qquad 
g = \exp \left(\frac12\tr [\tens{G} \Log \tens{C} ]\right) ,
\eeq
and the remaining three moduli follow from the identities   
\beq{644}
 \mathfrak{l}_1(\sb{a}) = \tr [\tens{E}_1 \Log \tens{C} ],
 \qquad
 \mathfrak{l}_2(\sb{a}) = \tr [\tens{E}_2 \Log \tens{C} ],
 \qquad
  \mathfrak{l}_3(\sb{a}) = \frac12 \tr [(\tens{E}_3+\tens{E}_4) \Log \tens{C} ].
\eeq
Thus, 
\beq{}
\hspace*{-4mm}
\begin{cases}\displaystyle{%
a =  \delta \frac{\cosh(\theta + \phi \cosh\theta)}{\cosh\theta},
\
b =  \delta \frac{\cosh(\theta - \phi \cosh\theta)}{\cosh\theta},
\
c=  \delta \frac{\sinh( \phi \cosh\theta)}{\cosh\theta}},  & \text{if } 
\phi \ne 0, \\ \ \\
a =  \delta e^\psi ,
\quad
b =  \delta e^{-\psi} ,
\quad
c=  0,  & \text{if } \phi = 0,
\end{cases}
\eeq
where 
\begin{subequations}
\begin{align}
&\phi = \frac12\tr [(\tens{E}_3+\tens{E}_4)\Log \tens{C} ],
 \qquad
 &\psi = \frac12\tr [(\tens{E}_1-\tens{E}_2)\Log \tens{C} ],
 \qquad
 \\
 &\delta = \exp \left(\frac12\tr [(\tens{E}_1+\tens{E}_2)\Log \tens{C} ]\right),
 \qquad
 &\theta = \log\left[ \frac{\psi}{\phi} + \left(\frac{\psi^2}{\phi^2} + 1\right)^{1/2}\right].
  \qquad
\end{align}
\end{subequations}

\section{Application and numerical examples}\label{numerical}


We illustrate the methods developed above by considering an example of a general elasticity tensor, with no assumed symmetry.  This type of data raises a  problem typically encountered, i.e.,   find the optimal orientation of the symmetry axes  in addition to finding the closest elastic tensors for a given symmetry axes or planes.   We first present the data, and the isotropic approximations, and then consider the question of orientation.

\subsection{Example}

A complete set of 21 elastic constants were determined ultrasonically by 
Fran\c{c}ois {\it et al.\/} \cite{fgm}.  The reader is referred to their paper for details of the measurement technique.  The  raw moduli are 
\beq{638}
{\bd C}  = \begin{pmatrix} 
243 & 136 & 135 & 22 & 52 & -17  \\ 
136 & 239 & 137 & -28 & 11 & 16  \\ 
135 & 137 & 233 & 29 & -49 & 3   \\ 
22 & -28 & 29 & 133 & -10&  -4   \\ 
52 & 11 & -49 & -10 & 119 & -2    \\ 
-17 & 16 & 3 & -4 & -2 & 130 
\end{pmatrix} \quad ({\rm GPa}). 
\eeq
We first search for the presence of symmetry planes, which are an indicator of underlying symmetry. 
Following Cowin and Mehrabadi \cite{cowin95}, define $\bd A$ and $\bd B$ by
\beq{338}
A_{ij} = C_{ijkk},
\qquad
B_{ij} = C_{ikjk}.
\eeq
If $\bd A$ and $\bd B$ have no common eigenvectors then there are no planes of reflection symmetry and the material has no effective elastic symmetry \cite{cowin87}.  We find that the two sets of eigenvectors are not coincident, and the smallest angle between  any pair from the two sets of eigenvectors is $16^\circ$.   Materials with symmetry higher than orthorhombic have 
five or fewer distinct eigenvalues $\Lambda_I$ \cite{c3}.  For the given moduli, we find 
$\Lambda_I = 47,   79,  244,  285,  312, 512$,  
consistent with the material having no symmetry plane.  

We next consider  isotropic approximations to the 21 moduli.  The Euclidean and log-Euclidean approximations follow from eqs. \rf{032a} and \rf{519}, respectively. 
The numerical procedure for finding the closest, in the Riemannian norm, isotropic tensor
$\tens{C}^{\rm iso}_R = 3 \kappa_R  \tens{J} + 2 \mu_R  \tens{K}$  is as follows.  Define the function
\[
F(3\kappa_R, 2\mu_R) = \tr [\Log (\tens{C}^{-1} (3 \kappa_R  \tens{J} + 
2 \mu_R  \tens{K})) \tens{J}].
\]
By the equal determinant rule (\ref{det-iso}), we have
$3\kappa_R (2\mu_R)^5 = \det \tens{C}$, and therefore
we  need only solve the equation for one of $\kappa_R$  or $\mu_R$.  In practice, we  solve 
\[
F(3\kappa_R, (\tfrac{\det \tens{C}}{3\kappa_R})^{1/5}) = 0,
\] 
using Newton's method.

The isotropic approximations are
\beq{2352}
\begin{array}{l|cc}
 & \kappa & \mu 
\\
\noalign
{\hrule width 6.1cm}
 {\rm Euclidean}, {\tens C} & 170.11   &      96.87 \\
 {\rm Euclidean},{\tens S} &       169.33   &      55.81\\
  {\rm log-Euclidean} &      169.84   &      75.91 \\
  {\rm Riemannian}   &    169.69    &     75.92
   \end{array}
  \eeq
The Euclidean projection is obviously not invariant under inversion.  Another way to see this is to  
consider the product 
\beq{501}
       \whbf{S}_{\rm iso}\whbf{C}_{\rm iso} \approx  1.00\whbf{I} + 0.74\whbf{K}, 
\eeq
where  $\whbf{C}_{\rm iso}$ and 
$\whbf{S}_{\rm iso}$ are the isotropic projections for the stiffness $\whbf{C}$ and
compliance $\whbf{S}=\whbf{C}^{-1}$, respectively.   
The product \rf{501} is an isotropic tensor, as expected, but not the identity.  The analogous product $\whbf{S}_{\rm sym}\whbf{C}_{\rm sym}$ of the Euclidean stiffness and compliance projections becomes closer to the identity as the symmetry is reduced, although not uniformly.  As one measure, we note that  the parameter $\det (\whbf{S}_{\rm sym}\whbf{C}_{\rm sym})$, which is unity for the full triclinic matrices, takes the values 4.2, 4.6, 6.6, 4.5, 8.1, 5.3, 15.8 for mon, ort, trig, tet, hex, cub and iso, respectively.

\subsection{Orientation effects}

This set of moduli present the general problem of finding the optimal orientation of the symmetry axes, for given symmetries.   This can be done by a ``brute force" approach of searching over all possible orientations of the basis vectors.  In practice this  is  achieved  using 
 Euler angles  $(\theta_1 ,\, \theta_2 ,\, \theta_3)$  to transform from 
$	\{ {\bf e}_1, {\bf e}_2, {\bf e}_3\} 
\rightarrow 
	\{ {\bf e}_1', {\bf e}_2', {\bf e}_3' \}
$
by  first rotating about the ${\bf e}_3$ axis by $\theta_1$, then about the intermediate ${\bf e}_1''$ axis by $\theta_2$, and finally about the ${\bf e}_3'$ axis by $\theta_3$. 
The moduli transform as ${\bd C}\rightarrow {\bd C}'$, 
 and for each triple $(\theta_1 ,\, \theta_2 ,\, \theta_3)$ the closest elasticity tensors ${\tens C}_{\rm sym}$ of different symmetries are found. Numerically, we fix the symmetries by reference to the original basis, with  
$	\{ {\bf a}, {\bf b}, {\bf c}\} = 	\{ {\bf e}_1, {\bf e}_2, {\bf e}_3\} $.

Let d be any of the three distance functions, and define
\beq{0531}
\rho_{\rm sym}(\theta_1 , \theta_2 , \theta_3) = \frac{\dis^2 \left({\tens C}_{\rm sym},\, {\tens C}_{\rm iso} \right)}
{\dis^2 \left({\tens C},\, {\tens C}_{\rm iso} \right)}.   
\eeq
Thus, 
 $0< \rho_{\rm sym} \le 1$, with equality only if the rotated ${\bd C}'$ has symmetry sym.  The preliminary analysis above indicates the absence of any symmetry so we expect ${\rho_{\rm sym}}$ to be less than unity.  
We also define 
\beq{0533}
\rho_{\rm sym}^* = 
\max\limits_{\theta_1 , \theta_2 , \theta_3} \rho_{\rm sym}(\theta_1 , \theta_2 , \theta_3) . 
\eeq
Orientations at which $\rho_{\rm sym} = \rho_{\rm sym}^*$ are candidates for symmetry axes that best approximate the moduli.

\begin{center}
\begin{table}[htbp] 
$$\begin{array}{c|cccccc}
 \dis_F, {\tens C} & {\rho_{\rm cub}} & {\rho_{\rm hex}} & {\rho_{\rm tet}}  & {\rho_{\rm ort}} & {\rho_{\rm trig}} & {\rho_{\rm mon}} 
\\
\noalign
{\hrule width 8.5cm}
{\rm cub} &{\bf 0.91}    &      0.60     &     0.91 &         0.92    &      0.61  &        0.94 \\
{\rm hex} &0.85        &  {\bf 0.60}     &     0.91 &         0.87    &      0.62  &        0.93 \\
{\rm tet} &0.01        &  0.45         & {\bf 0.92} &         0.88    &      0.47  &        0.94 \\
{\rm ort} &0.03        &  0.54         & 0.90       &   {\bf 0.94}    &      0.58  &        0.95 \\
{\rm trig} &0.12       &   0.20        &  0.21      &    0.20         & {\bf 0.95}  &        0.22 \\
{\rm mon} &0.02         & 0.07         & 0.35       &   0.81         & 0.08        &  {\bf 0.98}
\end{array}
$$
\caption{$\rho_{\rm sym}$ calculated using the Euclidean projection of the stiffness $\tens C$.  The diagonal elements in bold are the values  of $\rho_{\rm sym}^*$ for the symmetries indicated in the left column.  The other numbers in each row are the values of $\rho_{\rm sym}$ evaluated at the same  orientation as $\rho_{\rm sym}^*$.}
\end{table}
\end{center}
\begin{center}
\begin{table}[htbp] 
$$\begin{array}{c|cccccc}
 \dis_F, {\tens S}  & {\rho_{\rm cub}} & {\rho_{\rm hex}} & {\rho_{\rm tet}}  & {\rho_{\rm ort}} & {\rho_{\rm trig}} & {\rho_{\rm mon}} 
 \\
\noalign
{\hrule width 8.5cm}
{\rm cub} & {\bf 0.82}  &        0.54     &     0.83     &     0.97      &    0.54     &     0.99 \\
{\rm hex} &  0.18       &   {\bf 0.71}    &      0.90    &      0.83     &     0.72     &     0.97 \\
{\rm tet} &  0.82       &   0.35          &{\bf 0.96}    &      0.97     &     0.35     &     0.98 \\
{\rm ort} &  0.82       &   0.35         & 0.96         & {\bf 0.97}     &     0.35     &     0.98 \\
{\rm trig} &  0.21      &    0.32         & 0.33        &  0.42          &{\bf 0.84}    &     0.43 \\
{\rm mon} & 0.82        &  0.54          &0.83          &0.97          &0.54       &   {\bf 0.99}
\end{array}
$$
\caption{$\rho_{\rm sym}$ calculated using the Euclidean projection of the compliance ${\tens C}^{-1}$.  The elements are determined in the same manner as in Table 1.}
\end{table}
\end{center}
Tables 1 and 2 list the results using the Euclidean projection for the stiffness and compliance, respectively. Table 3 gives the results for the log-Euclidean distance. 
The results in  Tables 1-3 were obtained using $60\times60\times60$ discretized  Euler angles. 
In general,  larger values give an indication of the proximity of the symmetry. 
These numbers taken together suggests  that the material is not well approximated by hex, but that cub and certainly tet are reasonable candidates for approximating symmetries. 
The analogous log-Euclidean computations,   given in Table 3, reinforce this view. 
Also, 
   $\rho_{\rm cub}^* = 0.95$ for the Riemannian norm, using the method below. 
 We focus on the cubic approximation for the remainder of this Section. 
 \begin{center}
\begin{table}[htbp] 
$$\begin{array}{c|cccccc}
 \dis_L   & {\rho_{\rm cub}} & {\rho_{\rm hex}} & {\rho_{\rm tet}}  & {\rho_{\rm ort}} & {\rho_{\rm trig}} & {\rho_{\rm mon}} 
 \\
\noalign
{\hrule width 8.5cm}
{\rm cub} &  {\bf 0.92}  &        0.68    &      0.94  &        0.96   &       0.69  &0.96 \\
{\rm hex} &  0.87      &    {\bf 0.69}    &      0.93   &       0.92   &       0.70   &0.96 \\
{\rm tet} &  0.11  &        0.47     &     {\bf 0.95}  &        0.71  &        0.48   &0.96 \\
{\rm ort} &   0.09   &       0.20      &    0.37      &    {\bf 0.96}  &        0.21  &0.98 \\
{\rm trig} &   0.26    &      0.43        &  0.43       &   0.44      &    {\bf 0.94}  &0.47 \\
{\rm mon} &    0.91    &      0.60        &  0.92       &   0.96      &    0.62    &{\bf 0.99}      
\end{array}
$$
\caption{$\rho_{\rm sym}$ calculated using the log-Euclidean distance function,  determined in the same manner as in Table 1.  In particular the diagonal elements are $\rho_{\rm sym}^*$. }
\end{table}
\end{center}

\subsection{Cubic approximations}

We note some properties of the optimally oriented cubic approximations. 
First, the decomposition \rf{1335} for the Euclidean cubic projection has an interesting implication.  Noting that 
 \beq{6060}
2\tens{L}-3\tens{M}  = 5\tens{J}+2\tens{K} -5 \left(
{\bd a}\otimes{\bd a}\otimes{\bd a}\otimes{\bd a} 
 + {\bd b}\otimes{\bd b}\otimes{\bd b}\otimes{\bd b} 
 +{\bd c}\otimes{\bd c}\otimes{\bd c}\otimes{\bd c}\right) ,
 \eeq
 and using \rf{053}, 
we may write the cubic length 
\beq{0532}
\|\tens{C}_{\rm cub} \|^2 = 9\kappa^2 + 20 \mu^2  + \tfrac56
\left( 3\kappa +4\mu - c_{aa}- c_{bb}-c_{cc}\right)^2,   
\eeq
where $\kappa$ and $\mu$ are Fedorov's  isotropic moduli, and $c_{aa} = \langle \tens{C}, {\bd a}\otimes{\bd a}\otimes{\bd a}\otimes{\bd a}\rangle
= C_{ijkl}a_ia_ja_ka_l$, etc. 
Consider $\|\tens{C}_{\rm cub} \|$ as a function of the orientation of the cube axes.  Since 
$\kappa$ and $\mu$ are isotropic invariants, and $c_{aa}>0, c_{bb}>0, c_{cc}>0$ on account of the positive definite nature of $\tens{C}$, it follows that the largest length occurs  when 
$c_{aa}+ c_{bb}+c_{cc}$  achieves it smallest value.  This implies that {\it the best cubic approximation in the Euclidean sense occurs in the coordinate system with smallest value of $(c_{11}'+c_{22}'+c_{33}')$.}

The closest cubic material for log-Euclidean distance function is that which 
minimizes $\dis_L( \tens{C},\, \tens{C}_{\rm cub})$.  Using the  Euclidean property, see \rf{739}, that 
\beq{702}
\dis_L^2 \left( \tens{C},\, \tens{C}_{\rm cub}\right)
= \dis_L^2 \left( \tens{C},\, \tens{C}_{\rm iso}\right) - 
\dis_L^2 \left( \tens{C}_{\rm cub},\, \tens{C}_{\rm iso}\right), 
\eeq
and the fact that  $\dis_L( \tens{C},\, \tens{C}_{\rm iso})$ is unchanged under rotation, it follows that the optimal $\tens{C}_{\rm cub}$ {\it maximizes} 
$\dis_L \left( \tens{C}_{\rm cub}, \tens{C}_{\rm iso}\right)$. 
Let $\mu_{Li}$ denote the isotropic modulus, from eq. $\rf{519}_2$, then since the bulk modulus $\kappa_L$ is the same for isotropy and cubic symmetry, it follows that 
\beq{703}
\dis_L \left( \tens{C}_{\rm cub},\, \tens{C}_{\rm iso}\right)
=  \sqrt{\tfrac{15}2} \, | \log \frac {\mu_{L}}{\mu_{Li}}|
\eeq
Therefore, the optimal orientation is that for which $\frac12(\frac {\mu_{L}}{\mu_{Li}}+ \frac {\mu_{Li}}{\mu_{L}} )$ achieves its largest value. 

Alternatively, if we define the Euclidean cubic approximation for ${\tens C}$ as
${\tens C}_F^{\rm cub} = 3\kappa_F {\tens J} + 2\mu_F {\tens L}+ 2\eta_F {\tens L}$, then $\kappa_F = \kappa$ and the optimal  $\mu_F$ and $\eta_F$ maximize $|\mu_F-\eta_F|$  subject to the additive  constraint $\frac35\mu_F+\frac25\eta_F = \mu$, where $\mu$ is the isotropic (Fedorov) shear modulus. 
By comparison, the optimal log-Euclidean moduli maximize $|\log \mu_L-\log\eta_L|$ subject to the constraint $\frac35\log\mu_L+\frac25\log\eta_L = \log\mu$. 

The closest cubic tensor   in the Riemannian norm,  
$\tens{C}^{\rm cub}_R = 3 \kappa_R  \tens{J} + 2 \mu_R  \tens{L} + 
2 \eta_R  \tens{M}$, is obtained using a numerical scheme similar to that for the isotropic case. 
Define 
the functions
\begin{align*}
F_1(3\kappa_R, 2\mu_R, 2\eta_R) & = \tr [\Log (\tens{C}^{-1} (3 \kappa_R  
\tens{J} + 2 \mu_R  \tens{L} + 2 \eta_R  \tens{M})) \tens{J}], \\
F_2(3\kappa_R, 2\mu_R, 2\eta_R) & = \tr [\Log (\tens{C}^{-1} (3 \kappa_R  
\tens{J} + 2 \mu_R  \tens{L} + 2 \eta_R  \tens{M})) \tens{L}].
\end{align*}
Using the condition (\ref{det-cub}) of equality of determinants we can 
eliminate one of the three unknowns, and therefore
 only need to solve two equations for $\kappa_R$ and $\mu_R$ 
\[
F_1(3\kappa_R, 2\mu_R, (\tfrac{\det \tens{C}}{3\kappa_R (2\mu_R)^3})^{1/2}) 
= 0, \qquad
F_2(3\kappa_R, 2\mu_R, (\tfrac{\det \tens{C}}{3\kappa_R(2\mu_R)^3})^{1/2})= 0.
\]
These equations are solved by the two-dimensional Newton's method.

The numerical search for the optimal cubic approximation indicates that the orientation which yields the maximum $\rho_{\rm cub}^*$ coincides for the three distance functions for this set of moduli.  The optimal cubic moduli are as follows 
\beq{2357}
\begin{array}{l|ccc}
 & \kappa & \mu & \eta 
\\
\noalign
{\hrule width 6.3cm}
 {\rm Eucl}, {\tens C} & 170.1 & 139.7 &  32.6 \\
 {\rm Eucl},{\tens S} &   169.3 & 135.1  & 29.7 \\
  {\rm log-Eucl} &    169.7  &137.5  & 31.2 \\
  {\rm Riemannian}   &  169.8  & 138.1  & 30.9  \\
   \end{array}
  \eeq  
The Euclidean moduli of the first row are in agreement with Fran\c{c}ois {\it et al.\/} \cite{fgm}, who used a method based on  transformation groups to effect the projection.    Note that the Euclidean projection for compliance again gives a different answer than for stiffness.  Also, the bulk modulus for the log-Euclidean distance is the same as for the isotropic approximation, see \rf{2352}.  

This example illustrates the practical application of all three distance functions.  The procedure is similar for materials in which the closest or best symmetry for approximation might be, for instance,  transversely isotropic.   The Euclidean   projection follow from eq. \rf{032c} and the  log-Euclidean moduli from eqs. \rf{643} and \rf{644}.  The closest   transversely isotropic
tensor in the Riemannian norm can be found in a similar manner as above - in this case 
a system of four nonlinear equations for four unknowns is obtained, which 
can  be solved by Newton's method.   

\section{Conclusions}\label{Conclusions}

Three distance functions for elastic tensors have been introduced and their application to elasticity examined.  The Euclidean distance function is the most commonly used, but it lacks the  property of invariance under inversion.  Two distance functions with this important property have been described in detail: the log-Euclidean and the Riemannian norms.  For  each of the three distance functions we have developed a coordinate-independent  procedure to find the closest elasticity tensor of a given symmetry to a set of moduli of a lower symmetry. 

For the Euclidean distance function we have described a projection scheme using basis tensors, similar to vector space projections. Explicit forms for the projection operator have been given for isotropic, cubic, transversely isotropic, tetragonal and trigonal symmetries. The projection method was compared with the group transformation scheme and with the 21-dimensional approach.   We  also described the  form of the projection operator using $6\times 6$ matrices, which is easily implemented on a computer,  and  derived explicit expressions for the tensor complements/residues, and for the lengths of the projections.   This allows one to decompose, in a Pythagorean sense, the elastic stiffness or compliance, although with different results for each. 

Detailed and practical results have been presented for applying the  logarithmic based distance functions to the same problem of approximating using a higher elastic symmetry.  Both the log-Euclidean and the Riemannian distance functions are invariant under inversion.  As such they offer an unambiguous method for determining the closest moduli of a given symmetry, providing unique results regardless of whether  stiffness or compliance is considered. 

Calculation of the logarithmic distance functions requires  evaluation of the logarithm, square root and exponential of tensors.  Semi-analytical procedures for obtaining these for the most important, higher, elastic symmetries have been derived.  For the first time, practical expressions are available for computing the logarithmic distances for symmetries higher than orthotropic, that is: isotropic, cubic, hexagonal (transversely isotropic), tetragonal, and trigonal.  The numerical example considered illustrates how a data set with no symmetry can be reduced to find the optimally fitting cubic moduli.   The Euclidean/Frobenius norm yields different results based on whether stiffness or compliances is use, but the log-Euclidean and the Riemannian distance functions provide unique answers.

\newpage
\appendix  

\Appendix{The closest tensors using the Euclidean norm}\label{closesteuclidean}

Results for the Euclidean projection are summarized. We first give the explicit form of the projection for the various symmetries.  

\subsection{Summary of the Euclidean projections}

The projections onto the symmetry classes are expressed in terms of the elements of the six-dimensional second-order tensor $\whbf{C}$ of eq. \rf{a1}.  
For the purpose of subsequent calculations, we take $\{{\bd a}, {\bd b}, {\bd c}\}= \{{\bd e}_1, {\bd e}_2, {\bd e}_3\}$. The  symmetry classes are  characterized by a distinct direction, the direction of the monoclinic symmetry plane, the direction perpendicular to the plane spanned by the normals to the planes of reflection symmetry for trigonal and tetragonal symmetry, the axis of transversely isotropic symmetry, a cube axis.  In each case we take this direction as ${\bf e}_3$. 

\subsubsection{Monoclinic and orthorhombic symmetry}

\beq{mon}
 \whbf{C}_{\rm mon}   = 
\begin{pmatrix}
 \hat{c}_{11} &  \hat{c}_{12} &  \hat{c}_{13} & 0 & 0 &  \hat{c}_{16} \\
     &  \hat{c}_{22} &  \hat{c}_{23} & 0 & 0 &  \hat{c}_{26} \\ 
   ~  &   ~     &  \hat{c}_{33} & 0 & 0 &  \hat{c}_{36} \\ 
  ~  &  ~ & ~ &  \hat{c}_{44} &  \hat{c}_{45}  &0  \\
 {\rm S}  &  {\rm Y} & {\rm M} & ~&  \hat{c}_{55} &0 \\
~ & ~ & ~ & ~ & ~ &  \hat{c}_{66}
\end{pmatrix},
\eeq
\beq{orth}
 \whbf{C}_{\rm ort}   = 
\begin{pmatrix} 
 \hat{c}_{11} &  \hat{c}_{12} &  \hat{c}_{13} & 0 & 0 & 0 \\
 ~    &  \hat{c}_{22} &  \hat{c}_{23} & 0 & 0 & 0 \\ 
   ~  &   ~     &  \hat{c}_{33} & 0 & 0 & 0 \\ 
  ~  &  ~ & ~ &  \hat{c}_{44} & 0  &0  \\
 {\rm S}  &  {\rm Y} & {\rm M} & ~&  \hat{c}_{55} &0 \\
~ & ~ & ~ & ~ & ~ &  \hat{c}_{66}
\end{pmatrix}.
\eeq

\subsubsection{Tetragonal symmetry} 

$\tens{C}_{\rm tet}$ as defined  in \rf{deftet} becomes in $6\times 6$ notation
\beq{tetrag}
 \whbf{C}_{\rm tet}   = 
\begin{pmatrix}
\frac12 (b +q) & \frac12 (b -q) & \frac1{\sqrt{2}} c & ~0~ & ~0~ & \frac{r}{\sqrt{2}} \\
 ~    & \frac12 (b +q) & \frac1{\sqrt{2}} c & 0 & 0 & -\frac{r}{\sqrt{2}} \\ 
   ~  &   ~     &  a & 0 & 0 & 0 \\ 
  ~  &  ~ & ~ &  g & 0  &0  \\
 {\rm S}  &  {\rm Y} & {\rm M} & ~&  g &0 \\
~ & ~ & ~ & ~ & ~ & p
\end{pmatrix},  
\eeq
and the norm is 
\beq{t9}
 \|\tens{C}_{\rm tet} \|^2 = a^{2} +b^{2} + 2c^{2}+ p^{2} +q^{2}
+2r^2 +2g^2   \, .
\eeq
 Performing the inner products using \rf{m1}, 
with $N=7$ and 
$\{\tens{V}_1,\tens{V}_2,\tens{V}_3,\tens{V}_4,\tens{V}_5,\tens{V}_6 ,\tens{V}_6 \} = \{\tens{E}_1,\tens{E}_2,(\tens{E}_3+\tens{E}_4),\tens{F}_1 ,\tens{F}_2,(\tens{F}_3+\tens{F}_4), \tens{G}\}$
 gives 
 \beq{312}
 \begin{split}
 &a=\hat{c}_{33}, \quad 
b=\frac12 (\hat{c}_{11}+\hat{c}_{22}+ 2\hat{c}_{12}), \quad  
c=\frac1{\sqrt{2}} ( \hat{c}_{13}+\hat{c}_{23} ), \quad 
g= \frac12 (\hat{c}_{44}+\hat{c}_{55}) ,
 \\ 
& p=\hat{c}_{66}, \quad
q=\frac12 (\hat{c}_{11}+\hat{c}_{22}- 2\hat{c}_{12}), \quad 
r= \frac1{\sqrt{2}} ( \hat{c}_{16}-\hat{c}_{26} ). 
\end{split}
\eeq
In summary, the projection is 
\beq{tetra}
 \whbf{C}_{\rm tet}   = 
\begin{pmatrix}
\frac12 (\hat{c}_{11}+\hat{c}_{22}) & \hat{c}_{12} & \frac12 (\hat{c}_{13}+\hat{c}_{23}) & 0 & 0 & \frac12(\hat{c}_{16} - \hat{c}_{26})\\
 ~    & \frac12 (\hat{c}_{11}+\hat{c}_{22}) & \frac12 (\hat{c}_{13}+\hat{c}_{23}) & 0 & 0 & \frac12(\hat{c}_{26} - \hat{c}_{16})\\ 
   ~  &   ~     & \hat{c}_{33} & 0 & 0 & 0 \\ 
  ~  &  ~ & ~ & \frac12 (\hat{c}_{44}+\hat{c}_{55}) & 0  &0  \\
   &  {\rm SYM} &   & ~& \frac12 (\hat{c}_{44}+\hat{c}_{55}) &0 \\
~ & ~ & ~ & ~ & ~ & \hat{c}_{66}
\end{pmatrix} . 
\eeq

\subsubsection{Trigonal symmetry} 

$\tens{C}_{\rm trig}$ as defined  in \rf{defhex} becomes 
\beq{trigon}
 \whbf{C}_{\rm trig}   = 
\begin{pmatrix}
\frac12 (b +p) & \frac12 (b -p) & \frac1{\sqrt{2}} c & \frac1{\sqrt{2}} r & \frac1{\sqrt{2}} s &~0~ \\
 ~    & \frac12 (b +p) & \frac1{\sqrt{2}} c & -\frac1{\sqrt{2}} r & -\frac1{\sqrt{2}} s & 0 \\ 
   ~  &   ~     &  a & 0 & 0 & 0 \\ 
  ~  &  ~ & ~ &  q & 0  & -s  \\
    &  {\rm SYM} &   & ~&  q & r \\
~ & ~ & ~ & ~ & ~ & p
\end{pmatrix},  
\eeq
and the norm is 
\beq{904}
 \|\tens{C}_{\rm tet} \|^2 = a^{2} +b^{2} + 2c^{2}+ 2p^{2} +2q^{2}
+4r^2 +4s^2   \, ,
\eeq
where 
\beq{3112}
 \begin{split}
 &a=\hat{c}_{33}, \quad 
b=\frac12 (\hat{c}_{11}+\hat{c}_{22}+ 2\hat{c}_{12}), \quad  
c=\frac1{\sqrt{2}} ( \hat{c}_{13}+\hat{c}_{23} ), \quad 
p=\hat{c}_{66}, 
 \\ 
& q= \frac12 (\hat{c}_{44}+\hat{c}_{55}) ,\quad
r=\frac1{2\sqrt{2}} ( \hat{c}_{14} -\hat{c}_{24} ) +\frac12 \hat{c}_{56} , \quad 
s= \frac1{2\sqrt{2}} ( \hat{c}_{15} -\hat{c}_{25} ) -\frac12 \hat{c}_{46} . 
\end{split}
\eeq

In this case,  
\beq{Trig}
 \whbf{C}_{\rm trig}   = 
 \begin{pmatrix}
 \hat{c}_{11}^*  &  \hat{c}_{11}^* -\hat{c}_{66}^* & \tfrac12 (\hat{c}_{13}+\hat{c}_{23}) & \tfrac1{\sqrt{2}}\hat{c}_{56}^* & - \tfrac1{\sqrt{2}}\hat{c}_{46}^* & 0 \\
 ~    &  \hat{c}_{11}^* & \frac12 (\hat{c}_{13}+\hat{c}_{23}) & -\frac1{\sqrt{2}}\hat{c}_{56}^* &  \frac1{\sqrt{2}}\hat{c}_{46}^* & 0 \\ 
   ~  &   ~     & \hat{c}_{33} & 0 & 0 & 0 \\ 
  ~  &  ~ & ~ & \frac12 (\hat{c}_{44}+\hat{c}_{55}) & 0  &  \hat{c}_{46}^*  \\
  &  {\rm SYM} &  & ~& \frac12 (\hat{c}_{44}+\hat{c}_{55}) & \hat{c}_{56}^* \\
~ & ~ & ~ & ~ & ~ &  \hat{c}_{66}^*
\end{pmatrix} , \qquad
\eeq
where
\begin{subequations}\label{616}
\begin{align}\label{616a}
 \hat{c}_{11}^* &=  \tfrac18 \left( 3\hat{c}_{11}+3\hat{c}_{22}+2\hat{c}_{12}+2\hat{c}_{66}\right), 
 \\
  \hat{c}_{66}^* &=  \tfrac14 \left( \hat{c}_{11}+\hat{c}_{22}-2\hat{c}_{12}+2\hat{c}_{66}\right), 
   \label{616b} \\
 \hat{c}_{46}^* &= \tfrac12 \left(  \hat{c}_{46}- \tfrac1{\sqrt{2}}\hat{c}_{15}+\tfrac1{\sqrt{2}} \hat{c}_{25} \right),
  \label{616c} \\
  \hat{c}_{56}^* &= \tfrac12 \left(  \hat{c}_{56}+ \tfrac1{\sqrt{2}}\hat{c}_{14}-\tfrac1{\sqrt{2}} \hat{c}_{24} \right).
  \label{616d} 
\end{align} 
\end{subequations}

\subsubsection{Transverse isotropy}

With $\tens{C}_{\rm hex}$ defined as in \rf{defti}, we have 
\beq{ti2}
 \whbf{C}_{\rm hex}   = 
\begin{pmatrix}
\frac12 (b +f) & \frac12 (b -f) & \frac1{\sqrt{2}} c & ~0~ & ~0~ & ~0~ \\
 ~    & \frac12 (b +f) & \frac1{\sqrt{2}} c & 0 & 0 & 0 \\ 
   ~  &   ~     &  a & 0 & 0 & 0 \\ 
  ~  &  ~ & ~ &  g & 0  &0  \\
   &  {\rm SYM} &   & ~&  g &0 \\
~ & ~ & ~ & ~ & ~ &  f
\end{pmatrix} ,  
\eeq
and 
\beq{t65}
\|\tens{C}_{\rm hex} \|^2 = a^2 +b^2 + 2c^2+ 2f^2 +2g^2\, . 
\eeq
The projection onto the basis tensors $\{\tens{V}_1,\tens{V}_2,\tens{V}_3,\tens{V}_4,\tens{V}_5 \} = \{\tens{E}_1,\tens{E}_2,(\tens{E}_3+\tens{E}_4),\tens{F} ,\tens{G}\}$, 
yields $a$, $b$, $c$ and $g$ as given in \rf{312} and 
\beq{313a}
f=\tfrac12 (p+q) = \tfrac14 (\hat{c}_{11}+\hat{c}_{22} +2\hat{c}_{66} - 2\hat{c}_{12}) \,  . 
\eeq
Thus, 
\beq{tisoa}
 \whbf{C}_{\rm hex}   = 
\begin{pmatrix}
 \hat{c}_{11}^* & \hat{c}_{11}^* - \hat{c}_{66}^*& \frac12 (\hat{c}_{13}+\hat{c}_{23}) & 0 & 0 & 0 \\
 ~    &  \hat{c}_{11}^* & \frac12 (\hat{c}_{13}+\hat{c}_{23}) & 0 & 0 & 0 \\ 
   ~  &   ~     & \hat{c}_{33} & 0 & 0 & 0 \\ 
  ~  &  ~ & ~ & \frac12 (\hat{c}_{44}+\hat{c}_{55}) & 0  &0  \\
   &  {\rm SYM} & & ~& \frac12 (\hat{c}_{44}+\hat{c}_{55}) &0 \\
~ & ~ & ~ & ~ & ~ & \hat{c}_{66}^*
\end{pmatrix} . 
\eeq

\subsubsection{Cubic symmetry} 

The 6-D matrix associated with $\tens{C}_{\rm cub}$ defined in \rf{gencubic} 
is 
\beq{cub1}
 \whbf{C}_{\rm cub}   = 
\begin{pmatrix}
\frac13 (a + 2c)& \frac13 (a -c) & \frac13 (a -c) & ~0~ & ~0~ & ~0~ \\
 ~    & \frac13 (a + 2c) & \frac13 (a -c) & 0 & 0 & 0 \\ 
   ~  &   ~     &  \frac13 (a + 2c) & 0 & 0 & 0 \\ 
  ~  &  ~ & ~ & b & 0  &0  \\
    &  {\rm SYM}  &   & ~& b &0 \\
~ & ~ & ~ & ~ & ~ & b
\end{pmatrix} , 
\eeq
with length  
\beq{t93}
\|\tens{C}_{\rm cub}\|^2 = a^2 +3b^2 + 2c^2\, . 
\eeq
The Euclidean projection onto cubic symmetry follows from \rf{m1} with 
 $N=3$ and $\{\tens{V}_1, \,\tens{V}_2, \,\tens{V}_3\}$ $= \{\tens{J},\, \tens{L},\, \tens{M}\}$,
and gives 
\begin{subequations}\label{341}
\begin{align}\label{341a}
&a =\tfrac13 (
\hat{c}_{11}+\hat{c}_{22}+\hat{c}_{33}+2\hat{c}_{12}+2\hat{c}_{13}+2\hat{c}_{23})
\\
 & b =  \tfrac13 (\hat{c}_{44}+\hat{c}_{55}+\hat{c}_{66})\, , 
\label{341b} \\
& c = \tfrac13 (\hat{c}_{11}+\hat{c}_{22}+\hat{c}_{33} - \hat{c}_{12}-\hat{c}_{23}-\hat{c}_{31}). 
\label{341c}
\end{align} 
\end{subequations}
Note that  $a = 3\kappa$ where $\kappa$ defined in eq. $\rf{km}_1$.
In summary, the projection is 
\beq{cub11}
  \whbf{C}_{\rm cub}   =  
\begin{pmatrix}
 \hat{c}_{11}^C & \hat{c}_{12}^C & \hat{c}_{12}^C & 0 & 0 & 0 \\
 ~    &  \hat{c}_{11}^C & \hat{c}_{12}^C & 0 & 0 & 0 \\ 
   ~  &   ~     & \hat{c}_{11}^C & 0 & 0 & 0 \\ 
  ~  &  ~ & ~ & \hat{c}_{66}^C & 0  &0  \\
    &  {\rm SYM} &  & ~& \hat{c}_{66}^C  &0 \\
~ & ~ & ~ & ~ & ~ & \hat{c}_{66}^C 
\end{pmatrix} , 
\eeq
where
\beq{6112}
  \hat{c}_{11}^C = \tfrac13 (\hat{c}_{11}+ \hat{c}_{22} + \hat{c}_{33}),
\quad
\hat{c}_{12}^C = \tfrac13 (\hat{c}_{12}+ \hat{c}_{13} + \hat{c}_{23}), 
\quad
\hat{c}_{66}^C = \tfrac13 (\hat{c}_{44}+ \hat{c}_{55} + \hat{c}_{66}). 
\eeq

\subsubsection{Isotropy}

With  $\tens{C}_{\rm iso}$ as given by \rf{gendec3},
we have 
\beq{iso1}
  \whbf{C}_{\rm iso}   =  
\begin{pmatrix}
 \kappa+\frac43 \mu & \kappa -\frac23 \mu  & \kappa -\frac23 \mu & 0 & 0 & 0 \\
 ~    &  \kappa+\frac43 \mu  & \kappa -\frac23 \mu & 0 & 0 & 0 \\ 
   ~  &   ~     & \kappa+\frac43 \mu  & 0 & 0 & 0 \\ 
  ~  &  ~ & ~ & 2\mu & 0  &0  \\
   &  {\rm SYM} &  & ~& 2\mu  &0 \\
~ & ~ & ~ & ~ & ~ & 2\mu 
\end{pmatrix}  ,
\eeq
where  $\kappa$ and $\mu$ are defined  in \rf{km}.  These follow from the general projection formula \rf{m1} with $N=2$ and $\{\tens{V}_1, \,\tens{V}_2\} = \{\tens{J},\, \tens{K}\}$.

\subsection{The space between isotropy  and transverse isotropy}

We use a Gram-Schmid approach to define an orthogonal subspace between isotropy and transverse isotropy (hexagonal symmetry). 
Noting that 
\beq{t91}
 \tens{J} = \tfrac13\, [\tens{E}_1 + 2\tens{E}_2 + \sqrt{2}(\tens{E}_3+\tens{E}_4)],
 \quad
 \tens{K} = \tfrac13\, [ 2\tens{E}_1 +\tens{E}_2 -\sqrt{2}(\tens{E}_3+\tens{E}_4) +3\tens{F}+3\tens{G} ]\, , \quad
\eeq
we introduce  
\begin{subequations}\label{l123}
\begin{align}
\tens{L}_1 &= \tens{G} -\tens{F}\, , 
\\
\tens{L}_2 &=2\tens{E}_1 - 2\tens{E}_2 + \frac1{\sqrt{2}}(\tens{E}_3+\tens{E}_4)\, , 
\\
\tens{L}_3 &=  8\tens{E}_1 +4\tens{E}_2 -4\sqrt{2}(\tens{E}_3+\tens{E}_4) -3\tens{F}-3\tens{G}.
\end{align}  
\end{subequations}
Thus,  $\langle \tens{L}_i , \tens{L}_j \rangle = 0$, $i\ne j$,  and $\langle \tens{L}_i ,\tens{J} \rangle=0$, $\langle \tens{L}_i ,\tens{K}\rangle =0$, 
$i=1,2,3$.  Hence $\{ \tens{V}_i, i=1,2,\ldots 5\} = \{\tens{J},\tens{K}, \tens{L}_1,\tens{L}_2,\tens{L}_3\}$ form an   basis for Hex, 
with $\bd{D} = \diag (1,5,4,9,180)$.    Equation \rf{0312}   gives 
\begin{align}\label{0613}
P_{\rm hex} &= \tens{J}\otimes \tens{J} + \frac15 \tens{K}\otimes \tens{K}
+ \frac14 \tens{L}_1\otimes \tens{L}_1 + \frac19 \tens{L}_2\otimes \tens{L}_2
+ \frac1{180} \tens{L}_3\otimes \tens{L}_3
\nonumber \\
&= P_{\rm iso} +P_{\rm hex/iso}\,. 
\end{align} 
Thus, we have the explicit decomposition into orthogonal complements
$\tens{C}_{\rm hex} = \tens{C}_{\rm iso} + \tens{C}_{\rm hex/iso}$, where 
\beq{4444}
\tens{C}_{\rm hex/iso} = d_1\tens{L}_1+ d_2\tens{L}_2+ d_3\tens{L}_3 ,\qquad 
\left\|\tens{C}_{\rm hex/iso}\right\|^2 = 4d_1^2+9d_2^2+180d_3^2 , 
\eeq
with 
\beq{3333}
d_1= \tfrac12 ( g - f ) , \quad
d_2= \tfrac19 ( 2a - 2b + \sqrt{2}c) ,  \quad
d_3= \tfrac1{90} ( 4a + 2b -4\sqrt{2} c -3f -3 g) , \quad
\eeq
and  $a,\ldots, g$ are given by \rf{312} and \rf{313a}.  Eliminating the latter gives
\begin{subequations}\label{317}
\begin{align}
d_1&= \tfrac1{8} \left[ 2( \hat{c}_{44} +  \hat{c}_{55}+\hat{c}_{12}  ) - ( \hat{c}_{11} +\hat{c}_{22}+2 \hat{c}_{66}) \right]\, , 
\\
d_2&= \tfrac19 \left[ \hat{c}_{13}+ \hat{c}_{23}  + 2\hat{c}_{33} - ( \hat{c}_{11} +\hat{c}_{22} +2 \hat{c}_{12})  \right]\, , 
\\
d_3&= \tfrac1{360} \left[ \hat{c}_{11} +\hat{c}_{22}  +14 \hat{c}_{12} +16 (\hat{c}_{33}
-\hat{c}_{13} - \hat{c}_{23}) - 6 (\hat{c}_{44} +  \hat{c}_{55} + \hat{c}_{66})
\right]\, .
\end{align}
\end{subequations}

\subsection{Complements}

Some orthogonal complements or residues \cite{Gazis63} between the symmetry classes are presented  here.  These lead to the explicit expressions for elastic lengths in the next subsection, eqs. \rf{leng} and \rf{03}.  
Thus, 
\begin{subequations}\label{p1eq}
\begin{align}
 \whbf{C}_{\perp \rm mon}  & = 
\begin{pmatrix}
~ 0~ & ~ 0~ & ~ 0~ & \hat{c}_{14} & \hat{c}_{15} &  0 \\
 ~    &  0 & 0 & \hat{c}_{24} &  \hat{c}_{25} &  0 \\ 
   ~  &   ~     &  0 & \hat{c}_{34} & \hat{c}_{35} &  0 \\ 
  ~  &  ~ & ~ &  0 &  0  & \hat{c}_{46}  \\
   {\rm S}  &  {\rm Y} & {\rm M}  & ~&  0 &\hat{c}_{56} \\
~ & ~ & ~ & ~ & ~ &  0 
\end{pmatrix}  ,  
\label{p1} \\ 
 \whbf{C}_{\rm mon/ort}   &= 
\begin{pmatrix}
0 &  0 &  0 & 0 & 0 &  \hat{c}_{16} \\
     & 0 &  0 & 0 & 0 &  \hat{c}_{26} \\ 
   ~  &   ~     &  0 & 0 & 0 &  \hat{c}_{36} \\ 
  ~  &  ~ & ~ &  0 &  \hat{c}_{45}  &0  \\
   {\rm S}  &  {\rm Y} & {\rm M} & ~&  0 &0 \\
~ & ~ & ~ & ~ & ~ &  0
\end{pmatrix},    
\label{p2} \\ 
\whbf{C}_{\rm mon/tet}   &=  
\begin{pmatrix}
\frac12(\hat{c}_{11} - \hat{c}_{22}) & 0  & \frac12(\hat{c}_{13} - \hat{c}_{23}) & 0 & 0 & \frac12(\hat{c}_{16} + \hat{c}_{26})  \\
 &  \frac12(\hat{c}_{22} - \hat{c}_{11}) & \frac12(\hat{c}_{23} - \hat{c}_{13}) & 0 & 0& \frac12(\hat{c}_{16} + \hat{c}_{26})  \\
 &   & 0 & 0 & 0 & \hat{c}_{36}  \\
& &  & \frac12(\hat{c}_{44} - \hat{c}_{55}) & \hat{c}_{45} & 0  \\
  &   {\rm SYM} &  &   & \frac12(\hat{c}_{55} - \hat{c}_{44}) & 0  \\
 &  &  &  &  & 0  
\end{pmatrix} ,  
\label{p46} \\ 
\whbf{C}_{\rm ort/hex}   &=  
\begin{pmatrix}
r_0+\frac12(\hat{c}_{11} - \hat{c}_{22}) & -r_0  & \frac12(\hat{c}_{13} - \hat{c}_{23}) & 0 & 0 & 0 \\
 & r_0+ \frac12(\hat{c}_{22} - \hat{c}_{11}) & \frac12(\hat{c}_{23} - \hat{c}_{13}) & 0 & 0& 0  \\
 &   & 0 & 0 & 0 & 0  \\
& &  & \frac12(\hat{c}_{44} - \hat{c}_{55}) & 0 & 0  \\
  &   {\rm SYM} &  &   & \frac12(\hat{c}_{55} - \hat{c}_{44}) & 0  \\
 &  &  &  &  & -2r_0  
\end{pmatrix} ,  
\label{p4} \\ 
 \whbf{C}_{\rm tet/hex}   &=  
\begin{pmatrix}
~r_0 & -r_0  & ~0 & ~0 & ~0 & \frac12(\hat{c}_{16} - \hat{c}_{26})  \\
 & ~r_0  & ~0 & ~0 & ~0 & \frac12(\hat{c}_{26} - \hat{c}_{16})  \\
 &   & ~0 & ~0 & ~0 & ~0  \\
& &  & ~0 & ~0 & ~0  \\
  {\rm S}&   {\rm Y} & {\rm M}&   & ~0 & ~0  \\
 &  &  &  &  & -2r_0  
\end{pmatrix} ,  
\label{p22} \\ 
 \whbf{C}_{\rm tet/cub}   &=  
\begin{pmatrix}
 r_1 & -2r_2 & r_2 & 0 & 0 & \frac12(\hat{c}_{16} - \hat{c}_{26}) \\
 ~    &  r_1 & r_2 & 0 & 0 & \frac12(\hat{c}_{26} - \hat{c}_{16}) \\ 
   ~  &   ~     & -2r_1 & 0 & 0 & 0 \\ 
  ~  &  ~ & ~ & r_3 & 0  &0  \\
  {\rm S}  &  {\rm Y} & {\rm M} & ~& r_3  &0 \\
~ & ~ & ~ & ~ & ~ & -2r_3 
\end{pmatrix} , \quad
\label{tisob} \\ 
 \whbf{C}_{\rm cub/iso}   &=   
\begin{pmatrix}
 2r & -r& -r & 0 & 0 & 0 \\
 ~    &  2r & -r & 0 & 0 & 0 \\ 
   ~  &   ~     & 2r & 0 & 0 & 0 \\ 
  ~  &  ~ & ~ & -2r & 0  &0  \\
  {\rm S}  &  {\rm Y} & {\rm M} & ~& -2r  &0 \\
~ & ~ & ~ & ~ & ~ & -2r 
\end{pmatrix} , 
\label{p7}  
\end{align}
\end{subequations}
where 
\begin{subequations}\label{617}
\begin{align}
&r_0 = 
\tfrac18 (\hat{c}_{11}+\hat{c}_{22}  - 2\hat{c}_{12}-2\hat{c}_{66}),
\qquad
r_1 =  \tfrac16 (\hat{c}_{11}+ \hat{c}_{22} -2 \hat{c}_{33}),
& \\
&r_2 =  \tfrac16 (\hat{c}_{13}+ \hat{c}_{23} -2 \hat{c}_{12}),
\qquad\qquad\quad
r_3 =  \tfrac16 (\hat{c}_{44}+ \hat{c}_{55} -2 \hat{c}_{66}),
& \\
&r =  \tfrac1{15} \left( \hat{c}_{11}+ \hat{c}_{22} + \hat{c}_{33}
- \hat{c}_{12}- \hat{c}_{13} - \hat{c}_{23}   -\hat{c}_{44}- \hat{c}_{55} - \hat{c}_{66} \right) .
& \label{617b}
\end{align}
\end{subequations}
Also, $\whbf{C}_{\rm hex/iso}$ follows from \rf{317} and \rf{4444} but is not given explicitly because of its length, and neither are the complements involving $\whbf{C}_{\rm trig}$. 

\subsection{The Euclidean distances}

Three projections of the elastic length are as follows 
\beq{leng}
\begin{split}
\left\|\tens{C} \right\|^2 &=  
\left\|\tens{C}_{\rm iso}\right\|^2 
+ {\left\|\tens{C}_{\rm cub/iso}\right\|^2} 
+ {\left\|\tens{C}_{\rm tet/cub}\right\|^2} 
+ {\left\|\tens{C}_{\rm mon/tet}\right\|^2}
+ {\left\|\tens{C}_{\perp \rm mon}\right\|^2}
\\ 
&=
\left\|\tens{C}_{\rm iso}\right\|^2 
+  {\left\|\tens{C}_{\rm hex/iso}\right\|^2} 
+  {\left\|\tens{C}_{\rm tet/hex}\right\|^2} 
+ {\left\|\tens{C}_{\rm mon/tet}\right\|^2}
+ {\left\|\tens{C}_{\perp \rm mon}\right\|^2}\, 
\\ 
&=
\left\|\tens{C}_{\rm iso}\right\|^2 
+  {\left\|\tens{C}_{\rm hex/iso}\right\|^2} 
+  {\left\|\tens{C}_{\rm ort/hex}\right\|^2} 
+ {\left\|\tens{C}_{\rm mon/ort}\right\|^2}
+ {\left\|\tens{C}_{\perp \rm mon}\right\|^2}\, 
. 
\end{split}
\eeq
These can be explicitly calculated using  
\begin{subequations}\label{03}
\begin{align}
\left\| \whbf{C}_{\perp \rm mon} \right\|^2 =& 
2\, \left(  \hat{c}_{14}^2+ \hat{c}_{15}^2+  \hat{c}_{24}^2+ \hat{c}_{25}^2+  \hat{c}_{34}^2+ \hat{c}_{35}^2+
\hat{c}_{46}^2+ \hat{c}_{56}^2\right) \, , 
\\
\left\| \whbf{C}_{\rm mon/ort} \right\|^2 =& 
2\, \left( \hat{c}_{16}^2+ \hat{c}_{26}^2+  \hat{c}_{36}^2+  \hat{c}_{45}^2 \right) \, ,
\label{03b}
\\
\left\| \whbf{C}_{\rm ort/hex} \right\|^2 =& 
\tfrac18 \left( \hat{c}_{11} +\hat{c}_{22}-2\hat{c}_{12} - 2\hat{c}_{66}\right)^2 +
\tfrac12  (\hat{c}_{11}-\hat{c}_{22})^2 +  (\hat{c}_{13}-\hat{c}_{23})^2 + \tfrac12 (\hat{c}_{44}-\hat{c}_{55})^2 
 \, , 
\label{03cx} \\
\left\| \whbf{C}_{\rm mon/tet} \right\|^2 =& 
\tfrac12  (\hat{c}_{11}-\hat{c}_{22})^2 
+  (\hat{c}_{13}-\hat{c}_{23})^2 + 
\tfrac12 (\hat{c}_{44}-\hat{c}_{55})^2 +
(\hat{c}_{16}+\hat{c}_{26})^2
+2\hat{c}_{45}^2+2\hat{c}_{36}^2
 \, , 
\label{03c} \\
{\left\| \whbf{C}_{\rm tet/cub} \right\|^2} =& 
\tfrac16 
(\hat{c}_{11}+ \hat{c}_{22} -2 \hat{c}_{33})^2
+\tfrac13  (\hat{c}_{13}+ \hat{c}_{23} -2 \hat{c}_{12})^2
\nonumber \\ &  +\tfrac16 (\hat{c}_{44}+ \hat{c}_{55} -2 \hat{c}_{66})^2  
 +(\hat{c}_{16}-\hat{c}_{26})^2 , \quad 
\label{03f}\\
{\left\|\whbf{C}_{\rm cub/iso}\right\|^2} =&  
\tfrac{2}{15}\left( \hat{c}_{11}+ \hat{c}_{22} + \hat{c}_{33}
- \hat{c}_{12}- \hat{c}_{13} - \hat{c}_{23}  -\hat{c}_{44}- \hat{c}_{55} - \hat{c}_{66}\right)^2 ,
\label{03g}\\
{\left\| \whbf{C}_{\rm tet/hex} \right\|^2} =&  
\tfrac18 \left( \hat{c}_{11} +\hat{c}_{22}-2\hat{c}_{12} - 2\hat{c}_{66}\right)^2 
+(\hat{c}_{16}-\hat{c}_{26})^2 
\label{03d} \\
{\left\|\whbf{C}_{\rm hex/iso}\right\|^2}=& 
\tfrac1{16} \left[ 2( \hat{c}_{44} +  \hat{c}_{55}+\hat{c}_{12}  ) - ( \hat{c}_{11} +\hat{c}_{22}+2 \hat{c}_{66}) \right]^2
\nonumber \\ 
& + \tfrac19 \left[ \hat{c}_{13}+ \hat{c}_{23}  + 2\hat{c}_{33} - ( \hat{c}_{11} +\hat{c}_{22} +2 \hat{c}_{12})  \right]^2 
\nonumber \\ 
& +  \tfrac1{720} \left[ \hat{c}_{11} +\hat{c}_{22}  +14 \hat{c}_{12} +16 (\hat{c}_{33}
-\hat{c}_{13} - \hat{c}_{23}) - 6 (\hat{c}_{44} +  \hat{c}_{55} + \hat{c}_{66})
\right]^2, 
\label{03h}
\end{align}
\end{subequations}
and   eqs. \rf{km} and \rf{053} give
\beq{629}
\begin{split}
\left\|\tens{C}_{\rm iso}\right\|^2  =&
\tfrac19\left( 
\hat{c}_{11}+\hat{c}_{22}+\hat{c}_{33}+2\hat{c}_{12}+2\hat{c}_{13}+2\hat{c}_{23}\right)^2 
\\
& + \tfrac1{45}\left[ 2(\hat{c}_{11}+\hat{c}_{22}+\hat{c}_{33} - \hat{c}_{12}-\hat{c}_{23}-\hat{c}_{31}) +3(\hat{c}_{44}+\hat{c}_{55}+\hat{c}_{66}) \right]^2. 
\end{split}
\eeq
All versions of the elastic length in \rf{leng} are the sum of 21 positive numbers, the same as the raw form of the squared length in any rectangular coordinate system using the Voigt elements.  However, the 
three sets of 21 positive numbers in \rf{leng} contain far more information about  the underlying elastic symmetry of the material.

\Appendix{Exponential, logarithm and square root of Hermitian matrices}\label{appexp}

Let $A$ be the $2\times 2$ Hermitian matrix
\beq{1a1}
A = \begin{bmatrix} a & c+id \\ c-id & b \end{bmatrix},
\eeq
where $a$, $b$, $c$ and $d$ are  real numbers.
The matrix  can be decomposed into the sum of an  isotropic part and a deviatoric part, 
\beq{1a2}
A = A_{\rm iso} + A_{\rm dev} =
\begin{bmatrix} \alpha & 0 \\ 0 & \alpha \end{bmatrix}
+  \begin{bmatrix} \beta & c+id \\ c-id & -\beta \end{bmatrix},
\eeq
where  $\alpha = \ds{(a+b)/2}$, $\beta = \ds{(a-b)/2}$ and 
$\gamma = \sqrt{\beta^2 + c^2 + d^2}$.
Note that  $A_{\rm dev}$ vanishes if and only if $\gamma=0$, 
i.e., $a=b$ and $c=d=0$, and that $A_{\rm dev}^2 = \gamma^2 I$.
The eigenvalues of $A$ are real and  given by
\beq{1a3}
\lambda_{\pm} = \alpha \pm \gamma.
\eeq
For convenience we introduce the matrix 
 \beq{1a4}
 A' = \begin{cases} I & \text{ if } \gamma = 0,\\
\frac1\gamma A_{\rm dev} & \text{ if } \gamma > 0, \end{cases}
\eeq so that  $A'^2 = I$.   The exponential of $A$ is then
 \beq{1a5}
\exp A = e^\alpha (\cosh \gamma I  + 
\sinh  A') = \begin{cases}
e^\alpha I & \text{ if } \gamma = 0, \\
e^\alpha (\cosh \gamma I + 
\ds{\frac{\sinh \gamma}{\gamma}} A_{\rm dev}) & \text{ if } \gamma > 0.
\end{cases}
\eeq
When $ab -c^2 \ne 0$ the matrix $A$ is invertible and its inverse is
\begin{align}
A^{-1} &= \frac1{ab-c^2-d^2}\begin{bmatrix} b & -(c+id) \\ -(c-id) & 
a \end{bmatrix}= \frac1{ab-c^2-d^2}(\alpha I - \gamma A') 
\nonumber \\ & = 
\frac1{ab-c^2-d^2}(\alpha I - A_{\rm dev}).
\end{align}
If $a>0$, $b>0$ and $ab-c^2-d^2>0$, then $A$ is positive definite. Let 
$\delta = \sqrt{ab - c^2 - d^2}$, then the positive-definite square root of $A$ is 
 \beq{1a6}
A^{1/2} = \frac1{\sqrt{2(\alpha + \delta)}}
\begin{bmatrix} a + \delta & c+id \\ c-id & b + \delta \end{bmatrix} =
\frac1{\sqrt{2(\alpha + \delta)}} \left[ \left(\alpha + \delta \right) 
I + A_{\rm dev} \right],
\eeq
and the inverse of its square root is given by
 \beq{1a7}
A^{-1/2} = \frac1{\delta \sqrt{2(\alpha + \delta)}}
\begin{bmatrix} b + \delta & -(c+id) \\ -(c-id) & a + \delta \end{bmatrix} =
\frac1{\delta \sqrt{2(\alpha + \delta)}} \left[ \left(\alpha + \delta 
\right) I - A_{\rm dev} \right].
\eeq
The  logarithm is
 \beq{1a8}
\Log A =\ln \delta I +  \ln \sqrt{\frac{\alpha + \gamma}{\alpha - \gamma}} A' 
= \begin{cases}
\ln \delta I & \text{ if } \gamma = 0,\\ 
\ln \delta I + \ds{\frac1{\gamma} \ln \sqrt{\frac{\alpha + \gamma}{\alpha - 
\gamma}} A_{\rm dev}} & \text{ if } \gamma > 0. 
\end{cases}
\eeq

More generally, for any analytic function $f$ we have
 \beq{1a9}
f(A) = \begin{cases} f(\alpha) I & \text{ if } \gamma = 0,\\
\ds{\frac{f(\alpha+\gamma) + f(\alpha-\gamma)}2 I + \frac{f(\alpha+\gamma) - f(\alpha-\gamma)}{2\gamma} A_{\rm dev}} & \text{ if } \gamma > 0.
\end{cases}
\eeq 
The Cayley-Hamilton theorem states that
 \beq{1a11}
\det(A) I - \tr(A) A + A^2 = 0.
\eeq
Therefore, any analytic function of $A$ can be written as a linear
combination of $I$, $A$ and $A^2$, or, if $A$ is invertible, as a linear
combination of $I$, $A$ and $A^{-1}$. Using the fact that $2A_{\rm dev}
= A - \delta^2 A^{-1}$, we have the alternative expressions for the 
exponential, logarithm and square roots
\begin{subequations}
\begin{align}
& \exp A = e^\alpha \left[ \cosh \gamma I + \frac{\sinh \gamma}{2\gamma} (A
- \delta^2  A^{-1}) \right],\\
& \Log A = \ln \delta I + \frac1{4\gamma} \ln \frac{\alpha + \gamma}
{\alpha - \gamma} \left( A - \delta^2  A^{-1} \right),\\
& A^{1/2} = \frac1{\sqrt{2(\alpha+\delta)}}\left[ \delta I + A \right],\\
& A^{-1/2} = \frac1{\sqrt{2(\alpha+\delta)}}\left[ I + \delta A^{-1} \right].
\end{align}
\end{subequations}

We recall that for any $n\times n$ matrix $M$ we have 
\beq{1a14}
\det (\exp M) = e^{\tr M}.
\eeq
Therefore, when $M = \Log P$ for some matrix $P$ we get
\beq{1a15}
\det P = e^{\tr (\Log P)}.
\eeq

\newpage

\end{document}

%% file: pic.tex
\begin{figure}[t]

\begin{center}
\begin{picture}(115,70)(-5,-65)
  
    \gasset{linewidth=0.75,Nh=6,Nmr=3,Nadjust=w,Nadjustdist=3}
  \node(A)(50,0){triclinic 21}
  \gasset{linewidth=0.5}
  		\fmark[fangle=-30,flength=45,AHangle=30,AHLength=3](A)
			\fmark[fangle=-150,flength=15,AHangle=30,AHLength=3](A)
    
  \node(B)(25,-15){monoclinic 13}
  		\fmark[fangle=-30,flength=15,AHangle=30,AHLength=3](B)
			\fmark[fangle=-150,flength=15,AHangle=30,AHLength=3](B)
   		
  \node[Nh=6,Nmr=3](B)(0,-30){tetragonal 7}
  			\fmark[fangle=-30,flength=15,AHangle=30,AHLength=3](B)
  \node[Nh=6,Nmr=3](B)(50,-30){orthotropic 9}
  			\fmark[fangle=-150,flength=15,AHangle=30,AHLength=3](B)
  \node[Nh=6,Nmr=3](B)(100,-30){trigonal 7}
  		\fmark[fangle=-150,flength=15,AHangle=30,AHLength=3](B)
  		\fmark[fangle=-162,flength=84,AHangle=30,AHLength=3](B)
  		
  \node[dash={1.5}0,Nh=6,Nmr=3](B)(25,-45){tetragonal 6}
		  \fmark[fangle=-30,flength=15,AHangle=30,AHLength=3](B)
			\fmark[fangle=-150,flength=15,AHangle=30,AHLength=3](B)
	\node[dash={1.5}0,Nh=6,Nmr=3](B)(80,-45){trigonal 6}
			\fmark[fangle=-150,flength=15,AHangle=30,AHLength=3](B)
			 			
  \node[Nh=6,Nmr=3](B)(0,-60){cubic 3}
  			\fmark[fangle=-30,flength=15,AHangle=30,AHLength=3](B)
  \node[Nh=6,Nmr=3](B)(50,-60){hexagonal 5}
   			\fmark[fangle=-150,flength=15,AHangle=30,AHLength=3](B)
   			
  \node[linewidth=0.75,Nh=6,Nmr=3](B)(25,-75){isotropic 2}
  
 \end{picture}
 
 \bigskip   \bigskip  \bigskip  \bigskip

 \caption{The sequence of increasing elastic symmetries, from triclinic to isotropic.  
	The number of independent elastic constants are listed.  The dashed boxes for trigonal 6 and tetragonal 6 indicate that these are obtained from the lower symmetries by rotation, and do not represent new elastic symmetries. The true symmetries are the eight in solid boxes \cite{CVC}. }
		\label{fig1} 

		\end{center}  
	\end{figure}
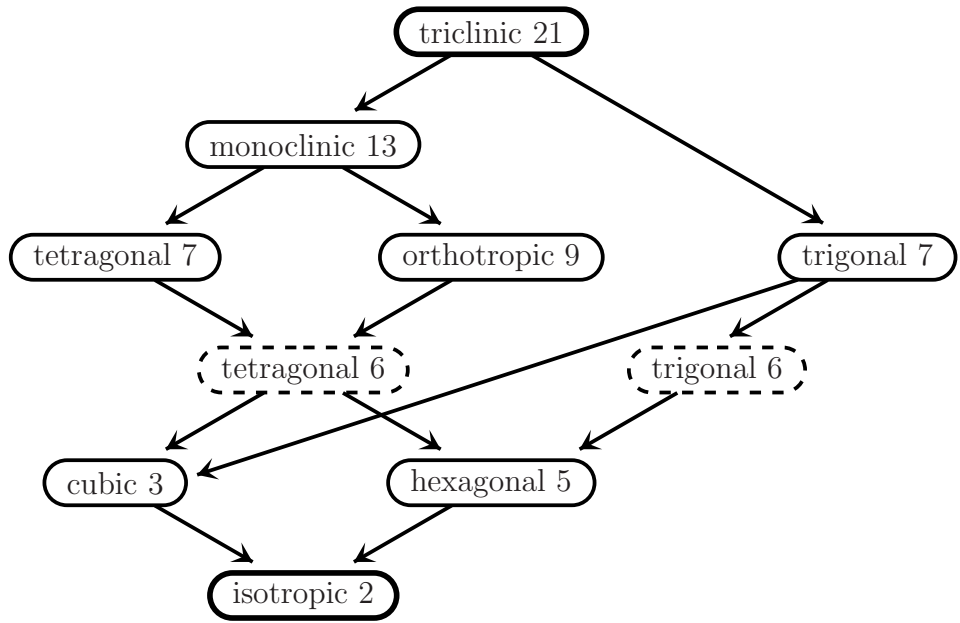
	

%% file: closest_anisotropic.bbl
\begin{thebibliography}{10}

\bibitem{Moakher06}
M.~Moakher.
\newblock On the averaging of symmetric positive-definite tensors.
\newblock {\em J. Elasticity}, 82:273--296, 2006.

\bibitem{Arsigny:MICCAI:05}
V.~Arsigny, P.~Fillard, X.~Pennec, and N.~Ayache.
\newblock Fast and simple calculus on tensors in the {L}og-{E}uclidean
  framework.
\newblock In J.~Duncan and G.~Gerig, editors, {\em Proc. 8th Int. Conf. on
  Medical Image Computing and Computer-Assisted Intervention - MICCAI 2005,
  Part I}, volume 3749 of {\em LNCS}, pages 115--122, Palm Springs, CA, 2005.
  Springer Verlag.

\bibitem{Gazis63}
D.~C. Gazis, I.~Tadjbakhsh, and R.~A. Toupin.
\newblock The elastic tensor of given symmetry nearest to an anisotropic
  elastic tensor.
\newblock {\em Acta Cryst.}, 16:917--922, 1963.

\bibitem{fed}
F.~I. Fedorov.
\newblock {\em Theory of Elastic Waves in Crystals}.
\newblock Plenum Press, New York, 1968.

\bibitem{Norris05g}
A.~N. Norris.
\newblock Elastic moduli approximation of higher symmetry for the acoustical
  properties of an anisotropic material.
\newblock {\em J. Acoust. Soc. Am.}, 119:2114--2121, 2006.

\bibitem{FV96}
S.~Forte and M.~Vianello.
\newblock Symmetry classes for elasticity tensors.
\newblock {\em J. Elasticity}, 43:81--108, 1996.

\bibitem{fgm}
M.~Francois, G.~Geymonat, and Y.~Berthaud.
\newblock Determination of the symmetries of an experimentally determined
  stiffness tensor, application to acoustic measurements.
\newblock {\em Int. J. Solids Struct.}, 35(31-32):4091--4106, 1998.

\bibitem{Arts91}
R.~J. Arts, K.~Helbig, and P.~N.~J. Rasolofosaon.
\newblock General anisotropic elastic tensors in rocks: approximation,
  invariants and particular directions.
\newblock In {\em Expanded Abstracts, 61st Annual International Meeting,
  Society of Exploration Geophysicists, ST2.4}, pages 1534--1537, Tulsa, OK,
  1991. Society of Exploration Geophysicists.

\bibitem{Helbig95}
K.~Helbig.
\newblock Representation and approximation of elastic tensors.
\newblock In E.~Fjaer, R.~M. Holt, and J.~S. Rathore, editors, {\em Seismic
  Anisotropy}, pages 37--75, Tulsa, OK, 1996. Society of Exploration
  Geophysicists.

\bibitem{Cavallini99}
F.~Cavallini.
\newblock The best isotropic approximation of an anisotropic {H}ooke's law.
\newblock {\em Bollettino di {G}eofisica {T}eorica e {A}pplicata}, 40:1--18,
  1999.

\bibitem{Gangi00}
A.~F. Gangi.
\newblock Fourth-order elastic-moduli tensors by inspection.
\newblock In L.~Ikelle and A.~F. Gangi, editors, {\em Anisotropy 2000:
  Fractures, converted waves and case studies. Proc. 9th International Workshop
  on Seismic Anisotropy (9IWSA)}, pages 1--10, Tulsa, OK, 2000. Society of
  Exploration Geophysicists.

\bibitem{ry}
J.~Rychlewski.
\newblock On {H}ooke's law.
\newblock {\em Prikl. Mat. Mekh.}, 48:303--314, 1984.

\bibitem{cowin92}
S.~C. Cowin and M.~M. Mehrabadi.
\newblock The structure of the linear anisotropic elastic symmetries.
\newblock {\em J. Mech. Phys. Solids}, 40:1459--1471, 1992.

\bibitem{Browaeys04}
J.~T. Browaeys and S.~Chevrot.
\newblock Decomposition of the elastic tensor and geophysical applications.
\newblock {\em Geophys. J. Int.}, 159:667--678, 2004.

\bibitem{Dellinger05}
J.~Dellinger.
\newblock Computing the optimal transversely isotropic approximation of a
  general elastic tensor.
\newblock {\em Geophysics}, 70:I1--I10, 2005.

\bibitem{Dellinger98}
J.~Dellinger, D.~Vasicek, and C.~Sondergeld.
\newblock Kelvin notation for stabilizing elastic-constant inversion.
\newblock {\em Revue de L'institut Francais du P\'etrole}, 53:709--719, 1998.

\bibitem{Norris05f}
A.~N. Norris.
\newblock The isotropic material closest to a given anisotropic material.
\newblock {\em J. Mech. Materials Struct.}, 1:223--238, 2006.

\bibitem{CVC}
P.~Chadwick, M.~Vianello, and S.~C. Cowin.
\newblock A new proof that the number of linear elastic symmetries is eight.
\newblock {\em J. Mech. Phys. Solids}, 49:2471--2492, 2001.

\bibitem{walpole84}
L.~J. Walpole.
\newblock Fourth rank tensors of the thirty-two crystal classes: multiplication
  tables.
\newblock {\em Proc. R. Soc. A}, A391:149--179, 1984.

\bibitem{c3}
M.~M. Mehrabadi and S.~C. Cowin.
\newblock Eigentensors of linear anisotropic elastic materials.
\newblock {\em Q. J. Mech. Appl. Math.}, 43:15--41, 1990.

\bibitem{Musgrave}
M.~J.~P. Musgrave.
\newblock {\em Crystal Acoustics}.
\newblock Acoustical Society of America, New York, 2003.

\bibitem{lang98}
Serge Lang.
\newblock {\em Fundamentals of Differential Geometry}.
\newblock Springer-Verlag, New York, 1998.

\bibitem{bhatia}
R.~Bhatia.
\newblock {\em Matrix Analysis}.
\newblock Springer, New York, 1997.

\bibitem{Moakher05b}
M.~Moakher.
\newblock A differential geometric approach to the geometric mean of symmetric
  positive-definite matrices.
\newblock {\em SIAM J. Matrix Anal. Appl.}, 26:735--747, 2005.

\bibitem{kelvin}
W.~Thomson.
\newblock Elements of a mathematical theory of elasticity.
\newblock {\em Phil. Trans. R. Soc. Lond.}, 146:481--498, 1856.

\bibitem{sut}
J.~Sutcliffe.
\newblock Spectral decomposition of the elasticity tensor.
\newblock {\em J. Appl. Mech. ASME}, 59:762--773, 1993.

\bibitem{TS}
P.~S. Theocaris and D.~P. Sokolis.
\newblock Spectral decomposition of the compliance fourth-rank tensor for
  orthotropic materials.
\newblock {\em Arch. Appl. Mech.}, 70:289--306, 2000.

\bibitem{Kunin}
I.~A. Kunin.
\newblock {\em Elastic {M}edia with {M}icrostructure}.
\newblock Springer-Verlag, Berlin, 1982.

\bibitem{sur}
Y.~Surrel.
\newblock A new description of the tensors of elasticity based upon irreducible
  representations.
\newblock {\em Eur. J. Mech. A/Solids}, 12:219--235, 1993.

\bibitem{FV}
S.~Forte and M.~Vianello.
\newblock Functional bases for transversely isotropic and transversely
  hemitropic invariants of elasticity tensors.
\newblock {\em Q. J. Mech. Appl. Math.}, 51:543--552, 1998.

\bibitem{Xiao95}
H.~Xiao.
\newblock Invariant characteristic representations for classical and micropolar
  anisotropic elasticity tensors.
\newblock {\em J. Elasticity}, 40:239 -- 265, 1995.


\bibitem{Xiao98}
H.~Xiao.
\newblock On symmetries and anisotropies of classical and micropolar linear
  elasticities:a new method based upon a complex vector basis and some
  systematic results.
\newblock {\em J. Elasticity}, 49:129--162, 1998.

\bibitem{Bona04}
A.~B\'ona, I.~Bucataru, and A.~Slawinski.
\newblock Characterization of elasticity-tensor symmetries using {SU}(2).
\newblock {\em J. Elasticity}, 75:267--289, 2004.

\bibitem{Bona04b}
A.~B\'ona, I.~Bucataru, and A.~Slawinski.
\newblock Material symmetries of elasticity tensors.
\newblock {\em Q. J. Mech. Appl. Math.}, 57:583--598, 2004.

\bibitem{cowin95}
S.~C. Cowin and M.~M. Mehrabadi.
\newblock Anisotropic symmetries of linear elasticity.
\newblock {\em Appl. Mech. Rev.}, 48:247--285, 1995.

\bibitem{backus}
G.~Backus.
\newblock A geometrical picture of anisotropic elastic tensors.
\newblock {\em Rev. Geophys. Space Phys.}, 8:633--671, 1970.

\bibitem{cowin89}
S.~C. Cowin.
\newblock Properties of the anisotropic elasticity tensor.
\newblock {\em Q. J. Mech. Appl. Math.}, 42:249--266, 1989.

\bibitem{baerheim}
R.~Baerheim.
\newblock Harmonic decomposition of the anisotropic elasticity tensor.
\newblock {\em Q. J. Mech. Appl. Math.}, 46:391--418, 1993.

\bibitem{zz}
W.~N. Zou and Q.~S. Zheng.
\newblock Maxwell's multipole representation of traceless symmetric tensors and
  its application to functions of high-order tensors.
\newblock {\em Proc. R. Soc. A}, A459:527--538, 2003.

\bibitem{bh}
R.~Baerheim and K.~Helbig.
\newblock Decomposition of the anisotropic elastic tensor in base tensors.
\newblock {\em Canadian J. Explor. Geophys.}, 29:41--50, 1993.

\bibitem{Smith57}
G.~F. Smith and R.~S. Rivlin.
\newblock Stress deformation relations for anisotropic solids.
\newblock {\em Arch. Rat. Mech. Anal.}, 1:107--112, 1957.

\bibitem{Tu68}
Y-O Tu.
\newblock The decomposition of an anisotropic elastic tensor.
\newblock {\em Acta Cryst.}, A24:273--282, 1968.

\bibitem{Spencer71}
A.~J.~M. Spencer.
\newblock Theory of invariants.
\newblock In A.~C. Eringen, editor, {\em Continuum Physics}, volume~1, New
  York, 1971. Academic Press.

\bibitem{Betten81}
J.~Betten.
\newblock Integrity basis for a second-order and a fourth-order tensor.
\newblock {\em Int. J. Math. Mathematical Sci.}, 5:87--96, 1981.

\bibitem{Xiao98b}
H.~Xiao.
\newblock On anisotropic invariants of a symmetric tensor: crystal classes,
  quasi-crystalclasses and others.
\newblock {\em Proc. R. Soc. A}, 454:1217--1240, 1998.

\bibitem{er}
D.~Elata and M.~B. Rubin.
\newblock A new representation for the strain energy of anisotropic elastic
  materialswith application to damage evolution in brittle materials.
\newblock {\em Mech. Materials}, 19:171--192, 1995.

\bibitem{fr}
M.~Francois, Y.~Berthaud, and G.~Geymonat.
\newblock Une nouvelle analyse des sym{\'e}tries d'un mat{\'e}riau
  {\'e}lastique anisotrope. exempled'utilisation {\`a} partir de mesures
  ultrasonores.
\newblock {\em C. R. Acad. Sci. Paris}, II/322:87--94, 1996.

\bibitem{cowin87}
S.~C. Cowin and M.~M. Mehrabadi.
\newblock On the identification of material symmetry for anisotropic elastic
  materials.
\newblock {\em Q. J. Mech. Appl. Math.}, 40:451--476, 1987.

\end{thebibliography}
